\documentclass[twocolumn,times]{aastex62}
\usepackage{url}
\usepackage{enumitem}
\usepackage{amsmath,amssymb}
\usepackage{booktabs}

\usepackage{natbib}
\usepackage{cleveref}
\hypersetup{breaklinks,colorlinks,citecolor=blue,linkcolor=blue}

\newcommand{\kms}{\ensuremath{\rm{km}\, \rm{s}^{-1}}}
\newcommand{\peryr}{\ensuremath{\rm{yr}^{-1}}}
\newcommand{\Msun}{\ensuremath{\rm{M}_{\odot}}}
\newcommand{\Mstar}{\ensuremath{\rm{M}_{*}}}
\newcommand{\Mmol}{\ensuremath{\rm{M}_{\rm{mol}}}}

\newcommand{\hd}{\hphantom{0}}

\newcommand{\Lref}{L_\mathrm{850\mu m,rest}}
\newcommand{\aco}{\alpha_\mathrm{CO}}
\usepackage[caption=false]{subfig}

\accepted{May 29, 2019}
\submitjournal{ApJ}

\shorttitle{The Molecular Gas Reservoirs of $z\sim 2$ Galaxies}
\shortauthors{Kaasinen et al.}

\graphicspath{.}

\begin{document}


\title{The Molecular Gas Reservoirs of $z\sim 2$ Galaxies: A comparison of CO(1-0) and dust-based molecular gas masses}


\author{M. Kaasinen}
\affil{Max-Planck-Institut f\"{u}r Astronomie, K\"{o}nigstuhl 17, D-69117 Heidelberg, Germany}
\affil{Universit\"{a}t Heidelberg, Zentrum f\"{u}r Astronomie, Institut f\"{u}r Theoretische Astrophysik, Albert-Ueberle-Straße 2, D-69120 Heidelberg, Germany}
\affil{ARC Centre of Excellence for All Sky Astrophysics in 3 Dimensions (ASTRO 3D)}
\author{N. Scoville}
\affil{Cahill Center for Astrophysics, California Institute of Technology, 1216 East California Boulevard, Pasadena, CA 91125}
\author{F. Walter}
\affil{Max-Planck-Institut f\"{u}r Astronomie, K\"{o}nigstuhl 17, D-69117 Heidelberg, Germany}
\affil{National Radio Astronomy Observatory, Pete V. Domenici Array Science Center, P.O. Box O, Socorro, NM 87801, USA}
\author{E. Da Cunha}
\affil{Research School of Astronomy and Astrophysics, Australian National University, Canberra, ACT 2611, Australia}
\author{G. Popping}
\affil{Max-Planck-Institut f\"{u}r Astronomie, K\"{o}nigstuhl 17, D-69117 Heidelberg, Germany}
\author{R. Pavesi}
\affil{Cornell University, Space Sciences Building, Ithaca, NY 14853, USA}
\author{B. Darvish}
\affil{Cahill Center for Astrophysics, California Institute of Technology, 1216 East California Boulevard, Pasadena, CA 91125}
\author{C. M. Casey}
\affil{University of Texas at Austin, 2515 Speedway Blvd Stop C1400, Austin, TX 78712, USA}
\author{D. A. Riechers}
\affil{Cornell University, Space Sciences Building, Ithaca, NY 14853, USA}
\affil{Max-Planck-Institut f\"{u}r Astronomie, K\"{o}nigstuhl 17, D-69117 Heidelberg, Germany}
\author{S. Glover}
\affil{Universit\"{a}t Heidelberg, Zentrum f\"{u}r Astronomie, Institut f\"{u}r Theoretische Astrophysik, Albert-Ueberle-Straße 2, D-69120 Heidelberg, Germany}



\begin{abstract}
	We test the use of long-wavelength dust continuum emission as a molecular gas tracer at high redshift, via a unique sample of 12, $z\sim 2$ galaxies with observations of both the dust continuum and  CO(1-0) line emission (obtained with the Atacama Large Millimeter Array and Karl G. Jansky Very Large Array, respectively).  Our work is motivated by recent, high redshift studies that measure molecular gas masses (\Mmol) via a calibration of the rest-frame $850\mu$m luminosity ($\Lref$) against the CO(1-0)-derived \Mmol\ of star-forming galaxies. We hereby test whether this method is valid for the types of high-redshift, star-forming galaxies to which it has been applied. We recover a clear correlation between the rest-frame $850\mu$m luminosity, inferred from the single-band, long-wavelength flux, and the CO(1-0) line luminosity, consistent with the samples used to perform the $850\mu$m calibration. The molecular gas masses, derived from $\Lref$, agree to within a factor of two with those derived from CO(1-0). We show that this factor of two uncertainty can arise from the values of the dust emissivity index and temperature that need to be assumed in order to extrapolate from the observed frequency to the rest-frame at 850$\mathrm{\mu m}$. The extrapolation to 850$\mathrm{\mu m}$ therefore has a smaller effect on the accuracy of \Mmol\ derived via single-band dust-continuum observations than the assumed CO(1-0)-to-\Mmol\ conversion factor. We therefore conclude that single-band observations of long-wavelength dust emission can be used to reliably constrain the molecular gas masses of massive, star-forming galaxies at $z\gtrsim2$.   
\end{abstract}


\section{Introduction}
	\label{sec:intro}


	Most star formation is observed to occur within the molecular phase of the interstellar medium (ISM), with observations demonstrating a strong correlation between the surface density of the star formation rate (SFR) and that of the molecular gas \citep[e.g.][]{2002ApJ...569..157W,2008AJ....136.2846B,2011ApJ...730L..13B,2008AJ....136.2782L}. On a cosmic scale, the SFR density peaked at $z\sim 2$, and has since declined exponentially \citep[see ][and references therein]{2014ARA&A..52..415M}. This decline can largely be attributed to the $10-100$ fold decrease in SFR of the dominant population of star-forming galaxies, i.e. those occupying the Main Sequence (MS, linear relation between the SFR, and stellar mass), from $z\sim 2$ to the present day \citep[e.g.][]{2007ApJ...670..156D,2011A&A...533A.119E,2014ApJS..214...15S,2014ApJ...795..104W}. Understanding the physical processes driving the decline in SFR requires the accurate measurement of the molecular gas content out to high redshift. But, measuring the molecular gas masses (\Mmol) of galaxies at $z>1$ remains a challenge.

	The emission from H$_2$, the most abundant component of the cold, dense phase of the ISM relevant to star formation, cannot be observed directly. Thus, \Mmol\ is usually measured via the emission from other, less abundant components of the ISM. Studies of local galaxies typically rely on the ground transition of CO ($J=1-0$) to measure \Mmol, converting the CO(1-0) line luminosity to a molecular gas mass via the application of an empirically-derived, CO-to-\Mmol\ conversion factor \citep[see][for a review]{2013ARA&A..51..207B}. However, the majority of high-redshift ($z>1$) observations are of higher-J CO transitions, which are brighter and more readily detected by millimeter interferometers operating in the 1-3mm atmospheric windows, i.e. the Atacama Large Millimeter Array (ALMA) and the NOrthern Extended Millimeter Array (NOEMA). The use of higher-J, CO lines requires an additional correction for the (a priori unknown) excitation of the gas, increasing the uncertainty of the derived gas masses by at least a factor of two (see e.g. the variation in the CO(3-2)-to-CO(1-0) line ratio in \citealt{2009ApJ...695.1537I,2010ApJ...724.1336M,2010ApJ...723.1139H,2011MNRAS.412.1913I,2011ApJ...739L..32R,2012MNRAS.426.2601P,2016ApJ...827...18S} and other issues discussed in \citealt{2013ARA&A..51..105C}). 


	To combat the difficulty of relying on faint and/or high excitation lines at high redshift, \cite{Scoville2012}, suggested the use of the long-wavelength Rayleigh–Jeans (RJ) tail of dust emission. Methods of measuring \Mmol\ from long-wavelength dust continuum emission were subsequently developed via empirical calibrations of rest-frame infrared (IR) luminosities against CO-derived molecular gas masses \citep{2012ApJ...761..168E,2013MNRAS.436..479B,2015ApJ...799...96G,2014ApJ...783...84S,2016ApJ...820...83S,2017ApJ...837..150S,2017MNRAS.468L.103H}. The rest-frame $850\mathrm{\mu m}$ luminosity ($\Lref$) was found to exhibit a particularly tight correlation with the CO(1-0) line luminosity \citep[e.g. ][]{2016ApJ...820...83S,2017MNRAS.468L.103H}. For brevity, we henceforth refer to the method calibrated against $\Lref$ as the RJ method. 

	Applying the RJ method requires the conversion of the observed emission to $\Lref$, either by fitting the IR portion of the spectral energy distribution (SED) \citep[e.g.][]{2017MNRAS.468L.103H}, or, extrapolating from a single-band measurement in the RJ tail, assuming the dust opacity coefficient and mean temperature of dust contributing to the RJ tail \citep[e.g.][]{2016ApJ...820...83S}. The single-band RJ method is particularly convenient at high redshift, where the required observations take of the order of a minute per source with ALMA \citep{2016ApJ...820...83S}, in contrast to the multiple hours required to observe CO emission \citep[e.g.][]{2013ApJ...768...74T}.  

	The variety of methods used to measure \Mmol\ complicate efforts to link the molecular gas contents and SFRs of galaxies. Like the SFR, the molecular gas mass fractions of star-forming galaxies appear to have declined since $z\sim 2$ \citep[e.g.][]{2010ApJ...713..686D,2010ApJ...724L.153R,2010MNRAS.407.2091G,2015ApJ...800...20G,2010Natur.463..781T,2018ApJ...853..179T,2015A&A...577A..50D,2016ApJ...833...70D,2016ApJ...833..112S,2017ApJ...837..150S,2018ApJ...860..111D}. But, there remains some tension between the gas scaling relations derived in these studies, especially regarding the contribution of the star formation efficiency to the declining SFRs of MS galaxies. 

	Most studies find a slight decline in the star formation efficiency of MS galaxies with redshift \citep{2010Natur.463..781T,2010MNRAS.407.2091G,2015ApJ...800...20G,2016ApJ...820...83S,2017ApJ...837..150S,2018ApJ...853..179T,2018ApJ...860..111D}, but the exact scaling varies. \cite{2015ApJ...800...20G} find that the increased SFRs of high-redshift (high-z), MS galaxies can be mainly attributed to higher gas mass fractions, whereas \cite{2016ApJ...820...83S} and \cite{2018ApJ...860..111D} also find a significantly more efficient mode of star formation in high-z galaxies ($5\times$ shorter gas depletion times). A similar inconsistency exists regarding the difference between MS and starburst galaxies. Whereas \cite{2016ApJ...820...83S} and \cite{2018ApJ...860..111D} conclude that the high SFRs of starburst galaxies relative to the MS are driven by both the higher molecular gas masses and star formation efficiencies of the former, \cite{2010ApJ...714L.118D} and \cite{2010MNRAS.407.2091G,2015ApJ...800...20G} mainly attribute the offset from the MS to higher star formation efficiencies. 

	\begin{table*}
	\begin{center}
	\caption{Source Information\label{tab:source_info}}
	\begin{tabular}{lccllccc}
	\toprule
	Galaxy ID & \multicolumn{2}{c}{Position (J2000)} &  z$_\mathrm{COSMOS}$\tablenotemark{a} & z$_\mathrm{flag}$\tablenotemark{b} & z$_\mathrm{CO(1-0)}$  \\
	\cmidrule{2-3}
			  & R.A. 			& Decl. 			 &     					& 			 & 					 \\
	\midrule
	\hd 1  		& $10^\mathrm{h}\,00^\mathrm{m}\,35.^\mathrm{s}29 $ 		& $2\arcdeg43\arcmin53.\arcsec2$ & 2.38 \hd 	& photometric		& 2.607 	 	\\
  		& 		&  & 2.608 \hd 	& 4 [MOSFIRE] \tablenotemark{c}			& 2.607 	 	\\
	\hd 2$^*$  	& $10^\mathrm{h}\,00^\mathrm{m}\,08.^\mathrm{s}91 $ 		& $2\arcdeg40\arcmin10.\arcsec3$ & 2.284	& 2 [MOSFIRE] \tablenotemark{c}			& -			 	\\
	 	& $ $ 		&  & 1.847 	& photometric 				& -			 	\\

	\hd 3   	& \hd $9^\mathrm{h}\,58^\mathrm{m}\,40.^\mathrm{s}28$ 	& $2\arcdeg05\arcmin14.\arcsec7$ & 2.416 	& 3	[DEIMOS] 	& 2.414		  	\\
	\hd 4   	& $10^\mathrm{h}\,00^\mathrm{m}\,31.^\mathrm{s}82 $ 		& $2\arcdeg12\arcmin43.\arcsec2$ & 2.104 	& 4 [MOSDEF] \tablenotemark{c}	& 2.104		 		\\
	\hd 5   	& $10^\mathrm{h}\,02^\mathrm{m}\,24.^\mathrm{s}77 $ 		& $2\arcdeg32\arcmin11.\arcsec6$ & 2.287 	& 4 [MOSFIRE]	& 2.287		 	\\
	\hd 6$^*$   	& $10^\mathrm{h}\,02^\mathrm{m}\,32.^\mathrm{s}09 $ 		& $2\arcdeg34\arcmin41.\arcsec4$ & 2.68 	&  photometric 	& - 		 	\\ 
	  	& 		&  & - 	&  [MOSDEF]\tablenotemark{c} 	& - 		 	\\ 
	\hd 7   	& $10^\mathrm{h}\,01^\mathrm{m}\,03.^\mathrm{s}55 $ 		& $1\arcdeg48\arcmin10.\arcsec6$ & 2.240 	& 4 [MOSFIRE]	& 2.240		 	 \\
	\hd 8   	& \hd $9^\mathrm{h}\,58^\mathrm{m}\,37.^\mathrm{s}34$ 	& $2\arcdeg42\arcmin58.\arcsec5$ & 2.11\hd	& photometric 			& 2.173 	 	\\
	\hd 9$^*$  	& $10^\mathrm{h}\,00^\mathrm{m}\,03.^\mathrm{s}89 $ 		& $2\arcdeg47\arcmin32.\arcsec4$ & 1.76 \hd 	& photometric			& -			 	\\
	  			& 		&  & 1.958 \hd 	& 4 [MOSFIRE] 			& -			 	\\
	10 			& $10^\mathrm{h}\,01^\mathrm{m}\,19.^\mathrm{s}52 $ 		& $2\arcdeg09\arcmin44.\arcsec7$ & 2.934 	& 3 [zDEEP]		& 2.934		 	\\
	11 			& \hd $9^\mathrm{h}\,59^\mathrm{m}\,57.^\mathrm{s}35$ 	& $2\arcdeg03\arcmin11.\arcsec3$ & 1.942 	& 4 [MOSFIRE]\tablenotemark{c}	& 1.941		 	\\
	12 			& $10^\mathrm{h}\,01^\mathrm{m}\,16.^\mathrm{s}28 $ 		& $2\arcdeg42\arcmin59.\arcsec4$ & 2.340 	& 4 [MOSFIRE] \tablenotemark{c}	& -			 	\\
	13 			& $10^\mathrm{h}\,01^\mathrm{m}\,58.^\mathrm{s}96 $ 		& $2\arcdeg06\arcmin58.\arcsec6$ & 2.400 	& 4 [MOSFIRE]	& 2.400		 	\\
	14$^*$ 			& $10^\mathrm{h}\,01^\mathrm{m}\,01.^\mathrm{s}24 $ 		& $2\arcdeg28\arcmin00.\arcsec6$ & 2.264 	& 2 [FMOS]		& -			 	\\
	15 			& $10^\mathrm{h}\,00^\mathrm{m}\,56.^\mathrm{s}68 $ 		& $2\arcdeg52\arcmin22.\arcsec5$ & 1.654 	& 3 [FMOS] 		& -			 	\\
	16 			& \hd $9^\mathrm{h}\,59^\mathrm{m}\,04.^\mathrm{s}39$ & $2\arcdeg13\arcmin12.\arcsec5$ & 1.779 	& 2 [zDEEP]		& 1.780		 	\\
	\bottomrule
	\end{tabular}
	\end{center}
	\vspace{-0.2cm}
	\tablenotetext{a}{Redshift provided in the COSMOS catalogue, as well as, spectroscopic redshifts obtained from our 2019 MOSFIRE observations (sources 1, 2, 4, 9, 11 and 12).}
	\tablenotetext{b}{Flag assigned to the quality of the spectroscopic redshift of column 4, where 4 is completely secure, 3 is secure but the classifier(s) recognise at least a remote possibility for error and 2 indicates that a significant possibility remains that the redshift is incorrect \citep{2007ApJS..172...70L}.  We reclassify sources 14 and 15 after inspecting the spectra ourselves. }
	\tablenotetext{c}{ We use new MOSFIRE spectroscopy, not yet included in the COSMOS catalogue, to analyse the data for sources 1, 2, 4, 6, 9, 11 and 12. No rest-frame optical emission lines were observed for source 6.}
	\tablecomments{Sources marked with $^{*}$ are deemed to have unreliable redshift estimates (insufficient to infer CO(1-0) upper limits or SED-based properties) and are therefore removed from the sample. Note that this includes sources 2 and 9, for which the VLA observations were optimised for the catalogued redshifts, which our 2019 MOSFIRE observations show to be incorrect. Hence, the observations do not encompass the CO(1-0) line for source 2, whereas for source 9 the CO line is expected on the edge of the observed frequency range, where the noise is the greatest.}
	\end{table*}

	The extent to which the differences between molecular gas scaling relations are affected by the assumptions used to infer molecular gas masses, from CO vs dust continuum emission, has not yet been quantified in detail. Recent studies have begun to address the need for consistency between dust and CO-based measurements. For example, \cite{2017MNRAS.468L.103H} compared the CO(1-0) line luminosities and $\Lref$ of local star-forming galaxies, whereas \cite{2018MNRAS.tmpL..76L} and \cite{2018arXiv180503649P} compared the RJ-based \Mmol\ with the ``true'' \Mmol\ of a set of simulated star-forming galaxies. Although these studies have investigated the use of $\Lref$ as a gas mass tracer, they do not confirm whether the gas masses, already determined for $\gtrsim 600$ high-z galaxies \citep{2016ApJ...820...83S,2017ApJ...837..150S,2016ApJ...833..112S,Miettinen_2017,2018ApJ...860..111D}, are equivalent to what would be derived using CO(1-0). 

	To compare the molecular gas masses, derived from the single-band RJ continuum and CO(1-0), we have assembled a unique sample of 16 unlensed, $z\sim 2$, star-forming galaxies with CO(1-0) observations from the VLA, and, dust continuum measurements from the ALMA. This sample represents a significant increase in the number of CO(1-0) detections at high redshift with $\sim 50$ sources at $z>1$ having been detected previously \cite[e.g. supplementary table of ][]{2013ARA&A..51..105C}, out of which $\lesssim 20$ sources are unlensed \citep[e.g][]{2013MNRAS.430.3465E,2014MNRAS.442..558A,2015ApJ...809..175B,2017MNRAS.467.1222H,2018ApJ...864...49P}. The fact that our sources are unlensed avoids potential complications involving differential lensing. Our sample consists of massive ($>2\times10^{10}\Msun$) galaxies both on and above the MS. We thereby focus on the galaxy population dominating the peak of the cosmic SFR density. 

	This paper is structured as follows. We present our sample and describe the observations and data reduction in Section \ref{sec:sample_and_obs}. In Section \ref{sec:derived_properties} we explain how we derive the molecular gas masses, SFRs and stellar masses. We present our results and discussion in Section \ref{sec:results}, comparing the CO(1-0) line luminosity and $\Lref$, as well as the molecular gas masses derived from these luminosities. Our work is summarised in Section \ref{sec:summary}. Throughout this paper we assume a $\Lambda CDM$ cosmology with $H_0 = 70 \kms \mathrm{Mpc}^{-1}$, $\Omega_M = 0.3$ and $\Omega_\Lambda = 0.7$. All stellar masses and SFRs are based on a \cite{2003PASP..115..763C} IMF. We use a CO-to-\Mmol\ conversion factor of $\aco = 6.5\, \Msun / (\mathrm{K}\, \kms \, \mathrm{pc}^{2})$ throughout this paper.


	\begin{figure*}
			\centering
			\includegraphics[width=0.95\columnwidth, trim={0.5cm 0.4cm 0.5cm 0.4cm},clip]{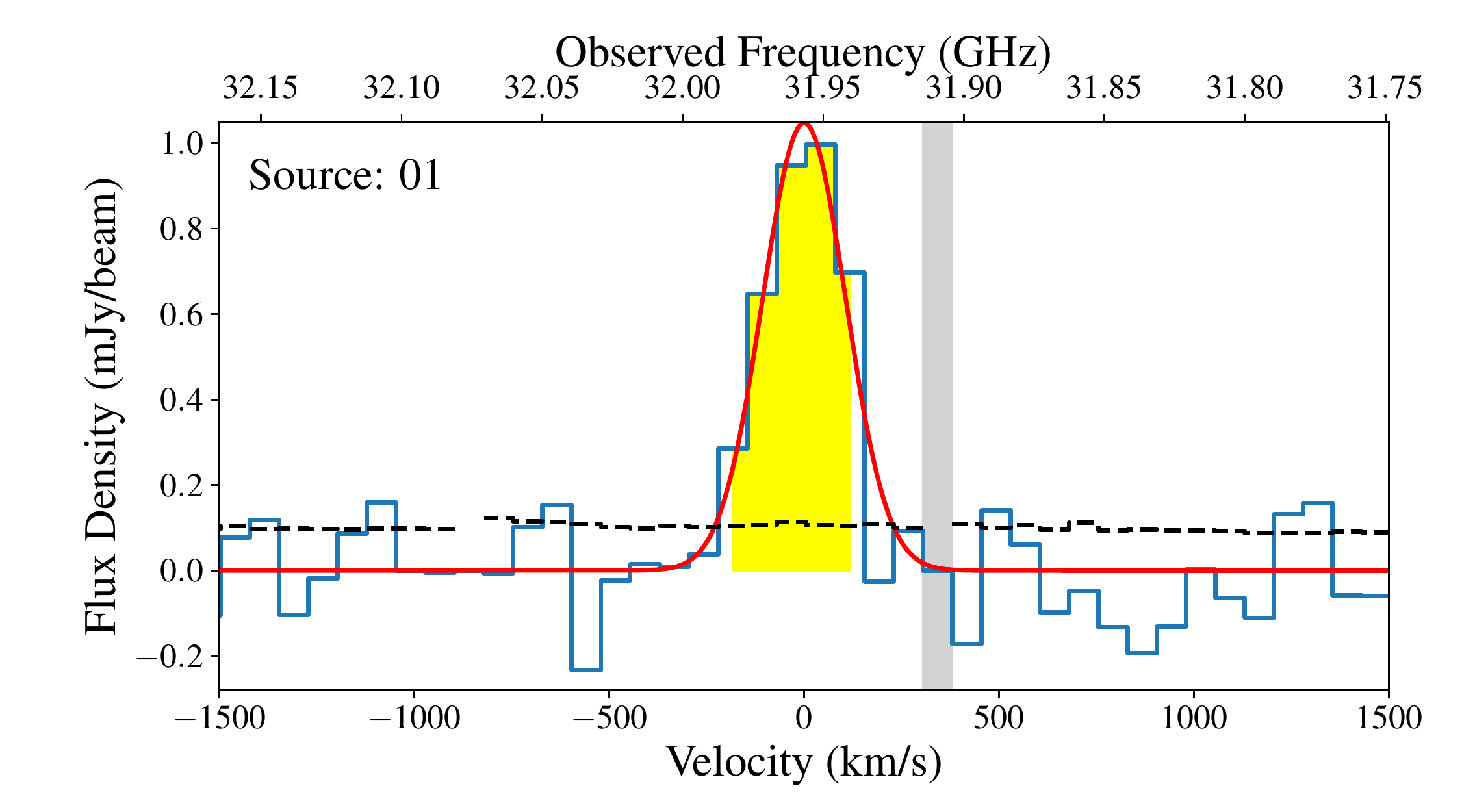}
			~
			\includegraphics[width=0.95\columnwidth, trim={0.5cm 0.4cm 0.5cm 0.4cm},clip]{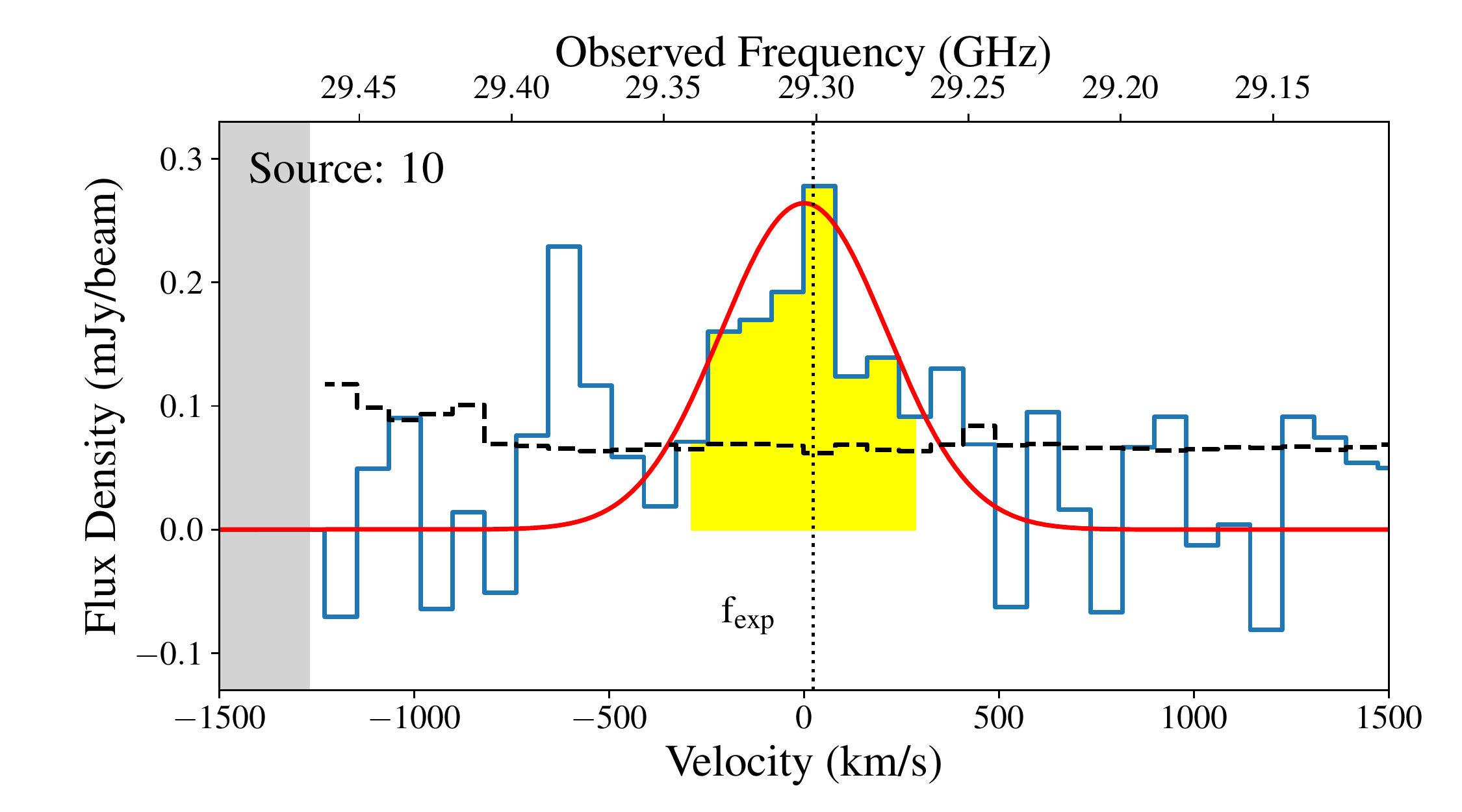}
			\caption{Examples of two CO(1-0) spectra (blue). The remaining spectra and line fits are shown in Figure \ref{fig:spectra}. The Gaussian line fits are shown for comparison (red) with the spectral region used to create the integrated maps in Figure \ref{fig:chmap_comp_ex} shaded in yellow. Flagged channels, not used for the line fits, are shaded in grey. The root-mean-square noise per channel is indicated by the black, dashed histogram. \label{fig:ex_spectra}}
	\end{figure*}

		\begin{figure*}
		\centering
		\includegraphics[width=0.29\textwidth, trim={0 1.2cm 3.cm 0.4cm},clip]{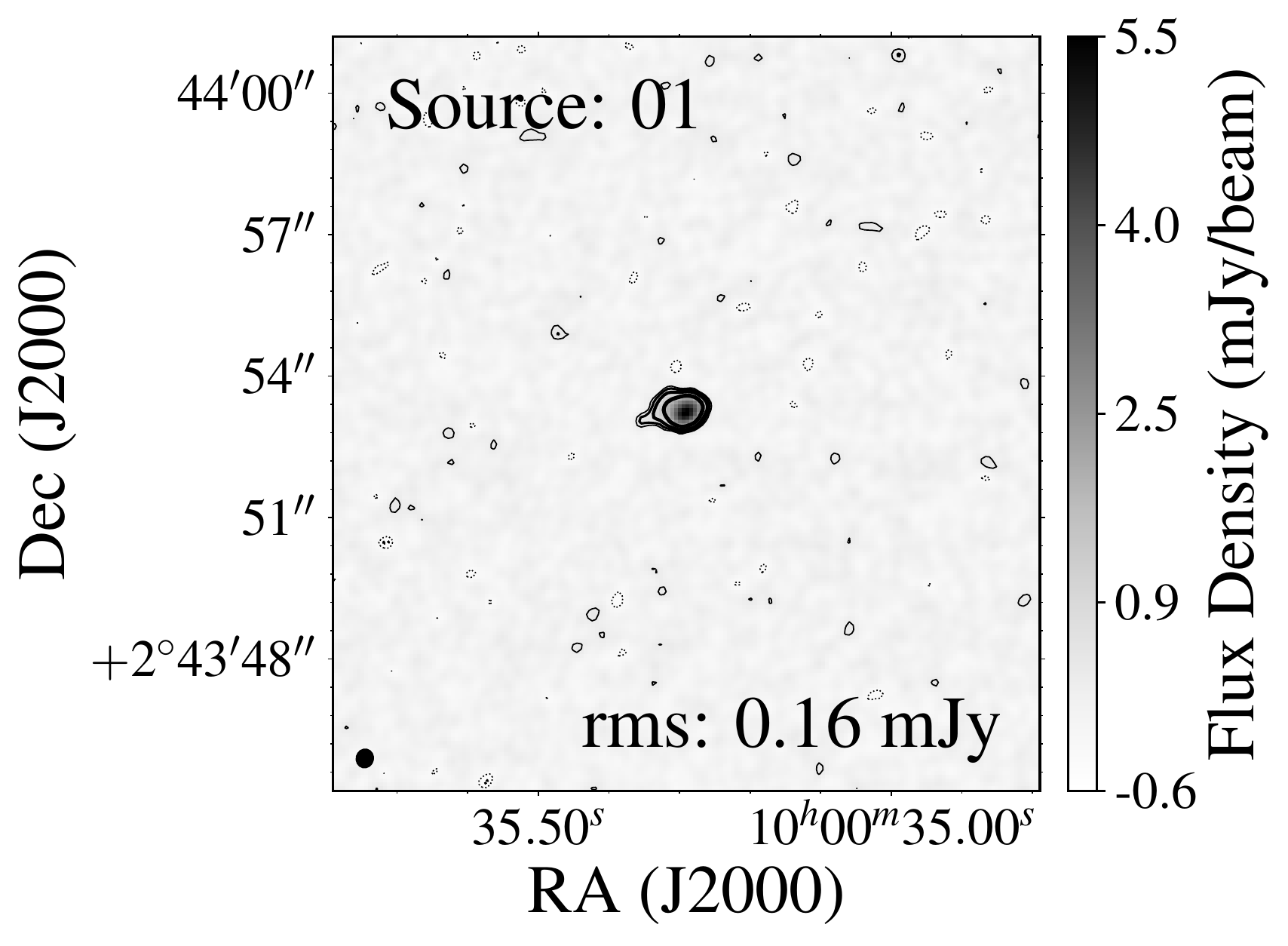}
		~
		\includegraphics[width=0.63\textwidth, trim={2.7cm 0.95cm 2cm 0.1},clip]{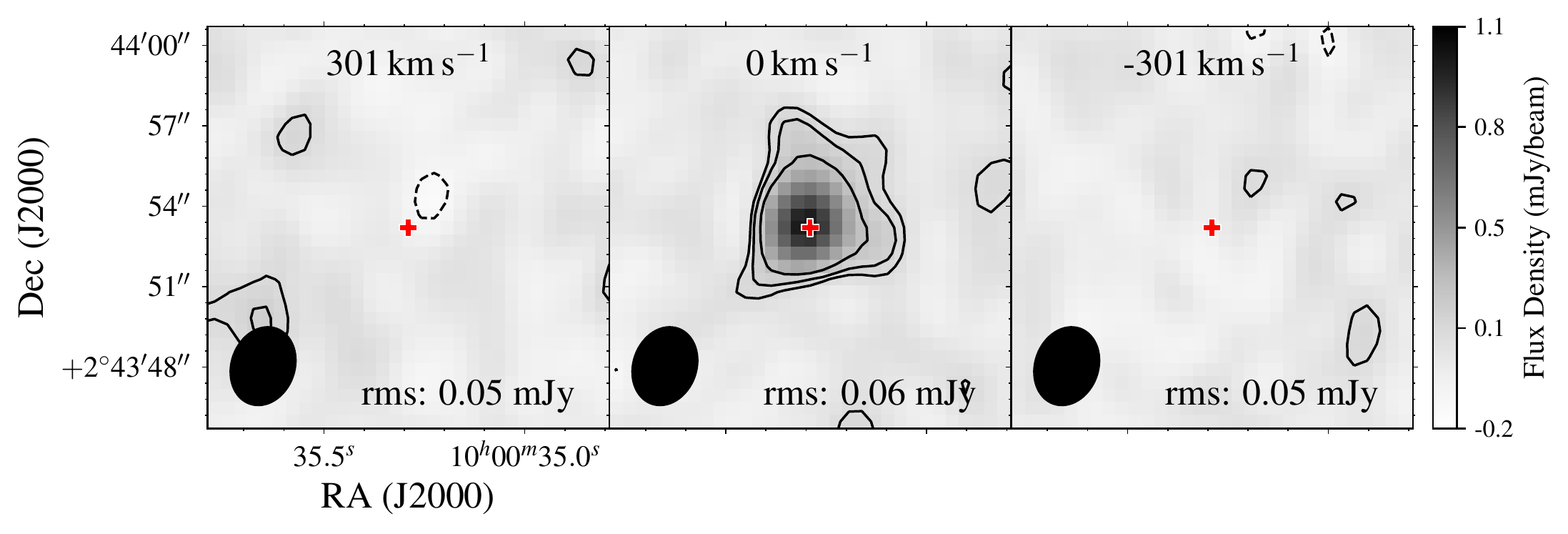}
		\\
		\includegraphics[width=0.29\textwidth, trim={0 -0.4cm 3.cm 0},clip]{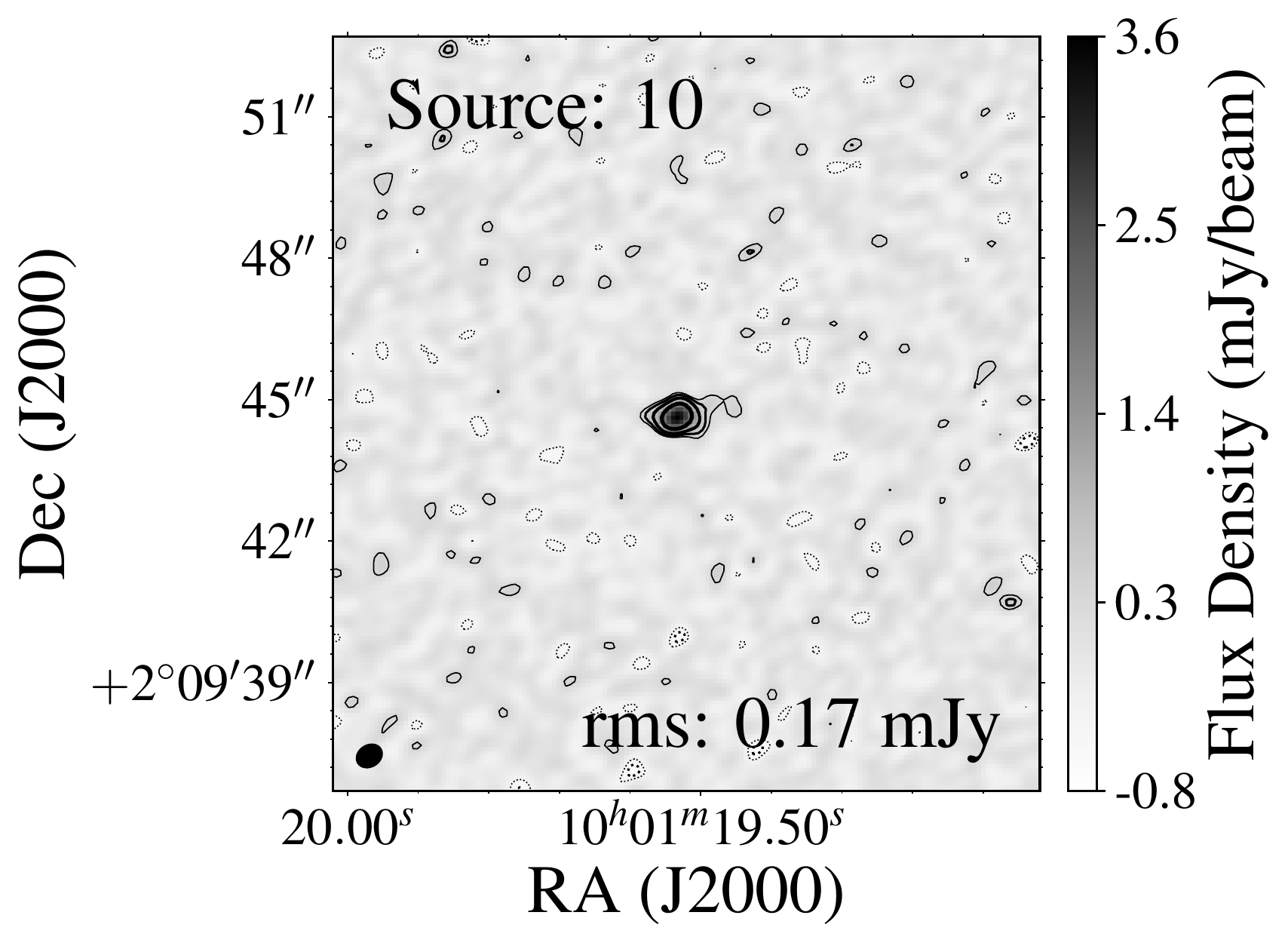}
		~
		\includegraphics[width=0.63\textwidth, trim={2.7cm 0.3 2cm 0},clip]{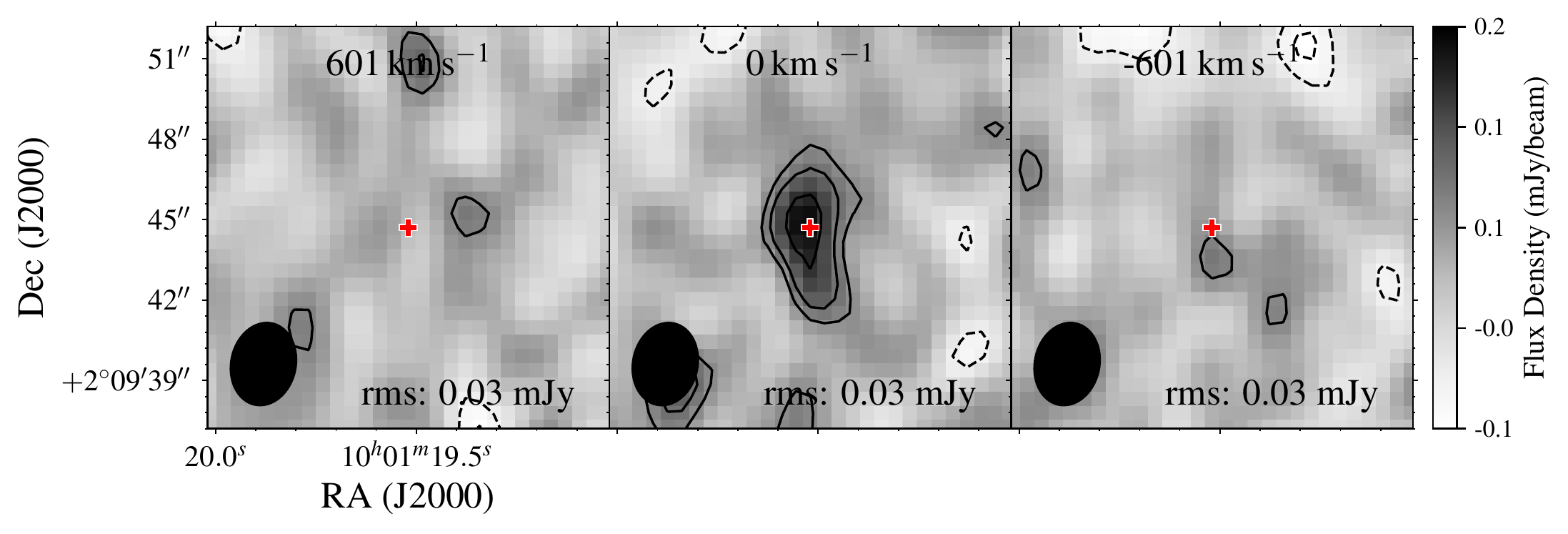}	
		\caption{Example comparison of the ALMA dust continuum (left map) and integrated VLA CO(1-0) maps for sources with CO(1-0) detections. The integrated maps of the remainder of the sample are shown in Figure \ref{fig:chmap_comp} (Appendix). The source number is labeled at the top left of the left hand panel in each row. The rms value is given in the bottom right corner of each map. Left column: ALMA observations at 343.5 GHz. Contours are shown for $\pm 2, 3, 5$ and $10\sigma$ (dashed contours for negative values).  Right columns: Channel maps around the measured (expected) CO(1-0) line. For each source, the central panel of the VLA channel map represents the moment zero map and is centred at the central velocity of the CO spectrum. The velocity width of the integrated maps is chosen to encompass the full source emission (1.2$\times$FWHM of the CO(1-0) line for sources other than 4 and 5). Contours are shown for $\pm 2, 3$ and $5\sigma$ (dashed contours for negative values). The red cross indicates the expected position of the source, at which the CO spectrum was extracted. The color shading indicates the flux density in mJy/beam.\label{fig:chmap_comp_ex}}	
	\end{figure*}

\newpage 

\section{Sample and Observations} 
	\label{sec:sample_and_obs}

	\subsection{Sample Selection} 
		\label{sub:sample}

		Our sample is comprised of 16, massive, star-forming galaxies at $z\sim 2$, with long-wavelength (rest-frame $\sim 250 \mu$m) dust continuum measurements from ALMA \citep[ALMA, ][]{2009IEEEP..97.1463W}. Our sample was selected from the ALMA-detected, IR-bright sample of \cite{2017ApJ...837..150S}. The parent sample of \cite{2017ApJ...837..150S} was chosen using the Herschel-based catalogue of far-IR sources in the COSMOS field \citep{2013ApJ...778..131L,2015ApJ...801...80L}. We therefore have photometric measurements of at least two of the five IR Herschel bands: the $100\mu$m and $160\mu$m bands from PACS \citep{2011A&A...532A..90L} and the $250\mu$m, $350\mu$m, and $500\mu$m bands from SPIRE \citep{2010A&A...518L...3G}, for all 16 sources discussed in this paper. Like the parent sample of \cite{2017ApJ...837..150S}, our sample is restricted to objects with $\Mstar > 2 \times 10^{10}\Msun$ (based on the COSMOS catalogue described in \citealt{2016ApJS..224...24L}). 

		To maximise the chances of detecting CO(1-0) emission, we selected the 16 galaxies with the highest ALMA Band 7 (343.5 GHz) fluxes. Based on the calibration presented in \cite{2014ApJ...783...84S,2016ApJ...820...83S,2017ApJ...837..150S} we expected CO(1-0) detections of $\mathrm{S_{CO}\Delta v} > 100\, \mathrm{mJy}\, \kms $, with the Very Large Array (VLA). Our sample is therefore intentionally biased to the types of high SFR ($\mathrm{SFR}>300\Msun \peryr$) sources to which the RJ method is applied \citep{2016ApJ...820...83S,2017ApJ...837..150S,2016ApJ...833..112S,Miettinen_2017,2018ApJ...860..111D}. 
		We discuss our sample with respect to the MS further in Section \ref{subsub:ms_offset}. Because of our selection criteria, our sample spans a wide redshift range of $1.6<z<2.9$. The coordinates and estimated redshifts of the full sample are provided in Table \ref{tab:source_info}. 



		\begin{table*}
	\begin{center}
	\caption{CO(1-0) and dust continuum data}
	\begin{tabular}{lccccccc}
	\toprule
	Galaxy 	&  \multicolumn{2}{c}{Beam} & PA & $\sigma$ & S$_\mathrm{peak, CO(1-0)} $&  S$_\mathrm{CO(1-0)}\Delta v $ &  S$_\mathrm{343.5 GHz,cont}$  \\
		\cmidrule{2-3}
			 	& 	  max (arcsec) & min (arcsec) & (deg) &  (mJy/beam) & mJy &  (mJy \kms)			  &	(mJy)	 			\\
	\midrule
	\hd 1  	& 2.96 & 2.34 & -19.6  				&	0.10 & 1.05 $\pm$ 0.08 &  279 $\pm$ 34	 	& 13.02 $\pm$ 0.46	  		\\
	\hd 3  	& 3.84 & 2.25 & 167.9  				&	0.15 & 0.50 $\pm$ 0.07 &  272 $\pm$ 64	 	& \hd 9.67  $\pm$ 0.47 	 	\\
	\hd 4  	& 2.75 & 1.98 & \hd \hd \hd -6.35  	&	0.09 & 0.29 $\pm$ 0.06 &  185 $\pm$ 55 		& \hd 8.23 $\pm$ 0.37	 	\\
	\hd 5  	& 2.49 & 2.05 & -179.0 				&	0.16 & 0.55 $\pm$ 0.13 &  158 $\pm$ 55	 	& \hd 6.34 $\pm$ 0.35	 	\\
	\hd 7  	& 2.67 & 1.90 & \hd \hd \hd 0.3 	&	0.20 & 0.79 $\pm$ 0.13 &  236 $\pm$ 59 	 	& \hd 7.96 $\pm$ 0.36	 	\\
	\hd 8  	& 2.50 & 2.23 & \hd -22.7 			&	0.13 & 0.44 $\pm$ 0.05 &  125 $\pm$ 22 	 	& \hd 6.08 $\pm$ 0.36	 	\\
	10 		& 2.54 & 1.19 & \hd -12.3 			&	0.07 & 0.26 $\pm$ 0.04 &  141 $\pm$ 31 	 	& \hd 4.91 $\pm$ 0.37	 	\\
	11 		& 2.25 & 1.77 & \hd \hd 18.7 		&	0.17 & 0.58 $\pm$ 0.07 &  368 $\pm$ 74 	 	& \hd 5.39 $\pm$ 0.47	 	\\
	12 		& 2.68  & 2.19 & 10.8 				&	-	 & - 			   & $\leq$ 80			& \hd 4.44 $\pm$ 0.40	 	\\
	13 		& 2.87 & 2.25 & \hd \hd 23.3 		&	0.15 & 0.74 $\pm$ 0.09 &  150 $\pm$ 28 	 	& \hd 4.29 $\pm$ 0.36	 	\\
	15 		& 2.24 & 1.81 & \hd \hd 16.7 		&	-	 & 	-		 		& $\leq$ 243 		& \hd 3.34 $\pm$ 0.29	 	\\
	16 		& 2.24 & 1.87 & \hd \hd 31.0		&	0.29 & 1.05 $\pm$ 0.16  &  112 $\pm$ 25 	& \hd 3.86 $\pm$ 0.81	 	\\
	\bottomrule
	\end{tabular}
	\label{tab:data}
	\end{center}
	\end{table*}

	\subsection{CO(1-0) Observations and Data Reduction} 
		\label{sub:co_dat}

		The VLA observations analysed here were taken during February and March 2017. Of the 16 galaxies in our sample, four were observed in the Q band ($40 - 50 \mathrm{GHz}$) and 12 in the Ka band ($26.5 - 40 \mathrm{GHz}$). Each target was observed for a total of four hours, including the time spent on bandpass, phase and amplitude calibration sources. 


		The raw VLA data were processed to produce clean images using the Common Astronomy Software Application (CASA)\footnote{\small{\url{https://casa.nrao.edu}}}, version 4.7.2. To calibrate the data, we applied the VLA calibration pipeline\footnote{\small{\url{https://science.nrao.edu/facilities/vla/data-processing/pipeline}}} without Hanning smoothing. We created the initial dirty images using CASA's \texttt{TCLEAN} algorithm. These dirty images are used to visually identify the presence of CO and ensure the source emission is found at the anticipated position. We then create cleaned image cubes for these sources via CASA's \texttt{TCLEAN}, applying a natural weighting scheme, a cleaning threshold of twice the root-mean-square (rms) noise level ($2\sigma$) and a circular mask with a radius of  5'', centred at the position of the source. 
		We optimise the spectral resolution to achieve higher signal-to-noise ratios (S/N). For sources 8 and 16 it was necessary to use the native resolution ($\sim 16 \kms$) to extract the spectra (see Figure \ref{fig:spectra}). We provide the smoothed, extracted spectra of our full sample in Figure \ref{fig:non_detections} (Appendix \ref{sec:App_data}).
		
		We detect CO(1-0) emission in 10 of the 16 sources in our sample. To classify a source as a CO detection we require the peak flux of the spectrum, extracted at the source position, and the peak flux in the moment zero map to be detected at $\geq 3\sigma$. For the CO-detected sources, we derive the CO(1-0) flux, central frequency, and spectroscopic redshift by fitting a single Gaussian to the CO(1-0) line emission via \texttt{Python}'s \texttt{scipy.optimize.curvefit} algorithm. We use a $1/\sigma^2$ weighting scheme, where $\sigma$ is the rms noise per channel, and take the $1\sigma$ errors estimated by our fitting routine as the uncertainties of the measured values. We provide the rms noise of the channel corresponding to the CO(1-0) peak, the peak flux and total line flux in Table \ref{tab:data}. The spectra and emission-line fits are shown in Figure \ref{fig:spectra} (Appendix \ref{sec:App_data}), with two examples shown in Figure \ref{fig:ex_spectra}. Although the spectra of sources 3 and 8 are best fit by double Gaussian profiles (based on the $\chi^2$ values) we provide the single Gaussian fits in this paper. We do not consider the double Gaussian profiles to be physical given the relative strengths of the rms and dip in flux. Our choice of line profile has no impact on our results. The fluxes derived from the double Gaussian line profiles are consistent, within uncertainties, with the single Gaussian fits shown in Figure \ref{fig:spectra} of the Appendix.
		
		The CO(1-0) line fluxes, and subsequent luminosities, quoted in this paper are based on our single Gaussian line profile fits. These values are consistent, within uncertainties, with the integrated line fluxes extracted from the moment zero maps, for which we tested two methods: (1) extracting the fluxes at the source position and (2) estimating the line flux via the 2D Gaussian fits of CASA's \texttt{imfit}. Our moment zero maps are shown in the central panel of the integrated maps in Figures \ref{fig:chmap_comp_ex} and \ref{fig:chmap_comp} (Appendix). We provide three channel maps in order to: (1) show the lack of continuum emission around CO(1-0), and, (2) highlight that we have captured all CO(1-0) emission in the moment zero maps (centre). The velocity widths of the moment zero (and adjacent, integrated) maps, indicated by the yellow shaded regions of Figures \ref{fig:ex_spectra} and \ref{fig:spectra} (Appendix), were selected to encompass the range of velocities at which the source emission was visible at $\geq 3\sigma$ in the cleaned data cubes. For most sources, the velocity widths of the integrated, moment zero maps are $\sim 1.2$ the full width at half maximum (FWHM) of the CO(1-0) line. However, for sources 4 and 5, where the spectra are not as well fit by Gaussian profiles, we create the integrated channel maps based on greater velocity widths. 

		Our CO-derived redshifts are consistent with the redshifts derived from rest-frame optical lines (flagged 3 or 4 in Table \ref{tab:source_info}). As for the rest-frame optical emission lines, the typical uncertainties of our CO-derived redshifts are $\sim 0.0002$. In contrast, the uncertainty of the photometric redshifts are $\sim 0.2$, on average, with values of $\sim 0.4$ for some COSMOS sources \citep{2016ApJS..224...24L}. For example, source 1 has a secure CO(1-0) detection at $z=2.607$ but a predicted photometric redshift of 2.38. 

		We remove four sources from our sample, which have unreliable redshifts (flagged in Table \ref{tab:source_info}). Accurate spectroscopic redshifts are required in order to estimate upper limits on the CO(1-0) emission and fit the correct SED models to infer stellar masses and SFRs (see Section \ref{sub:SED_props}). We relied on the COSMOS redshift catalogue to design our VLA observations but reobserved sources 1, 2, 4, 6, 9, 11 and 12 with the Multi-Object Spectrometer For Infra-Red Exploration (MOSFIRE, on Keck I), after our VLA observations were taken. Because we relied upon the catalogued redshifts when taking our VLA data, the observed frequency intervals for sources 2 and 9 are not optimised for the correct redshift (see note below Table \ref{tab:source_info}). Thus, the frequency range of the observations for source 2 does not encompass the CO(1-0) line, whereas for source 9, the expected position of CO(1-0) falls at the edge of the observed frequencies, where the spectral noise is greatest. For source 6 we observed neither CO(1-0) emission, nor any rest-frame optical emission lines. Thus, we conclude that the photometric redshift of $z=2.68$, on which we based both the VLA and MOSFIRE observations, is not reliable. For source 14, we observe no CO(1-0) emission and the rest-frame optical spectra exhibit only low S/N peaks. 
		Thus, we exclude sources 2, 6, 9 and 14 (starred in Table \ref{tab:source_info}) from further analysis.

		For the two sources without CO(1-0) detections but with reliable spectroscopic redshifts (sources 12 and 15) we estimate upper limits on the CO(1-0) emission based on the moment zero maps, centred on the expected frequency of the CO(1-0) line. We create each map using \texttt{TCLEAN}, with a channel width of $500\, \kms$ (consistent with the broader lines of our CO-detected sample). We measure the rms value in these single channel maps and use the $3\sigma$ value as our upper limit, consistent with our detection criteria. 
		It is likely that source 15 is not detected because it is just below the detection limit, i.e. the data here were too noisy to detect the predicted signal. However, the same is not true for source 12, for which we would expect to detect CO based on the predicted CO luminosity and noise limit. We can only speculate that one of the assumptions applied in the RJ method does not apply to this source, e.g. that the dust-to-gas ratio is greater than expected (see Section \ref{sub:single_band_method} for the assumptions).


		\begin{figure*}
			\centering
			\includegraphics[width=0.48\textwidth]{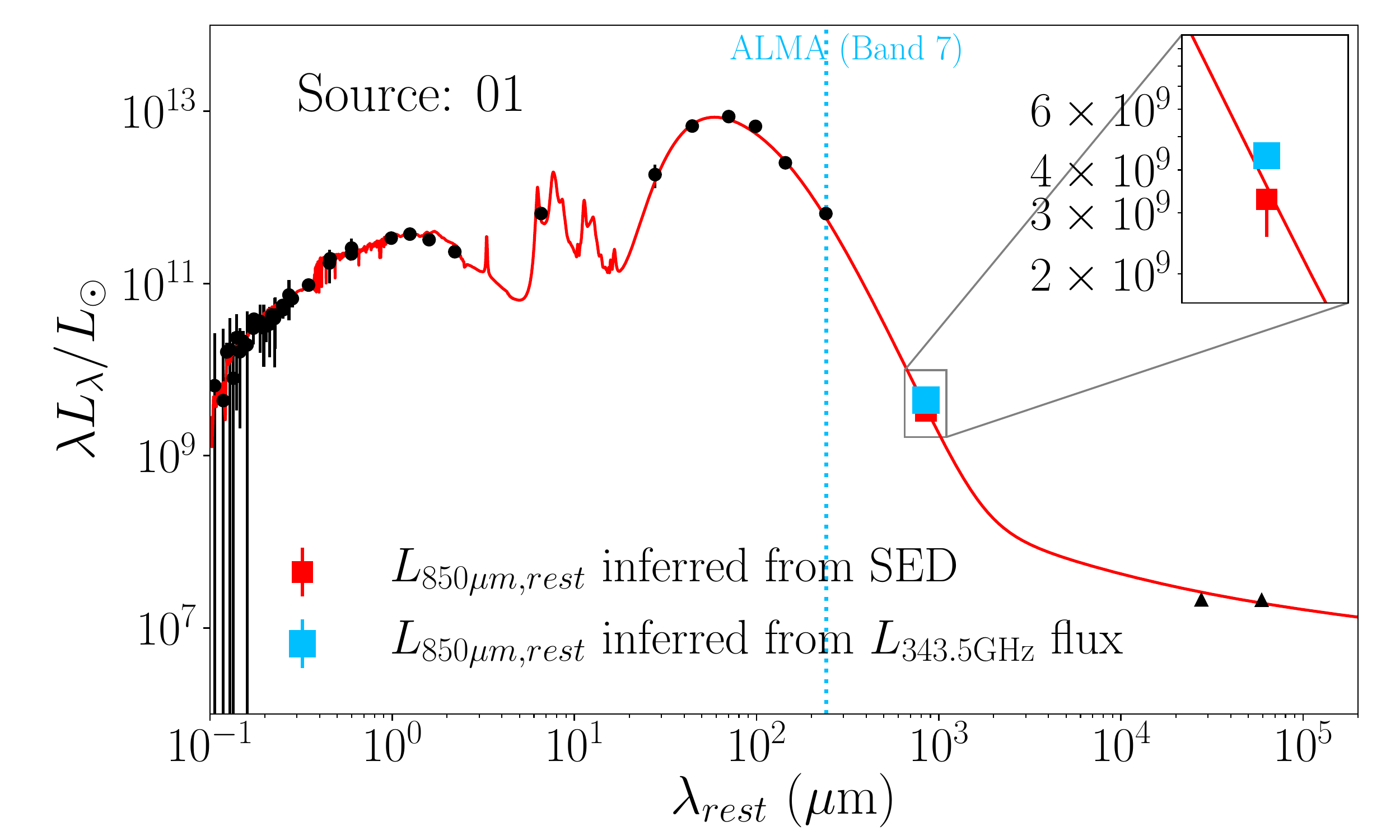}
			~
			\includegraphics[width=0.48\textwidth]{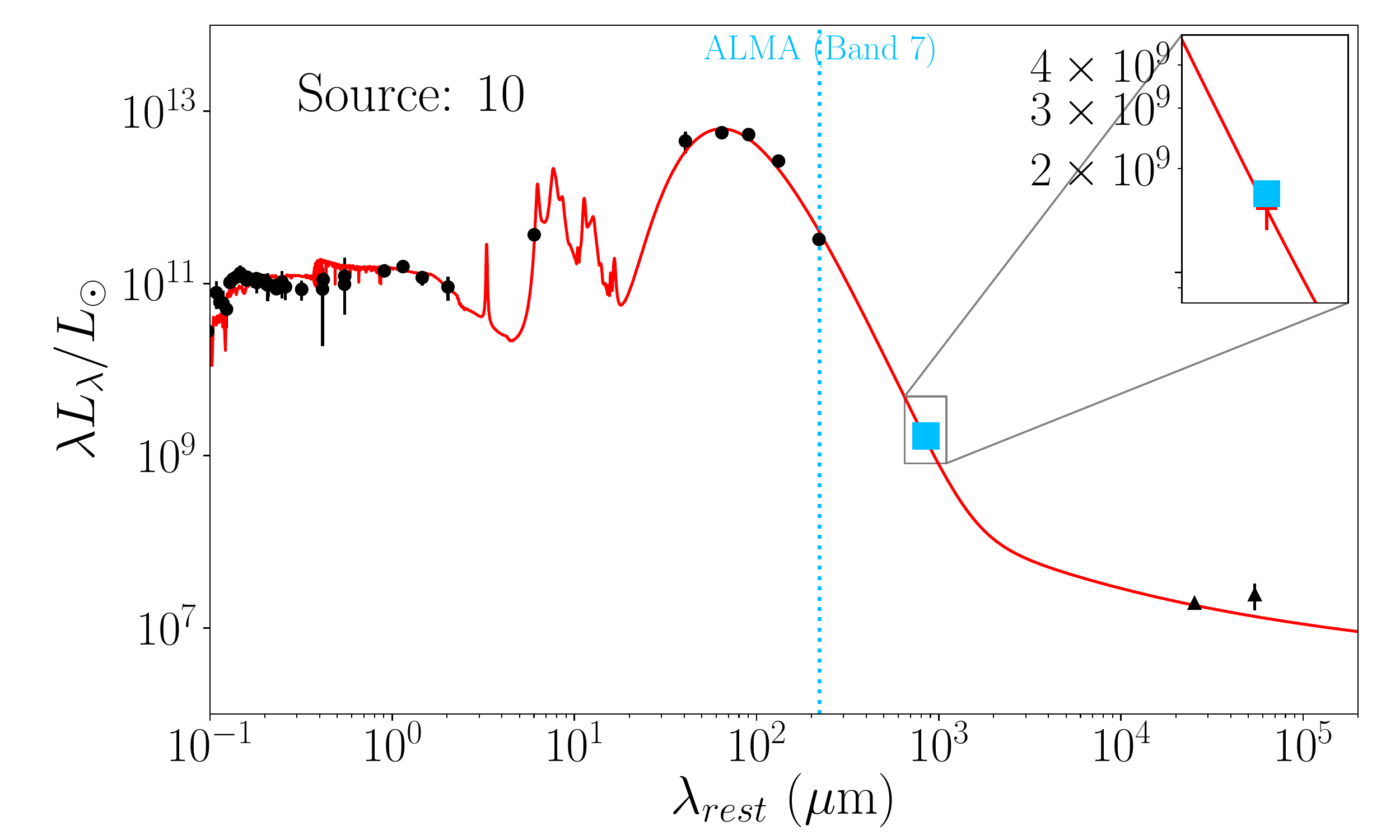}
			\caption{Examples of rest-frame SEDs. The inset focuses on the portion of the rest-frame spectrum around $850\mu m$, comparing the single-band (blue) and SED-derived (red) $\Lref$, where the single-band derived value is calculated from Equation \eqref{eq:L850}. Note that the best-fit SED and value of $\Lref$ shown are based on the standard \texttt{MAGPHYS} assumptions of $\beta=1.5$ and $2.0$ for the warm and cold components, respectively, with the temperature of the cold dust component as a free parameter.  The black filled circles represent the photometry used to fit the SEDs whereas the black triangles show the radio fluxes, which were not used to fit the SEDs. \label{fig:example_SEDs}
			 }
		\end{figure*}

	\subsection{Dust Continuum Data} 
		\label{sub:dust_continuum_data}

		We carefully measure the dust continuum flux of each source, examining the images in the left panels of Figures \ref{fig:chmap_comp_ex} and \ref{fig:chmap_comp}. We note that the FWHM of the synthesized ALMA beam is at least $6\times$ smaller than that of the VLA beam (Figures \ref{fig:chmap_comp_ex} and \ref{fig:chmap_comp}). Thus, unlike for the CO(1-0) data, some of our sources (particularly source 16) show resolved structure in the dust continuum emission. We therefore use CASA's \texttt{imfit} to fit elliptical Gaussians to the source emission, focussing on the region of interest within 5'' of the source. We fit two Gaussians to the emission of source 16 but fit all other sources with a single Gaussian element. We use the flux error returned by \texttt{imfit} as the uncertainty on our derived fluxes. The S/N of the continuum fluxes varies from 4 (Source 16) to 29 (Source 1). Our derived fluxes are mostly consistent with those predicted by the automated procedure of \cite{2017ApJ...837..150S}, except for source 5, for which our measured flux is 20\% lower. We provide our measured fluxes in Table \ref{tab:data}. 
	


\section{Derived Quantities} 
	\label{sec:derived_properties}

	\subsection{CO(1-0)-based Molecular Gas Masses} 
		\label{sub:co_gas_masses}

		\sloppy
		We use our measured CO(1-0) fluxes and upper limits to derive the CO(1-0) line luminosities, $L_\mathrm{CO(1-0)}^{\prime} (\mathrm{K}\, \kms\, \mathrm{pc}^2)$. For sources with a CO(1-0) detection we use the CO-derived spectroscopic redshift. For upper limits, we rely on the spectroscopic redshifts from the COSMOS catalogue. We calculate the CO(1-0) line luminosities, via Equation (3) of \cite{1992ApJ...398L..29S},
		\begin{align}
			L_\mathrm{CO}^{\prime} = 3.25 \times 10^7 S_\mathrm{CO} \Delta v \, \nu_\mathrm{obs}^{-2}  D_\mathrm{L}^2  (1+z)^{-3} \, ,
			\label{eq:Lco}
		\end{align}
		where the CO(1-0) line flux, $S_\mathrm{CO} \Delta v$, is in $\mathrm{Jy}\, \kms$, the observed frequency $\nu_\mathrm{obs}$ is in GHz, and the luminosity distance, $D_L$, is in Mpc. 

		\sloppy
		We derive the total molecular gas mass, based on the CO(1-0) line luminosities, via, 
		\begin{align}
			M_\mathrm{mol} = \aco L_\mathrm{CO(1-0)}^{\prime} \, ,
		\end{align}
		with $\aco = 6.5\, \Msun / (\mathrm{K}\, \kms \, \mathrm{pc}^{2})$. 
		The chosen value of $\aco$ is based on a standard Galactic conversion factor of $X_\mathrm{CO} = 3 \times 10^{20}\, \mathrm{cm}^{−2} (\mathrm{K}\, \kms)^{-1}$ and includes a factor of 1.36 to account for the associated mass of heavy elements (mostly He at 8\% by number). We provide the values of $L_\mathrm{CO(1-0)}^{\prime}$ and the CO(1-0)-based \Mmol\ in Table \ref{tab:derived_properties}. 

		\sloppy
		The CO- and $\Lref$-based \Mmol\ presented in this paper are calculated using the same CO-to-\Mmol\ conversion factor, $\aco=6.5\, \mathrm{\Msun/ (K\, \kms pc^{2})}$. We select this value solely in order to ensure consistency between the molecular gas masses we derive from CO(1-0) and $\Lref$, because the latter was calibrated against CO(1-0)-derived \Mmol\ using this conversion factor \citep{2016ApJ...820...83S}. Our adopted value is a factor of $\sim 1.5$ higher than the $4.36\, \mathrm{\Msun/ (K\, \kms pc^{2})}$ typically adopted by other studies, based on the recommendation of \cite{2013ARA&A..51..207B} \citep[e.g.][]{2013ApJ...768...74T,2015ApJ...800...20G,2018ApJ...853..179T}. Whereas the typically-adopted Milky Way value (corresponding to $X_{CO} = 2 \times 10^{20}\, \mathrm{cm}^{−2} (\mathrm{K}\, \kms)^{-1}$) is based on the compilation of results from gamma-ray observations, extinction measurements, dust emission and CO isotopologues, the value we have adopted here is derived from the correlation of the CO line luminosities and virial masses for resolved Galactic giant molecular clouds \citep{1987ApJ...319..730S,1987ApJS...63..821S}.

	\subsection{RJ-based Molecular Gas Masses} 
		\label{sub:dust_based_gas_masses}


		We derive the molecular gas masses from the inferred $\Lref$, the reference luminosity against which the RJ method is calibrated. We infer $\Lref$ via two methods: (1) from the single, ALMA band 7 flux only ($L_\mathrm{850\mu m,rest}^\mathrm{single-band}$), and, (2) from the best-fit model SEDs ($L_\mathrm{850\mu m,rest}^\mathrm{SED}$). We compare the values of $\Lref$ derived via these two methods in Section \ref{sub:comparison_of_lref}. 

		\subsubsection{Single-band Derived $\Lref$} 
			\label{sub:single_band_method}

				The first method we use to derive $\Lref$ follows the prescription described in detail in Appendix A.1 of \cite{2016ApJ...820...83S,2017ApJ...837..150S}. We outline their approach again here, for clarity. \cite{2016ApJ...820...83S,2017ApJ...837..150S} assume that the long-wavelength dust continuum emission can be described by a single modified blackbody, with a flux density given by:
				\begin{align}
					S_\mathrm{\nu} = \dfrac{\Mmol \kappa(\nu_\mathrm{rest}) B_\mathrm{\nu_{rest}}(T_\mathrm{dust}) (1+z) }{d_L^2} \, ,
					\label{eq:mod_BB}
				\end{align}
				where $\kappa(\nu_\mathrm{rest})$ is the absorption coefficient of dust per unit total mass of molecular gas, $B_\mathrm{\nu_\mathrm{rest}}$ is the Planck function and $d_L$ the luminosity distance. The mean, mass-weighted dust temperature, used to describe the modified blackbody emission, is assumed to be 25 K \citep[discussed in Appendix A.2 of][]{2016ApJ...820...83S}.  This choice is further justified by the recent work of \cite{2019arXiv190210727L}. In the long-wavelength regime, the flux is proportional to $\nu^2$ and Equation \eqref{eq:mod_BB} can be rewritten by including a correction for the departure in the rest frame of the Planck function from RJ approximation, $B_\mathrm{\nu_{rest}}/RJ_\mathrm{\nu_{rest}} = \Gamma_\mathrm{RJ}(\nu_\mathrm{obs},z)$. To relate the specific luminosity in the rest frame of the galaxy to $\Lref$, the long-wavelength dust opacity, $\kappa$, is described by a power law,
				\begin{align}
					\kappa(\nu_\mathrm{obs,rest}) = \kappa(\nu_\mathrm{850\mu m})(\lambda/ 850 \mu m)^{-\beta}
				\end{align}
				with a dust emissivity index of $\beta=1.8$ \citep[see Appendix A.3 of][for details]{2016ApJ...820...83S}. Note that because the dust opacity is defined relative to the molecular gas mass, there is an implicit assumption here of a constant gas-to-dust ratio. Combining the above assumptions, the reference luminosity can be calulated from the measured, RJ flux via,	
				\begin{align}
				 	L_\mathrm{850\mu m, rest}^\mathrm{single-band} = S_\mathrm{\nu}\left[\dfrac{\nu_\mathrm{850 \mu m}}{\nu_\mathrm{rest}} \right]^{3.8} %
				 	\dfrac{d_L^2}{1+z} \dfrac{\Gamma_\mathrm{RJ}(\nu_\mathrm{850\mu m},z)}{\Gamma_\mathrm{RJ}(\nu_\mathrm{obs},z)}  \, ,
				 	\label{eq:L850}
				\end{align} 
				where the exponent of 3.8 is the result of the $\nu^2$ dependence of the RJ flux combined with the dust emissivity index. We use Equation \ref{eq:L850} to derive the $\Lref^\mathrm{single-band}$ and provide the derived values in Table \ref{tab:derived_properties}.


	\begin{table*}
		\begin{center}
		\caption{Derived Properties}
		\begin{tabular}{lcccccccc}
		\toprule
		Galaxy 	& $L_\mathrm{CO(1-0)}^{\prime}$ & $\Lref^\mathrm{single-band}$\tablenotemark{*} & M$_\mathrm{mol, CO}$ & M$_\mathrm{mol,RJ}$\tablenotemark{*} & \Mstar 		& SFR 				& $T_\mathrm{cold\,dust}$ & $M_\mathrm{dust}$ \\
				 	& (10$^{10}$K \kms pc$^2$) & (10$^{31}$ erg s$^{-1}$ Hz$^{-1}$)	 &  ($10^{11}\Msun$) & ($10^{11}\Msun$) & 	($10^{11}\Msun$)	&  (\Msun/\peryr)   & (K) 			   & ($10^9\Msun$) \\
		\midrule
		\hd 1  	& 8.7 $\pm$ 1.1 &  	4.72 $\pm$ 0.16	& 5.7 $\pm$ 0.7 & 7.1 $\pm$ 1.8 & 2.6  \tiny{$^{+0.6\hd}_{-0.7}$} 			&  1028  \tiny{$^{+195}_{\hd-250}$} 			& 28.6\tiny{$^{+7.2}_{-1.4}$}   &	1.7\tiny{$^{+0.3}_{-0.4}$} \\
		\hd 3  	& 7.4 $\pm$ 1.8 &  	3.47 $\pm$ 0.15	& 4.8 $\pm$ 1.1 & 5.3 $\pm$ 1.4 & 1.8  \tiny{$^{+0.2\hd}_{-0.2}$} 			& \hd865 \tiny{$^{+230}_{-\hd60}$ }			& 31.3\tiny{$^{+8.0}_{-2.0}$}   &	1.5\tiny{$^{+0.1}_{-0.6}$} \\
		\hd 4  	& 4.0 $\pm$ 1.2	& 	2.90 $\pm$ 0.13	& 2.6 $\pm$ 0.8  	& 4.4 $\pm$ 1.1 	& 2.5  \tiny{$^{+0.6\hd}_{-0.3}$} 		& \hd948 \tiny{$^{+164}_{-294}$} 			& 26.3\tiny{$^{+0.6}_{-4.9}$}   &	1.6\tiny{$^{+0.1}_{-0.8}$} \\
		\hd 5  	& 3.9 $\pm$ 1.4 &  	2.26 $\pm$ 0.13	& 2.5 $\pm$ 0.9 & 3.4 $\pm$ 0.9 & 1.1  \tiny{$^{+0.1\hd}_{-0.2}$} 			& 1419   \tiny{$^{+50}_{-\hd294}$}			& 25.8\tiny{$^{+7.1}_{-1.6}$}   &	1.5\tiny{$^{+1.1}_{-0.0}$   } \\
		\hd 7  	& 5.7 $\pm$ 1.4 &   2.83 $\pm$ 0.13	& 3.7 $\pm$ 0.9 & 4.3 $\pm$ 1.1 & 1.0  \tiny{$^{+0.2\hd}_{-0.2}$} 			& 1089 \tiny{$^{+10}_{-600}$} 			& 28.3\tiny{$^{+13.}_{-2.9}$}  &	1.7\tiny{$^{+0.3}_{-0.5}$} \\
		\hd 8  	& 2.8 $\pm$ 5.0 &   2.15 $\pm$ 0.13	& 1.8 $\pm$ 0.3 & 3.3 $\pm$ 0.9 & 4.6  \tiny{$^{+0.9\hd}_{-1.2}$} 			& \hd395 \tiny{$^{+204}_{-132}$} 			& 29.1\tiny{$^{+7.9}_{-7.9}$}   &	1.1\tiny{$^{+0.2}_{-0.2}$} \\
		10 		& 5.4 $\pm$ 1.2 &   1.81 $\pm$ 0.14	& 3.5 $\pm$ 0.8 & 2.7 $\pm$ 0.7 & \hd 0.47 \tiny{$^{+0.01}_{-0.01}$} 		& \hd558 \tiny{$^{+15}_{-\hd13}$}			& 36.3\tiny{$^{+3.0}_{-2.3}$}   & 	0.8\tiny{$^{+0.2}_{-0.1}$} \\
		11 		& 6.9 $\pm$ 1.4 &   1.89 $\pm$ 0.16	& 4.5 $\pm$ 0.9 & 2.9 $\pm$ 0.8 & 2.0  \tiny{$^{+0.6\hd}_{-0.6}$} 			& \hd598 \tiny{$^{+268}_{-117}$}			& 30.2\tiny{$^{+7.2}_{-1.6}$}   &	1.0\tiny{$^{+0.1}_{-0.2}$} \\
		12	    & $\leq$ 2.1 	&   1.58 $\pm$ 0.13	& $\leq$ 1.3    & 2.4 $\pm$ 0.6 & 2.0 $\substack{+0.5\\-0.6}$ 		&  264 $\substack{+79\\-61}$ 	& 28.9$\substack{+15.1\\-2.5}$ &	0.9$\substack{+0.3\\-0.3}$ \\
		13 		& 4.1 $\pm$ 7.7 &   1.54 $\pm$ 0.13	& 2.6 $\pm$ 0.5 & 2.3 $\pm$ 0.6 & 2.4  \tiny{$^{+0.7\hd}_{-0.7}$} 			&  \hd449 \tiny{$^{+186}_{-170}$} 			& 28.1\tiny{$^{+15.}_{-3.7}$} 		& 	0.6\tiny{$^{+0.3}_{-0.1}$} \\
		15 		& $\leq$ 3.4 	&   1.14 $\pm$ 0.98	& $\leq$ 2.2 	& 1.7 $\pm$ 0.5 	& 1.0  \tiny{$^{+0.5\hd}_{-0.4}$} 		&  \hd391 \tiny{$^{+\hd90}_{-\hd99}$}		& 29.7\tiny{$^{+3.7}_{-3.5}$}  		& 	0.5\tiny{$^{+0.2}_{-0.1}$} \\
		16 		& 1.8 $\pm$ 4.1 &   1.34 $\pm$ 0.28	& 1.2 $\pm$ 0.3 & 2.0 $\pm$ 0.7 & 1.3  \tiny{$^{+0.1 \hd}_{-0.6}$} 		&  \hd391 \tiny{$^{+\hd\hd5}_{-\hd53}$}  	& 29.2\tiny{$^{+4.9}_{-4.1}$}  		&	0.4\tiny{$^{+0.2}_{-0.3}$} \\
		\bottomrule
		\end{tabular}
		\tablenotetext{*}{Derived from the dust continuum flux at 343.5 GHz using Equation \eqref{eq:L850}.}
		\label{tab:derived_properties}
		\end{center}
	\end{table*}

		\subsubsection{SED-derived $\Lref$} 
			\label{sub:sed_based_method}

				For our second method, we apply the updated version of the SED-fitting algorithm \texttt{MAGPHYS} high-z, which builds a likelihood distribution for $\Lref$ by marginalising over a library of models with varying dust emission parameters, such as dust temperatures \citep[see also][]{2013ApJ...765....9D}. \texttt{MAGPHYS} models the dust emission, from the rest-frame mid-infrared to millimeter wavelengths, as the sum of four dust components: (i) a component of polycyclic aromatic hydrocarbons (PAHs); (ii) a mid-infrared continuum that characterizes the emission from hot grains at temperatures in the range 130-250 K; (iii) a warm dust component in thermal equilibrium with temperatures in the range 30-80K; and (iv) and a component of cold grains in thermal equilibrium with adjustable temperature in the range 20-40 K. The emission from the warm and cold dust components are described by modified blackbodies.

				We choose not to analyse the SED-derived dust masses or temperatures due to degeneracies between these parameters. Our data do not sample far enough along the RJ tail to break the degeneracy between $T_\mathrm{dust}$ and $\beta$ \citep[see e.g.][]{2015ApJ...806..110D}. Moreover, the dust temperature relevant to the RJ emission tail is the mass-weighted dust temperature \citep[discussed in Appendix A.2. of ][]{2016ApJ...820...83S}. The total dust mass inferred by \texttt{MAGPHYS} is a sum of the masses of the modeled components, which are free parameters, and is therefore degenerate with the temperatures of the dust components. 
				
				To assess what impact variations in $\beta$ and $T_\mathrm{dust}$ have on the derived $\Lref$ we apply \texttt{MAGPHYS} with four sets of assumptions: 
				\begin{enumerate}[noitemsep,nolistsep,nosep]
					\item with $\beta=1.5$ and $2.0$ for the warm and cold components, respectively, and with the cold dust temperature as a free parameter (standard version),
					\item with $\beta=1.8$ for the warm and cold components and with the cold dust temperature as a free parameter,
					\item with $\beta=1.5$ and $2.0$ for the warm and cold components, respectively, but with the temperature of the cold component fixed to $25$K, and,
					\item with $\beta=1.8$ for the warm and cold components and the temperature of the cold component fixed to $25$K.
				\end{enumerate}
				We fit the rest-frame SEDs and build likelihood distributions for $\Lref^\mathrm{SED}$ for each of these four cases. The rest-frame SEDs and best fit, based on the standard version of \texttt{MAGPHYS}, are shown in Figures \ref{fig:example_SEDs} and \ref{fig:SEDs} (Appendix), with the single-band and SED-derived values of $\Lref$ compared in the zoom-in panels.


		\subsubsection{Converting $\Lref$ to \Mmol} 
			\label{subsub:conversion_to_Mmol}

				We convert both the single-band and SED-derived values of $\Lref$ to molecular gas masses by applying the conversion factor,
				\begin{align}
				\begin{split}
				 	\alpha_\mathrm{850\mu m} & = \dfrac{\Lref}{\Mmol} \\
				 						& = \dfrac{\Lref}{L_\mathrm{CO(1-0)}^{\prime}} \dfrac{1}{\aco} \, ,
				\end{split}
				 \label{eq:alpha_850}
				\end{align}  
				derived by \cite{2016ApJ...820...83S}. The calibration of $\alpha_\mathrm{850\mu m}$ is based on the Herschel SPIRE 350 and $500\mu$m data and CO(1-0)-derived \Mmol\ of 28 local star-forming galaxies, 12 low-redshift ultraluminous infrared galaxies (ULIRGs), and 30 $z\sim 2$ submillimeter galaxies (SMGs) from the literature \citep{2016ApJ...820...83S}. Thus, the single-band RJ method is suited to massive ($>2\times10^{10}\Msun$), star-forming galaxies for which the dust-to-gas ratios are $\sim1:100$ \citep[see discussion in ][]{2016ApJ...820...83S}.

				Applying the single-band method of Equation \eqref{eq:L850}, with the constant $\left<\alpha_\mathrm{850\mu m}\right>$ we find a range of \Mmol\ of $1.7-7.1 \times 10^{11}\Msun$ with a mean uncertainty of $0.9\times 10^{11}\Msun$. The uncertainties of these \Mmol\ values are derived from the uncertainty of both the measured dust continuum fluxes and $\left<\alpha_\mathrm{850\mu m}\right>$. Applying the standard version of \texttt{MAGPHYS} to derive $\Lref$, with the constant $\left<\alpha_\mathrm{850\mu m}\right>$, we find a range of \Mmol\ of $1.1-5.3 \times 10^{11}\Msun$ with a mean uncertainty of $0.8\times 10^{11}\Msun$. The uncertainty of the \texttt{MAGPHYS}-derived \Mmol\ is based on the 16th and 84th percentiles of the likelihood distribution of $\Lref$ and the uncertainty of $\left<\alpha_\mathrm{850\mu m}\right>$.


	\subsection{Stellar Masses and SFRs} 
		\label{sub:SED_props}

		We self-consistently derive the stellar masses, SFRs and reference luminosities, $\Lref^\mathrm{SED}$, upon which the RJ method is based (see Section \ref{sub:dust_based_gas_masses}) by fitting the SEDs of our sources. The stellar masses and SFRs of our sample have been previously derived by fitting the optical-to-infrared SEDs via various techniques \citep[see ][for details]{2013ApJ...778..131L,2015ApJ...801...80L,2017ApJ...837..150S}. The stellar masses, determined with the SED-fitting algorithm \texttt{LePhare} \citep{1999MNRAS.310..540A,Ilbert_2006}, are provided as part of the COSMOS catalogue \citep{2016ApJS..224...24L}. However, for this small sample we revisit the stellar mass and SFR fits, using the SED-fitting algorithm \texttt{MAGPHYS}, in order to: (1) self-consistently fit the stellar masses, SFRs and $\Lref^\mathrm{SED}$; (2) explore the effect of the assumptions used to derive $\Lref^\mathrm{SED}$; (3) use the CO(1-0)-based redshifts (where these are the most reliable redshifts), and; (4) validate the robustness of the SED-derived properties. We describe our application of \texttt{MAGPHYS} in the following subsection. Sources with uncertain redshift measurements (starred in Table \ref{tab:source_info}) have been excluded from this analysis. The derived stellar masses and SFRs are provided in Table \ref{tab:derived_properties}.

		The stellar masses and SFRs presented here are consistent with the previously-derived values. The stellar masses derived via \texttt{MAGPHYS} and \texttt{LePhare} are consistent within a factor of two (with no systematic offset) for all but source 15, for which the \texttt{MAGPHYS} derived \Mstar\ is a factor of five greater (we trust our SED fit in Figure \ref{fig:SEDs}). Similarly, our \texttt{MAGPHYS}-derived SFRs are consistent to within a factor of two with the values used in \cite{2015ApJ...801...80L} and \cite{2017ApJ...837..150S}, with no systematic offsets. Thus the conclusions we draw with respect to the global properties of our sample are not affected by the choice of SED-fitting algorithm. However, the scatter in values derived via different SED-fitting algorithms indicates an uncertainty of at least a factor of two in both the stellar masses and SFRs.

		\begin{figure}[t!]
			\includegraphics[width=\columnwidth,trim={0.1cm 0.3cm 1.2cm 1.2cm},clip]{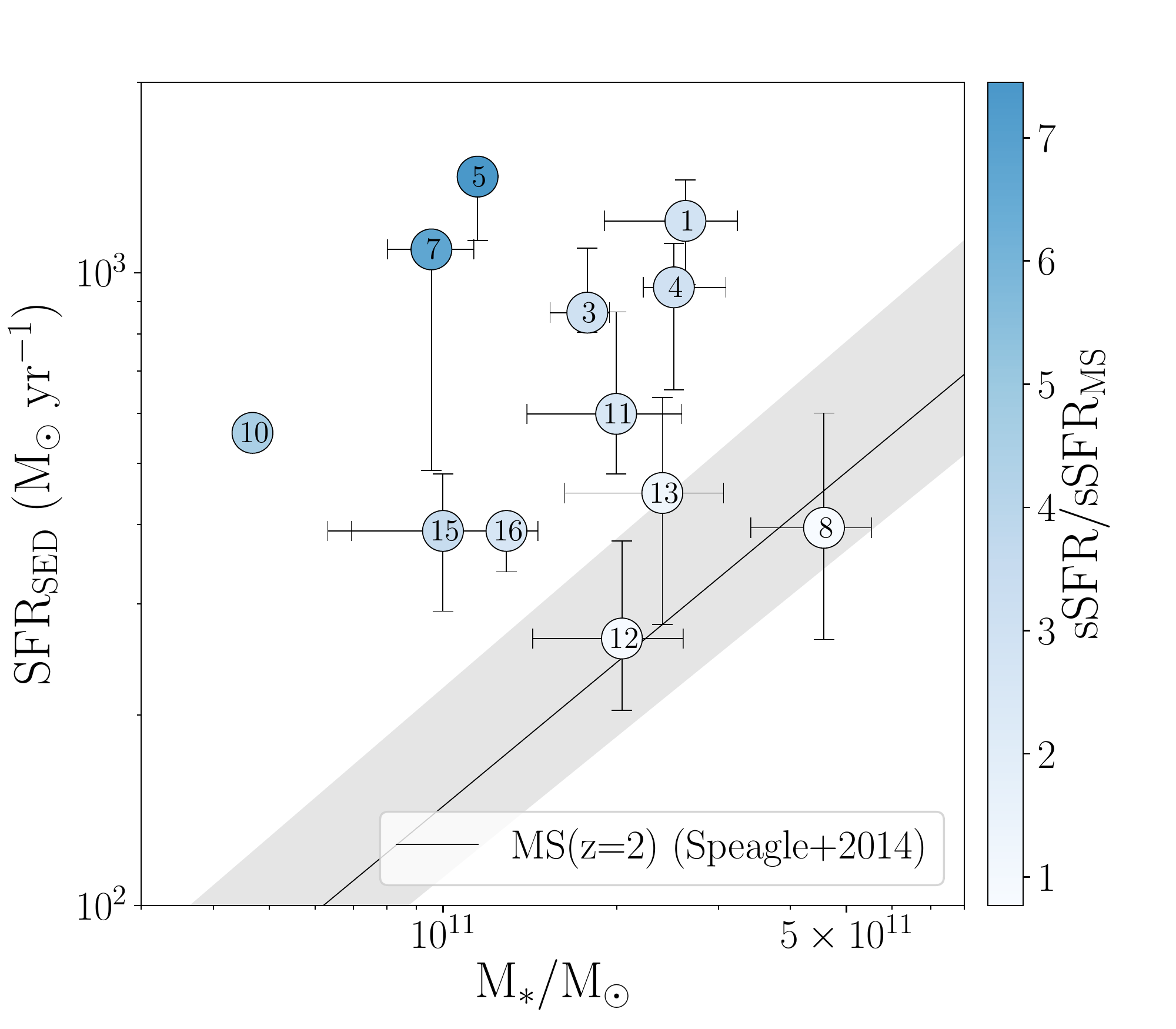}
			\caption{Position of our sample (labeled by their ID) on the stellar mass vs SFR plane relative to the predicted Main Sequence (based on Speagle et al. 2014). The solid black line represents the Main Sequence at $z\sim 2$, whereas the filled grey region indicates the expected scatter of Main Sequences encompassed by the redshift range of our sample.  Stellar masses were derived by fitting the galaxy SEDs with the high-z version of \texttt{MAGPHYS}. SFRs are derived from the total IR luminosity. The colour coding of the points (scalebar on the right) indicates the relative offset of the sources from the main sequence at the source redshift. }
			\label{fig:main_sequence}
		\end{figure}

	\subsubsection{Application of \texttt{MAGPHYS}} 
		\label{subsub:SED_fitting}

		We derive the stellar masses, SFRs and $\Lref^\mathrm{SED}$ of our sample via an updated version of the SED-fitting algorithm, \texttt{MAGPHYS}, that builds a likelihood distribution for $\Lref^\mathrm{SED}$, applying the version optimised for high-z galaxies \citep[][]{2015ApJ...806..110D}. 
		We fit the available COSMOS photometry, from the GALEX far-ultraviolet to Herschel PACS 500 $\mu$m filters (taken from the 2016 catalogue of \citealt{2016ApJS..224...24L}), in addition to the ALMA Band 7 (343.5 GHz) dust continuum measurements (described in Section \ref{sub:dust_continuum_data}). For source 16, no Herschel data is available in the \cite{2016ApJS..224...24L} catalogue. Instead, we use the Herschel data described in \cite{2013ApJ...778..131L}\footnote{provided via private correspondence}. 
		The details of the SED-fitting framework, stellar model libraries and treatment of dust are described in \cite{2008MNRAS.388.1595D,2015ApJ...806..110D}. Two example SED fits are shown in Figure \ref{fig:example_SEDs}, the rest are shown in Figure \ref{fig:SEDs} in the Appendix. We choose not to apply the 1.4 and 3 GHz continuum flux measurements, available for the majority of the sample (see overplotted black triangles in both figures) because the radio/FIR correlation (assumed in \texttt{MAGPHYS} \citealt{2015ApJ...806..110D}) is uncertain at these redshifts and may bias the results. 

		We use the median stellar masses and SFRs inferred via \texttt{MAGPHYS}, and quote errors as the 16th and 84th percentile ranges of the posterior likelihood distributions. These values are based on the spectroscopic redshifts, derived from either CO(1-0), or, infrared emission lines (see Table \ref{tab:source_info} for redshifts). The SFRs inferred by \texttt{MAGPHYS} represent the average of the star formation history (SFH) over the last 100 Myr \citep{2015ApJ...806..110D}.  The SFH library of \texttt{MAGPHYS} includes a wide range of continuous SFHs as well as accounting for stochasticity on the SFHs by superimposing star formation bursts of random duration and amplitude. Thus, \texttt{MAGPHYS} models the SFHs of starburst-like and more quiescently star-forming galaxies \citep[see][ for details]{2015ApJ...806..110D}. We use the SFH-based SFRs, rather than SFRs derived empirically from the IR luminosities here. However, we note that the SFRs derived empirically from the total IR luminosities are consistent to within a factor of $1.5$ for the entire sample, with the IR-inferred SFRs a factor of 1.2 higher on average.   

	\subsubsection{Main Sequence Offset}
		\label{subsub:ms_offset}

		To investigate the relationship between the derived gas masses, we derive the offset from the MS, defined as the specific SFR (sSFR) of the source relative to the sSFR expected for a MS galaxy of the given stellar mass and given epoch ($\mathrm{sSFR_{MS}}$). We define the MS using the best fit from \cite{2014ApJS..214...15S}:
		\begin{align}
		\begin{split}
		 	\log{\mathrm{SFR (\Mstar, t(z))}} = & (0.84 - 0.026 t(z))\log{\Mstar} \\ 
		 										& - (6.51 - 0.11 t(z)) \, ,
		\end{split} 	
		\label{eq:speagle_ms}
		\end{align} 
		where $t(z)$ is the age of the Universe at redshift, z, in Gyr. We show the position of our sample in the MS plane in Figure \ref{fig:main_sequence}. The points are colour-coded by the MS offset, $\mathrm{sSFR/sSFR_{MS}}$, based on the \cite{2014ApJS..214...15S} definition, where the SFRs are derived from the SED. 
		Our sample encompasses both MS and above-MS galaxies, spanning $1< \mathrm{sSFR/sSFR_{MS}} <7$.


\section{Results and Discussion} 
	\label{sec:results}

	\subsection{Comparison of single-band and SED-derived $\Lref$} 
		\label{sub:comparison_of_lref}

\begin{figure}[t!]
			\centering
			\includegraphics[width=0.85\columnwidth,trim={0.5cm 1.cm 0.5cm 0.0cm},clip]{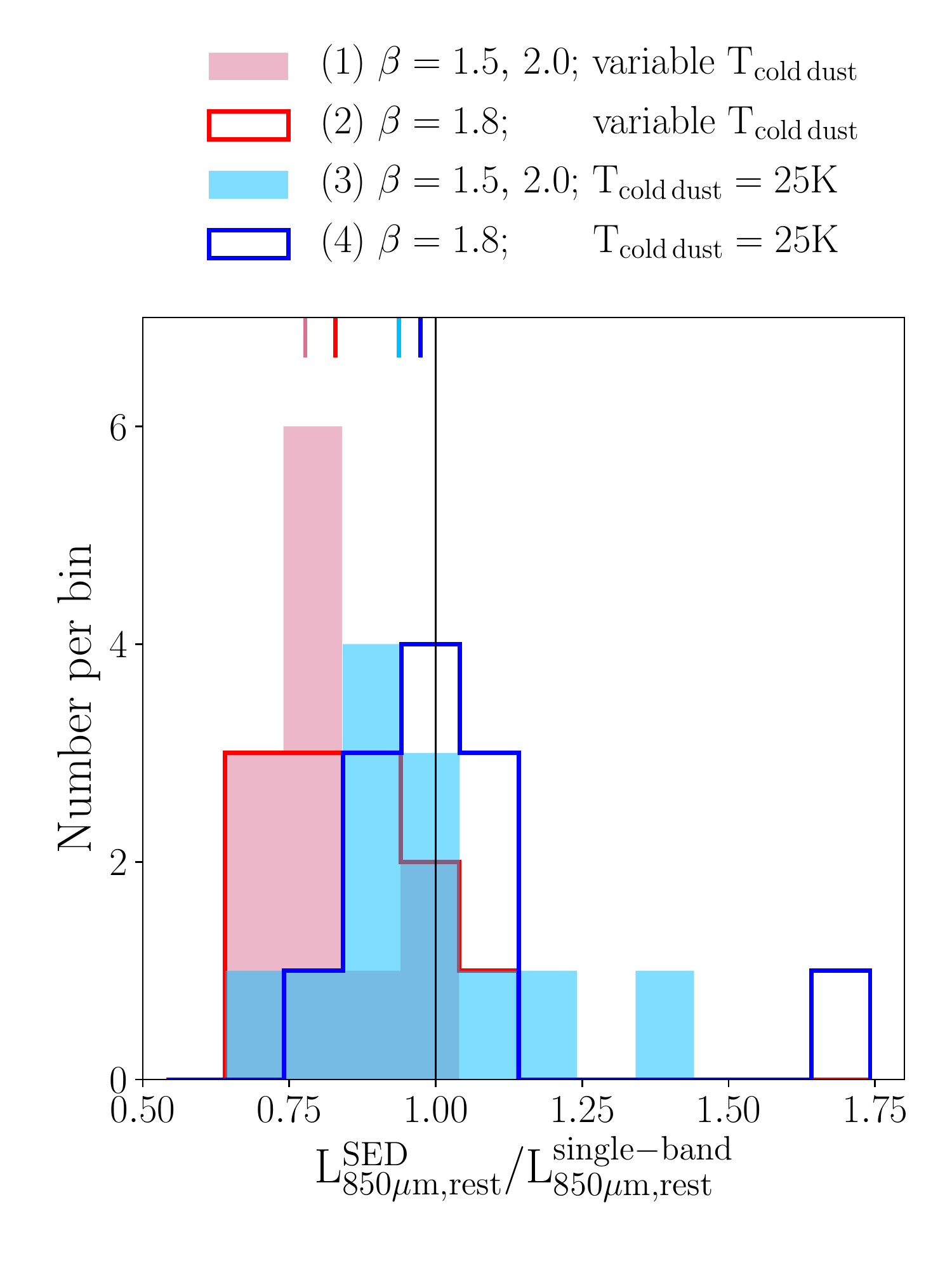}
			\caption{Histograms of the ratio of the single-band to SED-inferred $\Lref$ for the four sets of \texttt{MAGPHYS} assumptions (labeled at the top): (1) the standard \texttt{MAGPHYS} assumptions (filled pink), (2) $\beta=1.8$ for both dust components with the $T_\mathrm{cold\, dust}$ as a free parameter (red outline), (3) $\beta=1.5 (2)$ for the warm (cold) components and $T_\mathrm{cold\, dust}$ , and (4) replicating the assumptions of the single-band RJ calibration. The coloured solid lines at the top of the plot indicate the median values of the four distribution, with the colour matching that of the histogram (labeled at top). The black solid line at unity indicates where the SED- and single-band derived values are equivalent. \label{fig:L850_comp}}
		\end{figure}

\begin{figure*}
			\centering
			\includegraphics[width=\textwidth,trim={1.3cm 0.5cm 1.6cm 0cm},clip]{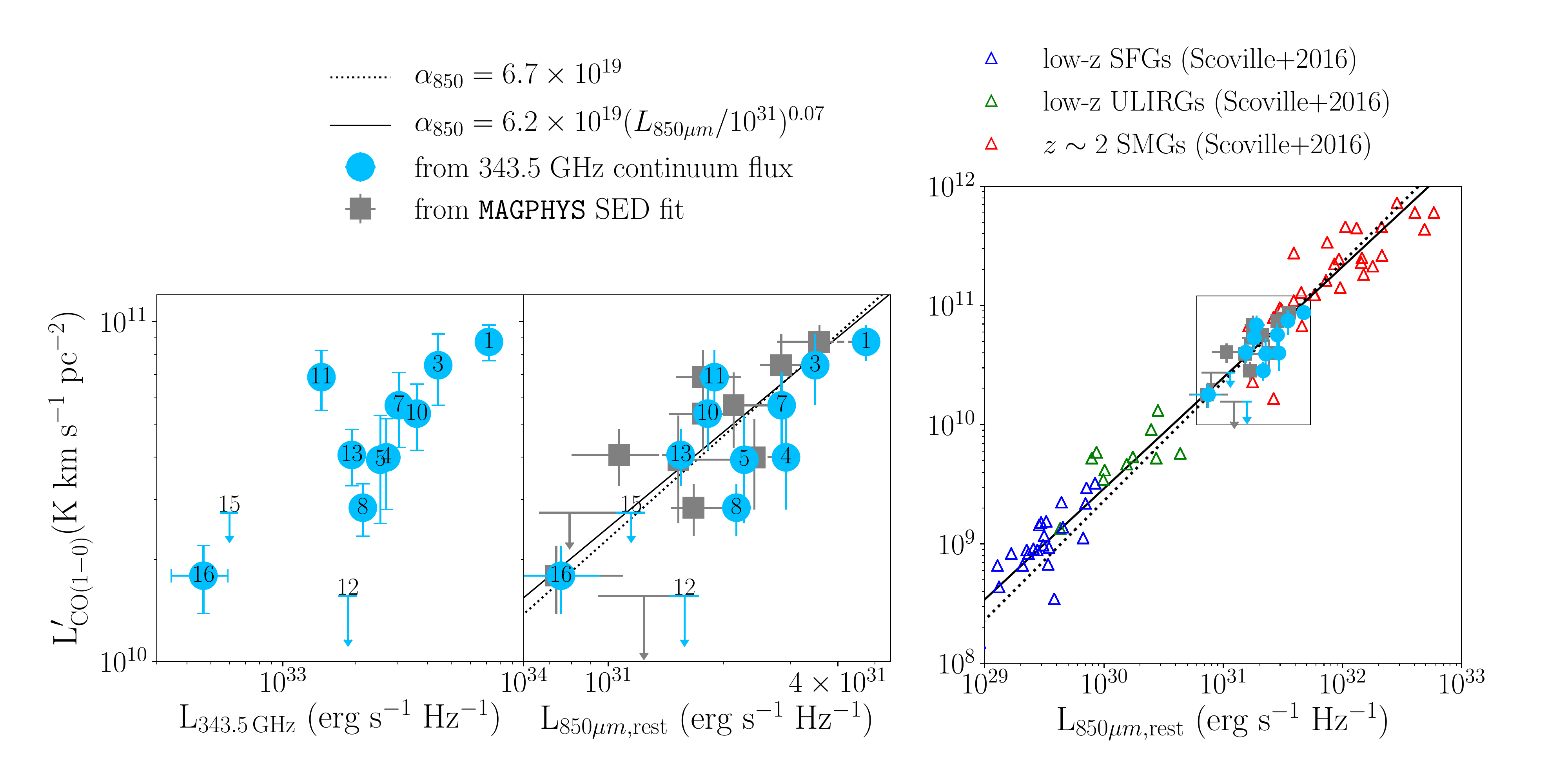}
			\caption{Comparison of CO(1-0) and dust continuum luminosities. Left panel: CO(1-0) luminosity vs. measured 343.5 GHz (ALMA Band 7) luminosity. Central panel: CO(1-0) luminosity vs. rest-frame 850$\mu$m luminosity from ALMA band 7 measurements (blue) and rest-frame, SED fits (grey; from the standard, high-z version of \texttt{MAGPHYS} with $\beta=1.5$ and $2$, respectively for the warm and cold dust components). Right panel: CO(1-0) luminosity vs. rest-frame 850$\mu$m luminosity of our sample (rectangular inset from central panel) compared to the calibration samples of Scoville et al. 2016. The dotted (solid) lines represent the constant (best-fit) $\alpha_\mathrm{850\mu m}$ fit to the calibration sample of \cite{2016ApJ...820...83S} \label{fig:lum_comp}}
		\end{figure*}

		To assess the extent to which the assumed values of $\beta$ and $T_\mathrm{dust}$ affect the derived $\Lref$, and thereby also \Mmol, we compare the single-band and SED-derived values of $\Lref$ via the ratio between the two in Figure \ref{fig:L850_comp}. Our results are based on the mean value, $\left<\alpha_\mathrm{850\mu m}\right>  = 6.7\pm 1.7 \times 10^{19} \mathrm{erg \, s^{-1}\, Hz^{-1}\, \Msun^{-1}} $ derived for the \cite{2016ApJ...820...83S} calibration sample. \cite{2016ApJ...820...83S} also provide a fit for the conversion factor that varies with $\Lref$, based on the best-fit relation between $\Lref$ and L$_{CO(1-0)}^\prime$ (solid line in Figure \ref{fig:lum_comp}). However, the molecular gas masses derived from the mean and best-fit $\alpha_\mathrm{850\mu m}$ are consistent for our sample. The comparison between the single-band and SED-derived values of $\Lref$ is of particular relevance given that there now exist a range of approaches to estimating $\Lref$. For example, \cite{2018MNRAS.474.3866H} fit their Herschel SPIRE $250-500 \mu$m and AzTEC 1.1 mm photometry with a modified blackbody \citep[based on Equation 14 of][]{2002ApJ...568...88Y} whereas \cite{2017MNRAS.468L.103H} extrapolate the value of $\Lref$ from the best-fit SEDs, modeled with \texttt{MAGPHYS}.

		We find that most values of $\Lref$ inferred via the SED fit of \texttt{MAGPHYS} are consistent, within errors, with the single-band derived values, once the dust emissivity and temperature of the cold dust are fixed to match the assumptions of the single-band RJ method (see Figures \ref{fig:L850_comp} and \ref{fig:lum_comp}). The standard version of \texttt{MAGPHYS}, which allows for variable dust temperatures for the hot and cold components, predicts values of $\Lref$ that are systematically lower than those inferred via Equation \eqref{eq:L850}, i.e. centred at 78\% of the single-band values (pink line and filled histogram). Fixing the dust emissivities of both the warm and cold components to $\beta=1.8$, but allowing the dust temperatures to vary, leads to slightly stronger agreement, with a median $\Lref$ ratio of 83\% (red line and outlined histogram). Fixing only the temperature of the cold dust component of \texttt{MAGPHYS} leads to stronger agreement with the single-band method (median of 94\% that of the single-band method) than when solely fixing the dust emissivities to match (see light blue line and filled histogram). Attempting to replicate the assumptions made in the single-band RJ method (by fixing the temperature of the cold dust component to $T_\mathrm{cold\,dust}=25$K, and assuming $\beta=1.8$ for both dust components), leads to values of $\Lref$ that are mostly consistent with those derived from the single-band measurement (blue line and outlined histogram). 

		We find a significant range in the relative reference luminosities, $\Lref^\mathrm{SED}/\Lref^\mathrm{single-band}$, derived for each set of \texttt{MAGPHYS} assumptions. Even when the assumptions on $\beta$ and $T_\mathrm{dust}$ are matched to the single-band method the ratio varies from  $0.7-1.6$ (Figure \ref{fig:L850_comp}). This distribution of values can be attributed mainly to the variation in the relative masses of the warm and cold dust components fit by \texttt{MAGPHYS}, but may also reflect small variations in the intrinsic gas-to-dust ratios of the sample.
		
		We find that mismatches between the assumed $\beta$ and $T_\mathrm{dust}$ vs the intrinsic properties of the sources can lead to up to a factor of two uncertainty on $\Lref$. The mean variation of values of $\Lref^\mathrm{SED}$ inferred per source, for different sets of \texttt{MAGPHYS} assumptions, was 20\%, with differences of up to 70\%. Likewise, the values of $\Lref^\mathrm{SED}$, derived from the standard \texttt{MAGPHYS} assumptions, differed from the $\Lref^\mathrm{single-band}$ values by $20\%$ on average, with differences of up to 70\%. These results are consistent with the comparison of the inferred and intrinsic $\Lref$ for the simulated galaxies of \cite{2018arXiv180503649P}, for which the extrapolation from the observed frequency to the rest frame at $850\mu \mathrm{m}$ resulted in errors of the order of $\sim 20\%$, with deviations of up to $50\%$ for some snapshots. 
		
		Our comparison of $\Lref^\mathrm{SED}$ indicates that the value of $\alpha_\mathrm{850}$ derived by calibrating the CO(1-0)-derived \Mmol\ against the $\Lref$, is systematically affected by the assumptions on the dust emissivity index and temperature. For example, \cite{2017MNRAS.468L.103H}, who apply the standard form of \texttt{MAGPHYS} to derive $\Lref$ from the SEDs of their sample, find a slightly lower $\alpha_\mathrm{850}$ than \cite{2016ApJ...820...83S}, based on the single-band conversion of Equation \eqref{eq:L850}. The lower values found by \cite{2017MNRAS.468L.103H} are consistent with our findings that the $\Lref^\mathrm{SED}$ estimated from the standard \texttt{MAGPHYS} SED models are 20\% lower, on average, than for the single-band method. However, based on the $\chi^2$ values of the \texttt{MAGPHYS} SED fits, none of the four sets of model assumptions (on $\beta$ and $T_\mathrm{cold\,dust}$) that we tested result in a better systematic fit to the photometry.


		\begin{figure}
			\centering
			\includegraphics[width=0.4\textwidth,trim={0.2cm 0cm 0.6cm 0.5cm},clip]{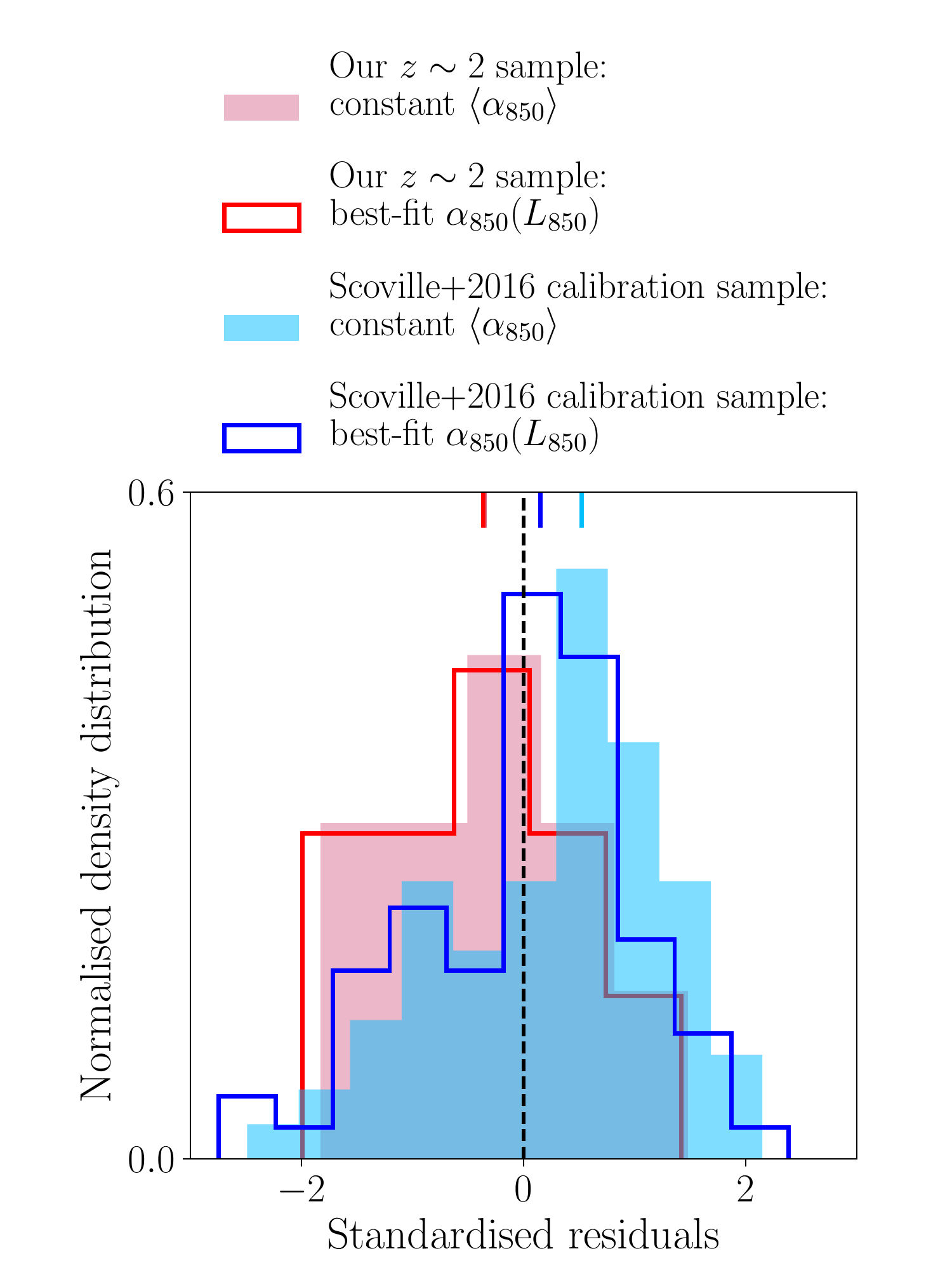}
			\caption{Comparison of the standardized residual distributions for our sample (red, solid and filled) and the \cite{2016ApJ...820...83S} calibration samples (blue, solid and filled). The standardized residuals represent the distance of the measured $L_\mathrm{CO(1-0)}^\prime$ from the mean (filled histograms) and best-fit (solid outline) $\alpha_\mathrm{850\mu m}$.   \label{fig:residual}}		
		\end{figure}

		\begin{figure}
			\centering
			\includegraphics[width=0.5\textwidth,trim={1.8cm 2.5cm 0.8cm 1.2cm},clip]{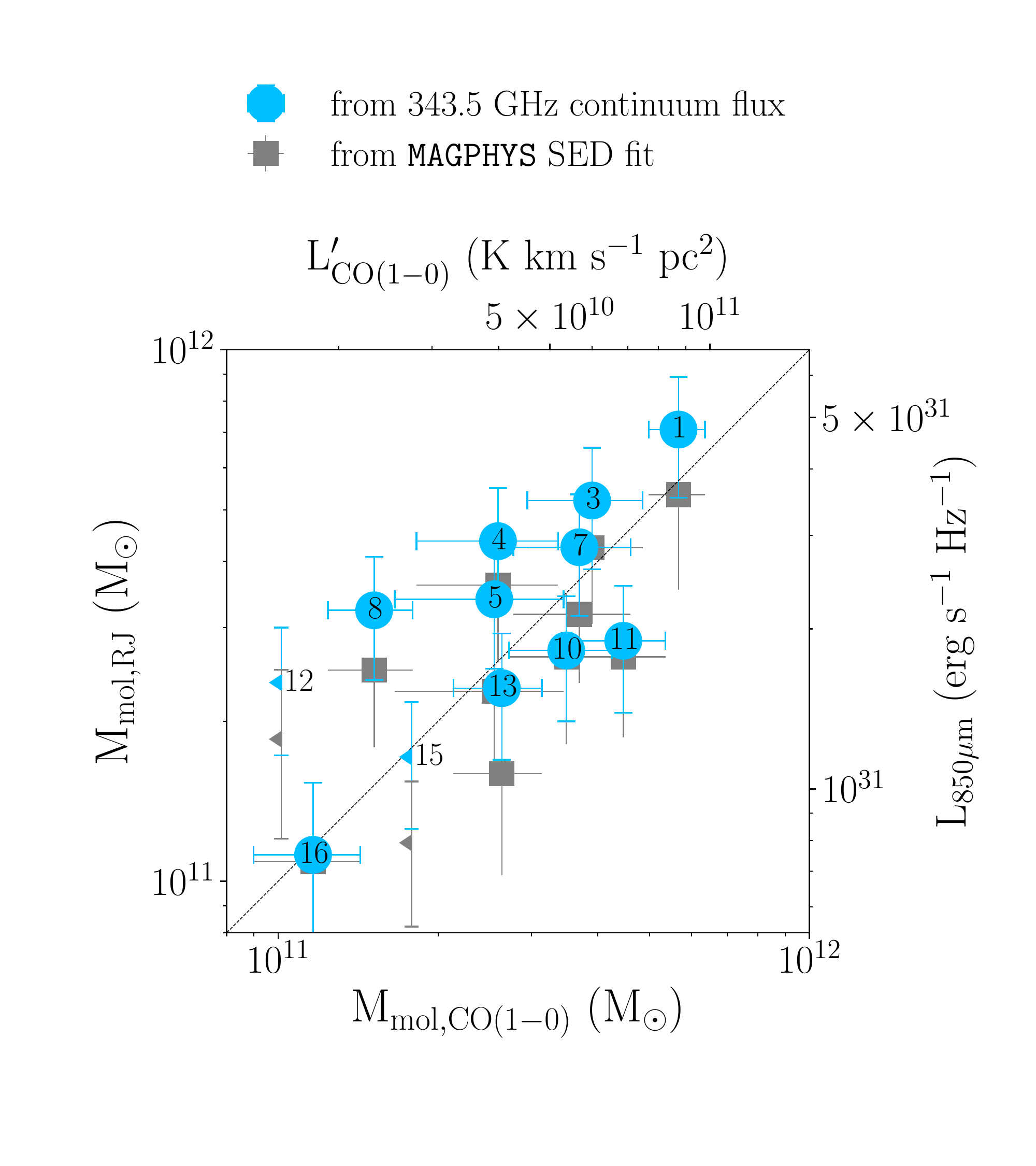}
			\caption{Comparison of CO(1-0) and RJ-based molecular gas masses derived from the single ALMA Band 7 flux measurement (filled sky blue circles) and $\Lref$ derived from the standard \texttt{MAGPHYS} SED fit (filled grey squares). The dotted black line is the line of equality. Note that the RJ-based molecular gas masses are the right y-axis, $\Lref$, divided by $\alpha_\mathrm{850 \mu m}=6.7\times 10^{19} \mathrm{erg \, s^{-1}\, Hz^{-1}\, \Msun^{-1}}$ whereas the CO(1-0) based molecular gas masses are the top x-axis, $L_\mathrm{CO(1-0)}^\prime$, multiplied by $\aco= 6.5\, \Msun / (\mathrm{K}\, \kms \, \mathrm{pc}^{2})$. \label{fig:Mmol_comp}}	
		\end{figure}

	\subsection{CO(1-0) vs. Dust Continuum Luminosity} 
		\label{sub:reference_luminosity}

		We compare the luminosities used to derive \Mmol\ in Figure \ref{fig:lum_comp}. For the CO-detected sources, the observed-frame luminosity (at 343.5 GHz) correlates strongly (Pearson rank correlation coefficient of 0.78) with the CO(1-0) luminosity (left hand panel). This strong correlation remains even after the 343.5 GHz flux is scaled to $\Lref$ (blue points, central panel), indicating that the dust continuum traces the same component of molecular gas as the CO(1-0) emission for our sample. The values of $\Lref$ derived from both the single-band fluxes and SEDs are mostly consistent with the predictions from the $\Lref$-to-\Mmol\ conversion factors derived by \cite{2016ApJ...820...83S} (solid and dotted lines in Figure \ref{fig:lum_comp}), with all but source 12, consistent to within a factor of two. 

		We compare the scatter of our sample, around the best-fit line of $L_\mathrm{CO(1-0)}^\prime$ vs $\Lref$, with the scatter of the calibration sample of \cite{2016ApJ...820...83S}. To quantify the scatter and the goodness of fit compared to the predictions from the mean and best-fit $\alpha_\mathrm{850 \mu m}$, we compute the standardized residuals. By comparing the standardized residuals of the two samples we account for the difference in samples sizes and dynamic ranges.
		
		We use the residuals, given by,
		\begin{align}
			e = L_\mathrm{CO(1-0),measured}^\prime - L_\mathrm{CO(1-0),predicted}^\prime \, ,
		\end{align}
		to calculate the standardized residuals, via, 
		\begin{align}
			e/\sigma_{e} \, ,
		\end{align}
		where $\sigma_{e}$ is the standard deviation of residuals. For each source, $L_\mathrm{CO(1-0),predicted}^\prime$ is inferred from the value of $\Lref$ derived from our Band 7 dust continuum flux measurements (via Equation \eqref{eq:L850}) as well as: (1) the mean, $\alpha_\mathrm{850 \mu m}$, and (2), the best-fit $\alpha_\mathrm{850 \mu m}(\Lref)$, derived for the \cite{2016ApJ...820...83S} calibration sample. We compare the distributions of the standardized residuals from our sample against the \cite{2016ApJ...820...83S} calibration sample in Figure \ref{fig:residual}. Note that for the chosen $\alpha_\mathrm{850 \mu m}$ to be a good fit, the distribution of standardized residuals should approximate a normal distribution, with 95\% between -2 and +2. Because the best-fit $\alpha_\mathrm{850 \mu m}$ was derived as a fit to the calibration data, the standardized residuals should be centred at 0, and, indeed we find a mean (median) standardized residual of 0.00 (0.15). 

		Our sample exhibits the same level of scatter as the calibration sample of \cite{2016ApJ...820...83S} and is well fit by the constant and best-fit $\alpha_\mathrm{850 \mu m}$, derived using the calibration sample. For both our sample, and the calibration sample, the standard deviation of the distribution of standardized residuals $\approx 1.0$, indicating that our sample exhibits the same variation around the line of best fit as the calibration sample. %
		Our sample exhibits a very small (and insignificant) offset from the best fit to the calibration sources, with a mean (median) standardized residual of -0.40 (-0.36). Moreover, the standardized residuals of our sample lie within, $|e/\sigma_{e}|<2$, indicating that the line that best fits the calibration data is also suitable for our sample. 

		We find no significant difference in the goodness of fit when applying the mean vs best-fit $\alpha_\mathrm{850 \mu m}$ of \cite{2016ApJ...820...83S}, with almost identical distributions of the standardized residuals for both cases (filled red vs solid red outlined histogram, Figure \ref{fig:residual}). Moreover, the mean $\alpha_\mathrm{850 \mu m}$ of our sample, $7.7\pm2.3 \times 10^{19} \, \mathrm{erg \, s^{-1}\, Hz^{-1}\, \Msun^{-1}} $, is consistent with the mean of the calibration samples. Thus, we conclude that the application of the constant value, $\left<\alpha_\mathrm{850 \mu m}\right> = 6.7\pm 1.7 \times 10^{19} \mathrm{erg \, s^{-1}\, Hz^{-1}\, \Msun^{-1}}$, is valid for our sample.    
		


		\begin{figure*}
			\centering
			\includegraphics[width=0.95\textwidth,trim={1.2cm 0.2cm 0.6cm 0.2cm},clip]{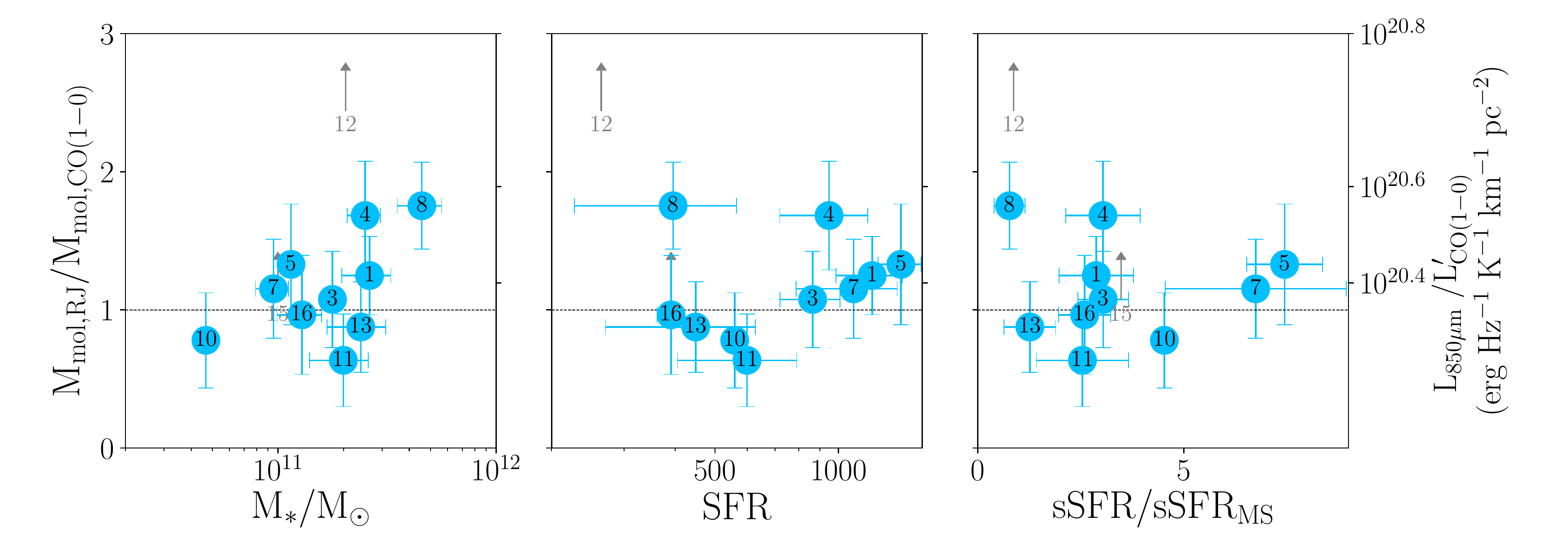}
			\caption{Ratio between the CO(1-0)-based to dust-based molecular gas masses as a function of stellar mass (left), SFR (center) and main sequence offset (right). \label{fig:Mmolratio_vs}}		
		\end{figure*}

	\subsection{The Accuracy of CO(1-0) and Dust Continuum-Based Molecular Gas Masses} 
		\label{sub:molecular_gas_masses}

		The molecular gas masses derived from the RJ dust continuum are consistent with those derived from CO(1-0) (Figure \ref{fig:Mmol_comp}).  For six of the 10 CO-detected sources, the values of \Mmol\ derived from the single-band RJ method are consistent within the errors, whereas for all CO-detected sources, the values of \Mmol\ are consistent to within a factor of 1.7. The main source of uncertainty in measuring \Mmol\ from the dust continuum is the conversion to a common rest-frame luminosity. As discussed in Section \ref{sub:comparison_of_lref}, the variation of $\beta$ and $<T_\mathrm{dust}>$ alone are sufficient to account for the differences in the derived $\Lref$ and thus also the differences between the RJ and CO-based \Mmol. 

	 	We investigate whether the scatter about the 1:1 line in Figure \ref{fig:Mmol_comp} correlates with any observed galaxy properties (Figure \ref{fig:Mmolratio_vs}). To quantify the scatter, we use the relative luminosities, $\Lref/L_\mathrm{CO(1-0)}^\prime$. Note that this is equivalent to the ratio between the value of \Mmol\ determined from the single-band RJ method and CO(1-0) ($\mathrm{M_{mol,RJ}}/\mathrm{M_{mol,CO(1-0)}}$). We find no statistically significant correlation with the stellar mass, SFR or main sequence offset within our sample (Spearman rank coefficients of $<0.3$ with p-values of $>0.5$), despite our sample spanning an order of magnitude in both \Mstar\ and SFR. 

	 	The robustness of $\Lref$ as a molecular gas tracer for our sample, appears to be consistent with recent studies based on cosmological zoom-in simulations. \cite{2018arXiv180503649P} and \cite{2018MNRAS.tmpL..76L} compared the RJ-derived \Mmol\ with the \Mmol\ of simulated massive star-forming galaxies, finding that the RJ-based \Mmol\ correlates well with ``true'' \Mmol\ of their simulated galaxies. The strong correlation only appeared to break down for sources with $L_{850\mu m,rest}<10^{28} $ erg s$^{-1}$ or $\log{(Z/Z_\odot)} <-0.8$ \citep{2018arXiv180503649P}, which are likely to have significantly higher gas-to-dust ratios than expected for our sample. 
	 	The fact that we find such strong agreement between the CO-derived \Mmol\ and RJ-based \Mmol, indicates that our sources have similar gas-to-dust ratios to the calibration samples of \cite{2016ApJ...820...83S}. 

	 	The assumption of a constant gas-to-dust abundance ratio is implicit in the RJ method, for which the calibration and application has so far been intentionally restricted to galaxies with $\Mstar>2\times10^{10}\Msun$ \citep{2016ApJ...820...83S,2017ApJ...837..150S}. These massive sources are expected to be high metallicity systems, with gas-to-dust ratios of the order of $\sim 100:1$ and little CO-dark gas \citep{2013ARA&A..51..207B}.  Based on the M$_\mathrm{dust}$ inferred via \texttt{MAGPHYS}, and the $\aco$ used here, our sample has gas-to-dust ratios of the order of $\sim 300:1$. Although there likely exists some variation in these values, which contributes to the scatter about the 1:1 line in Figure \ref{fig:Mmol_comp}, it is not sufficient to lead to any significant offset in \Mmol. We note that these results are consistent with the work of \cite{2015ApJ...799...96G}, who find that the total gas mass can be constrained to within a factor of two for massive galaxies.
	 	
	 	We do not find any major deviations between the CO- and RJ- derived gas masses of our sample. However, significant deviations are to be expected for low metallicity, and low \Mstar, galaxies. For example, \cite{2015ApJ...799...96G} and \cite{2018arXiv180503649P} show that for low metallicity (low \Mstar) galaxies, the calibration based on higher mass sources (such as the one tested here) would underpredict the observed $L_\mathrm{CO(1-0)}^\prime$, or simulated \Mmol, respectively. Similarly, \cite{2015A&A...577A..50D} find that low-mass, MS galaxies at $z\sim2$ show smaller $\alpha_\mathrm{850\mu m}$ than our sample by $\sim 0.4$ dex.  These low mass galaxies are likely to have significantly higher gas-to-dust ratios, of $\sim 1000:1$, \citep[e.g.][]{2017MNRAS.471.3152P} than the samples to which the RJ method has so far been applied (which have ratios of $\sim 100:1$). Hence, the constant gas-to-dust abundance ratio, implicit in the RJ method, is not applicable to low \Mstar\ sources. The underprediction of the CO-derived \Mmol\ for low \Mstar\ sources would be even more pronounced if one were to account for the expected variation of $\aco$ with metallicity \citep[e.g.][]{2006MNRAS.371.1865B,2010ApJ...716.1191W,2011MNRAS.412.1686S,2012ApJ...747..124F,2013ARA&A..51..207B}. 
	 	

		Our results indicate that for massive, star-forming galaxies, the main source of uncertainty in deriving \Mmol\ remains the conversion factor, $\aco$. The value of $\aco$ we have adopted does not affect the relative comparison of the \Mmol\ derived via CO(1-0) and the long-wavelength dust continuum, both of which depend on the same $\aco$. However, it has a significant impact on the absolute \Mmol\ inferred via either method. As argued in previous studies, the value of $\aco$ is likely to vary across the calibration samples, with values of $\sim 4$, $\sim 1.3$ and $\sim 0.8 \mathrm{\Msun/ (K\, \kms pc^{2})}$ derived for SFGs, ULIRGs and SMGs, respectively \citep[see list of references in][]{2013ARA&A..51..207B}. In addition to the $\aco$ adopted here, this variation can lead to a factor of $\lesssim 8$ uncertainty in the derived \Mmol, far more than the factor of $\sim 2$ uncertainty introduced by the assumptions of $\beta$ and $\left<T_\mathrm{dust}\right>$ in the RJ method (Section \ref{sub:comparison_of_lref}).



\section{Summary} 
	\label{sec:summary}

	We have presented a comparison of the molecular gas masses derived from dust continuum and CO(1-0) observations for a unique sample of 12, $z\sim 2$ star-forming galaxies. This is the first time that the Rayleigh-Jeans method of measuring molecular gas masses from single-band, dust continuum observations has been tested on the types of high-redshift sources to which it is typically applied \citep{2016ApJ...820...83S,2017ApJ...837..150S,2016ApJ...833..112S,Miettinen_2017,2018ApJ...860..111D}. 

	The dust-continuum method we have tested is based on the calibration of the rest-frame $850\mu m$ luminosity ($\Lref$) against the CO(1-0)-derived molecular gas masses, proposed by \cite{2016ApJ...820...83S}. We applied two techniques to derive the rest-frame $850\mu m$ luminosity: (1), using the calibration of \cite{2016ApJ...820...83S}, based on the observed dust-continuum emission at 343.5 GHz (ALMA Band 7), and (2) from the model fits to the rest-frame SEDs. 

	Our work can be summarised as follows:
	\begin{enumerate}[leftmargin=*,noitemsep,nolistsep,nosep] 
	 	\item We find that the molecular gas masses derived from single-band, long-wavelength flux measurements are consistent, within a factor of two, with the molecular gas masses derived from CO(1-0), for $\Mstar>2\times 10^{10}\Msun$ star-forming galaxies, when the same $\aco$, is assumed.
	 	\item This factor of $<2$ variation between the \Mmol\ derived from CO(1-0) vs the single-band dust continuum, can be accounted for by variations in the dust emissivity index and temperature, which are assumed when extrapolating from the observed flux to $\Lref$. Small variations in the gas-to-dust ratios are also likely to contribute to this scatter.
	 	\item The main source of uncertainty in deriving \Mmol, regardless of whether one uses CO(1-0) or the dust continuum, remains the CO-to-\Mmol\ conversion factor, $\aco$, which varies by a factor of up to $\sim 8$. 
	 \end{enumerate}  

	We conclude that single-band, dust-continuum observations can be used to constrain the molecular gas masses of massive ($\Mstar>2\times 10^{10}\Msun$), star-forming galaxies at high redshift as reliably as the CO(1-0) line. Thus, future single-band surveys with ALMA will provide important constraints on the physics of star formation in massive, high-z galaxies.



\acknowledgments

We thank the anonymous referee whose comments improved the clarity and details of this work. We thank Olivier Ilbert for the assistance provided in cross-checking the fits to the COSMOS photometry using \texttt{LePHARE}. We also thank Alberto Bolatto and Desika Narayanan for their helpful insights. SCOG acknowledges support from the DFG via SFB 881 ``The Milky Way System'' (sub-projects B1, B2 and B8). CMC thanks UT Austin College of Natural Sciences for support in addition to support from NSF grants AST-1714528 and AST-1814034. BD acknowledges financial support from the National Science Foundation, grant number 1716907.  DR and RP acknowledge support from the NSF, through grant AST-1614213. EdC gratefully acknowledges the Australian Research Council as the recipient of a Future Fellowship (project FT150100079). MK acknowledges support from the International Max Planck Research School for Astronomy and Cosmic Physics at Heidelberg University (IMPRS-HD). Parts of this research were conducted by the Australian Research Council Centre of Excellence for All Sky Astrophysics in 3 Dimensions (ASTRO 3D), through project number CE170100013. This paper makes use of the VLA data: VLA/17A-251, and, the ALMA data: 2013.1.00034.S, 2015.1.00137.S, 2013.1.00118.S, and 2013.1.00151.S. The Joint ALMA Observatory is operated by ESO, AUI/NRAO, and NAOJ. The National Radio Astronomy Observatory is a facility of the National Science Foundation operated under cooperative agreement by Associated Universities, Inc. The near IR spectroscopy presented herein were obtained at the W. M. Keck Observatory, which is operated as a scientific partnership among the California Institute of Technology, the University of California, and the National Aeronautics and Space Administration. The Observatory was made possible by the generous financial support of the W. M. Keck Foundation. The authors would like to recognize and acknowledge the very prominent cultural role and reverence that the summit of Mauna Kea has always had within the indigenous Hawaiian community. We are fortunate to have the opportunity to perform observations from this mountain.
 

\vspace{5mm}
\facilities{VLA, ALMA, Keck I}

\bibliographystyle{apj}
\bibliography{mybib}

\appendix

\section{Data}
\label{sec:App_data}

\begin{figure*}[h!]
		\centering
		\includegraphics[width=0.48\columnwidth, trim={0 0.4cm 0 0.4cm},clip]{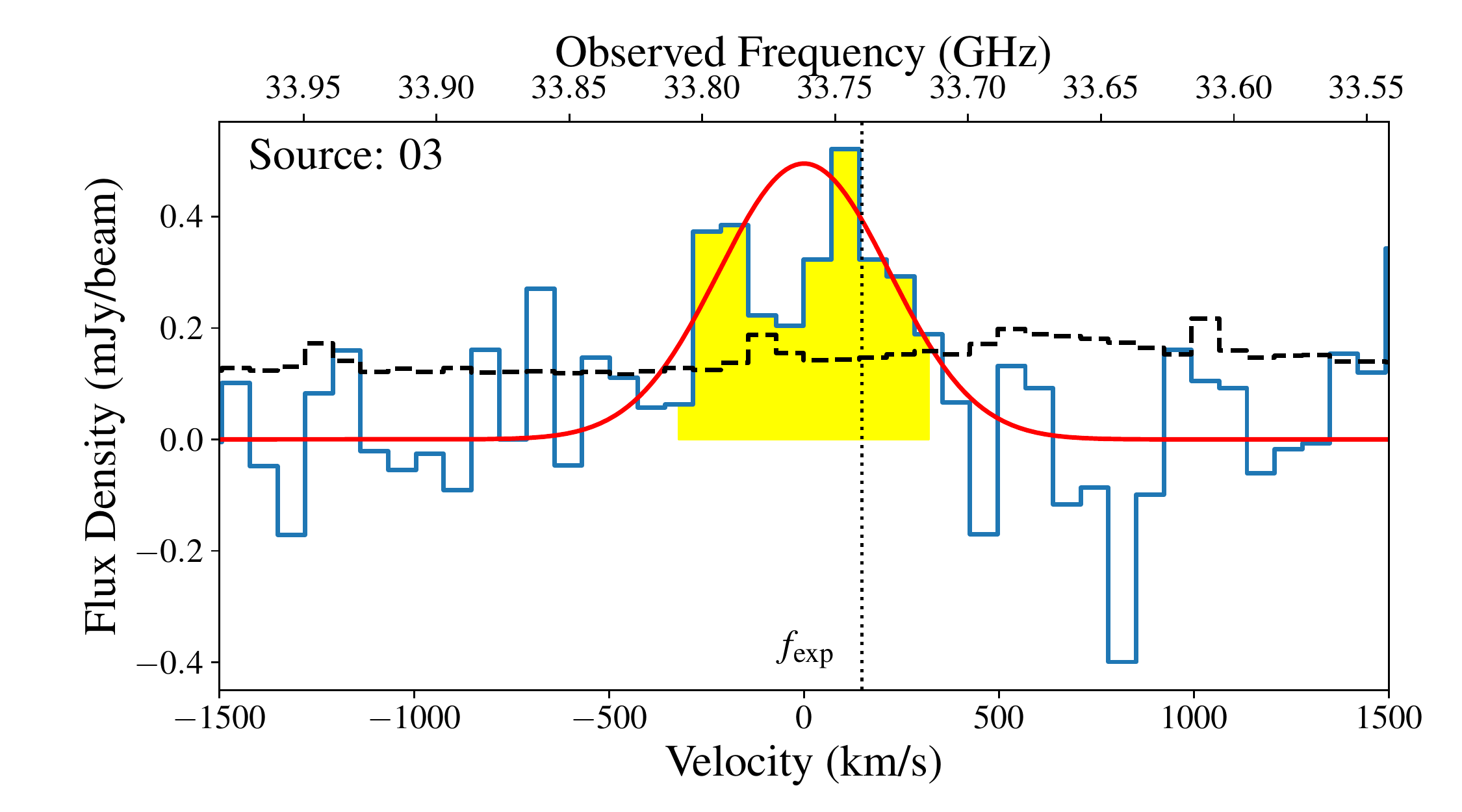}
		~
		\includegraphics[width=0.48\columnwidth, trim={0 0.4cm 0 0.4cm},clip]{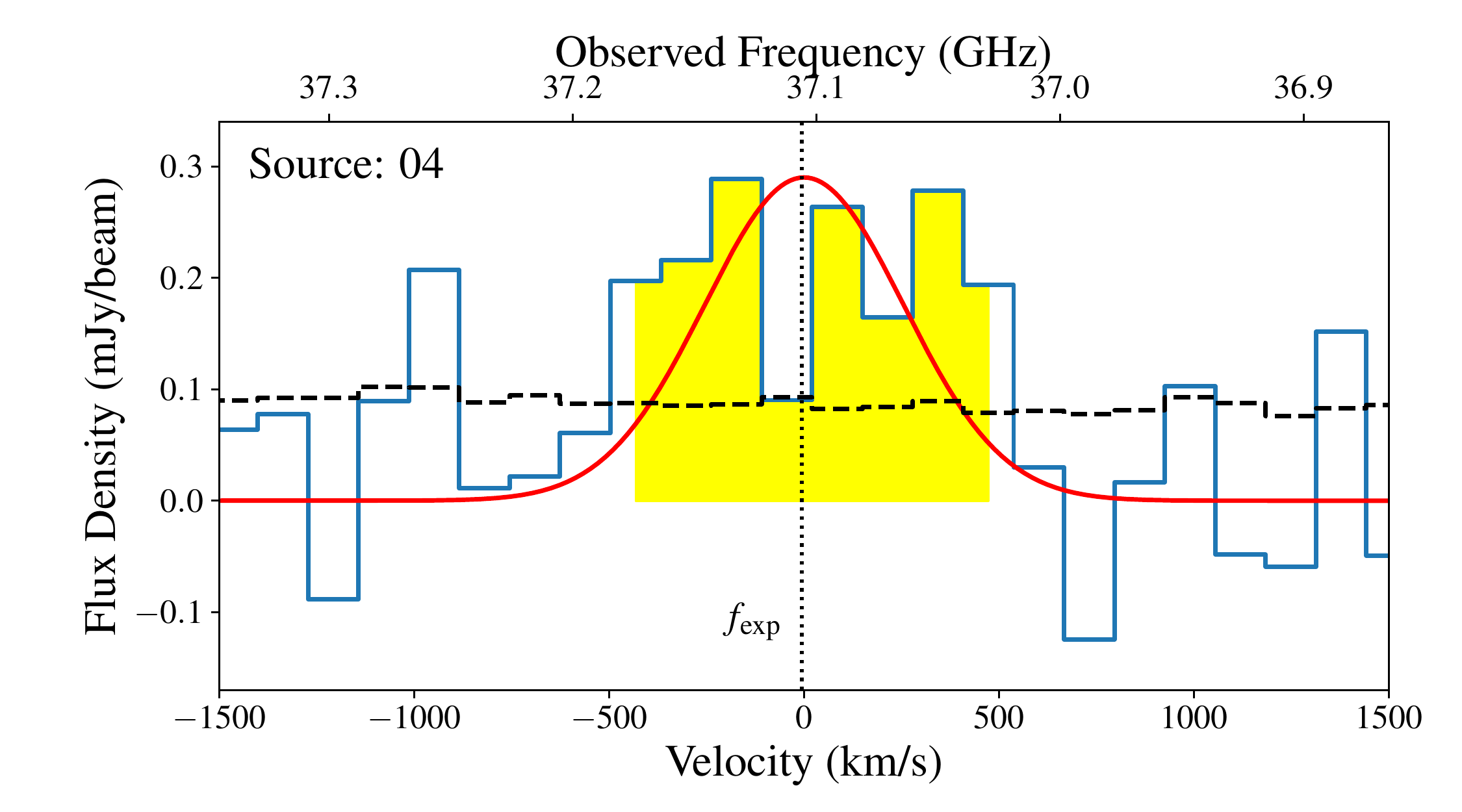}
		\\
		\includegraphics[width=0.48\columnwidth, trim={0 0.4cm 0 0.4cm},clip]{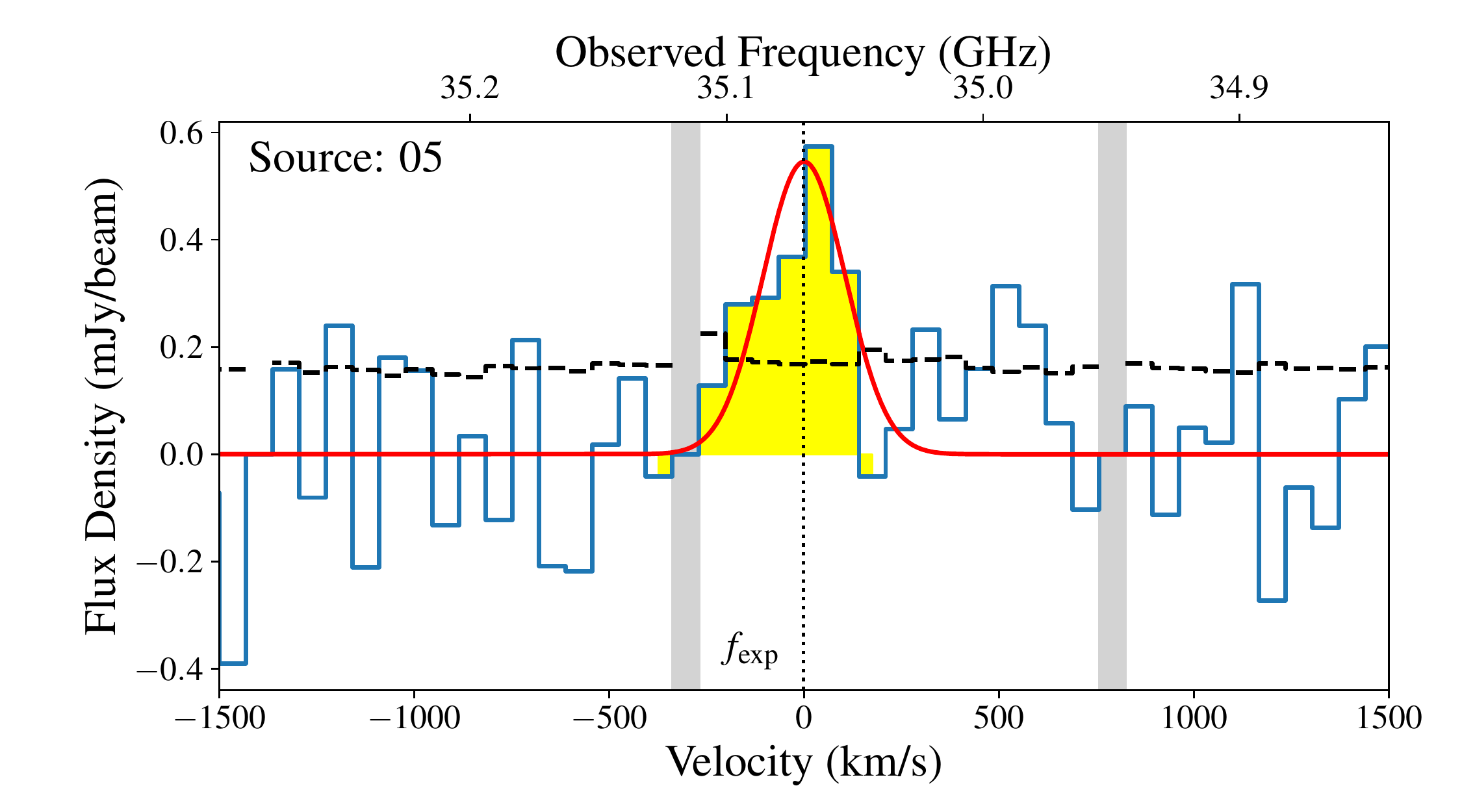}
		~
		\includegraphics[width=0.48\columnwidth, trim={0 0.4cm 0 0.4cm},clip]{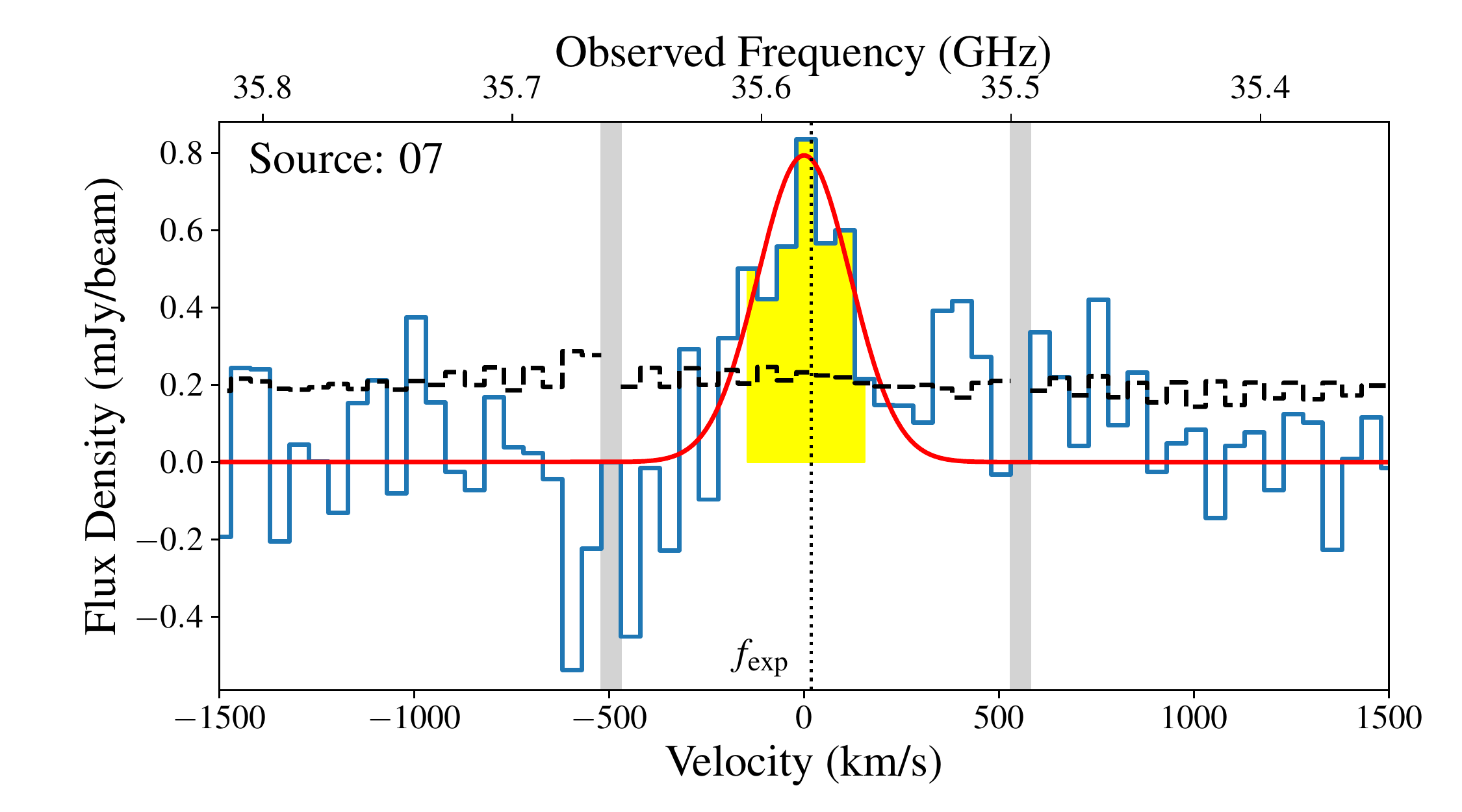}
		\\
		\includegraphics[width=0.48\columnwidth, trim={0 0.4cm 0 0.4cm},clip]{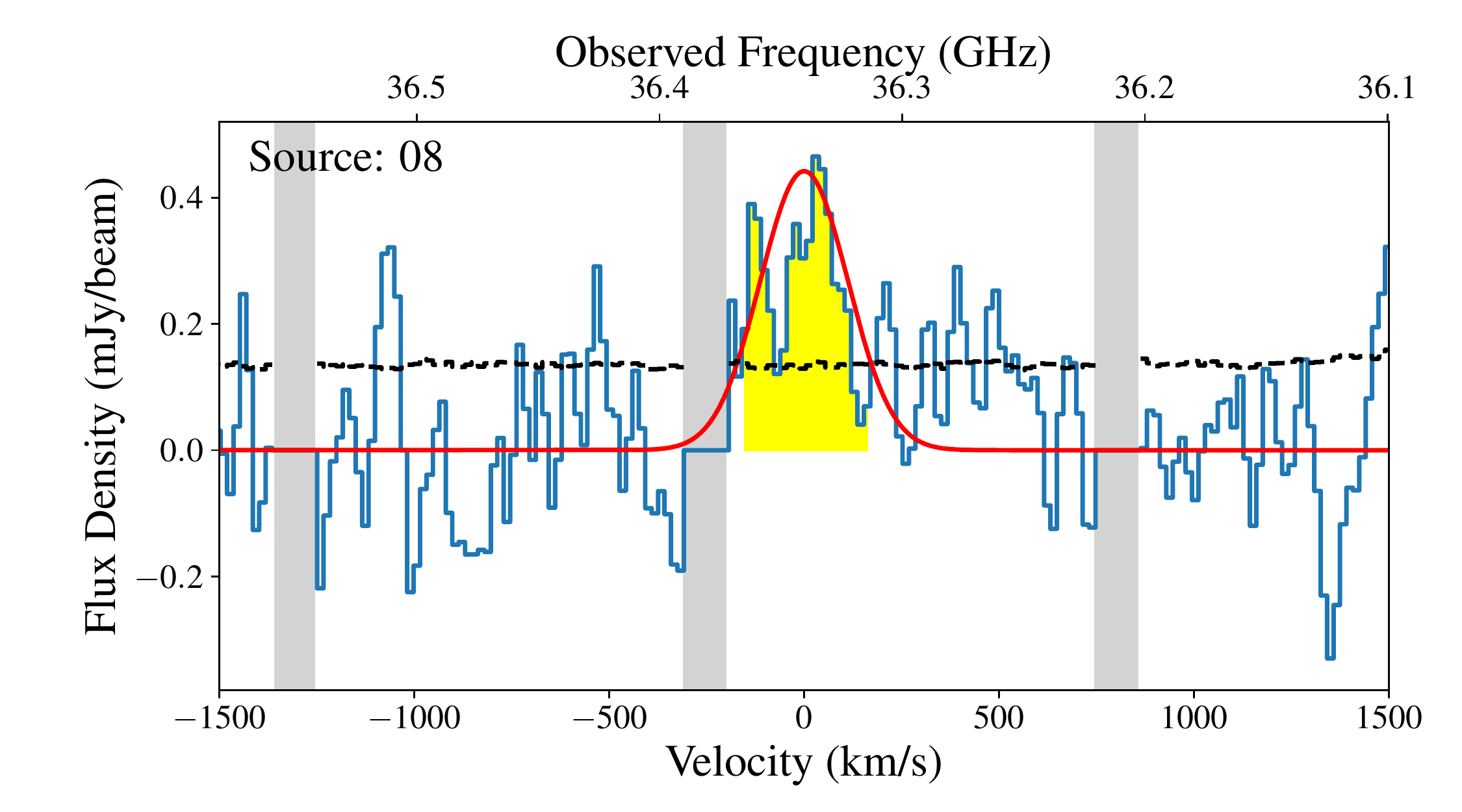}
		~
		\includegraphics[width=0.48\columnwidth, trim={0 0.4cm 0 0.4cm},clip]{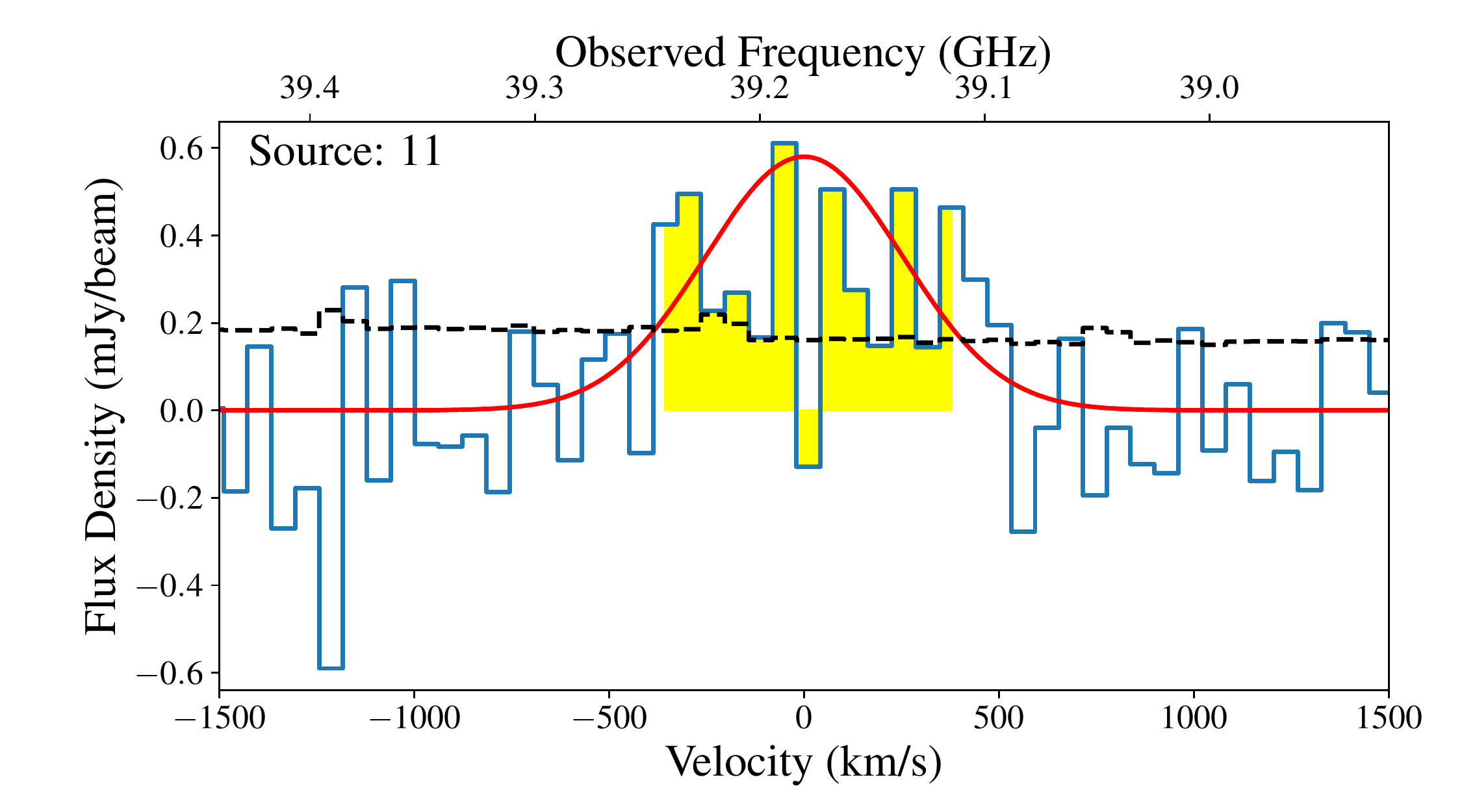}
		\\
		\includegraphics[width=0.48\columnwidth, trim={0 0.4cm 0 0.4cm},clip]{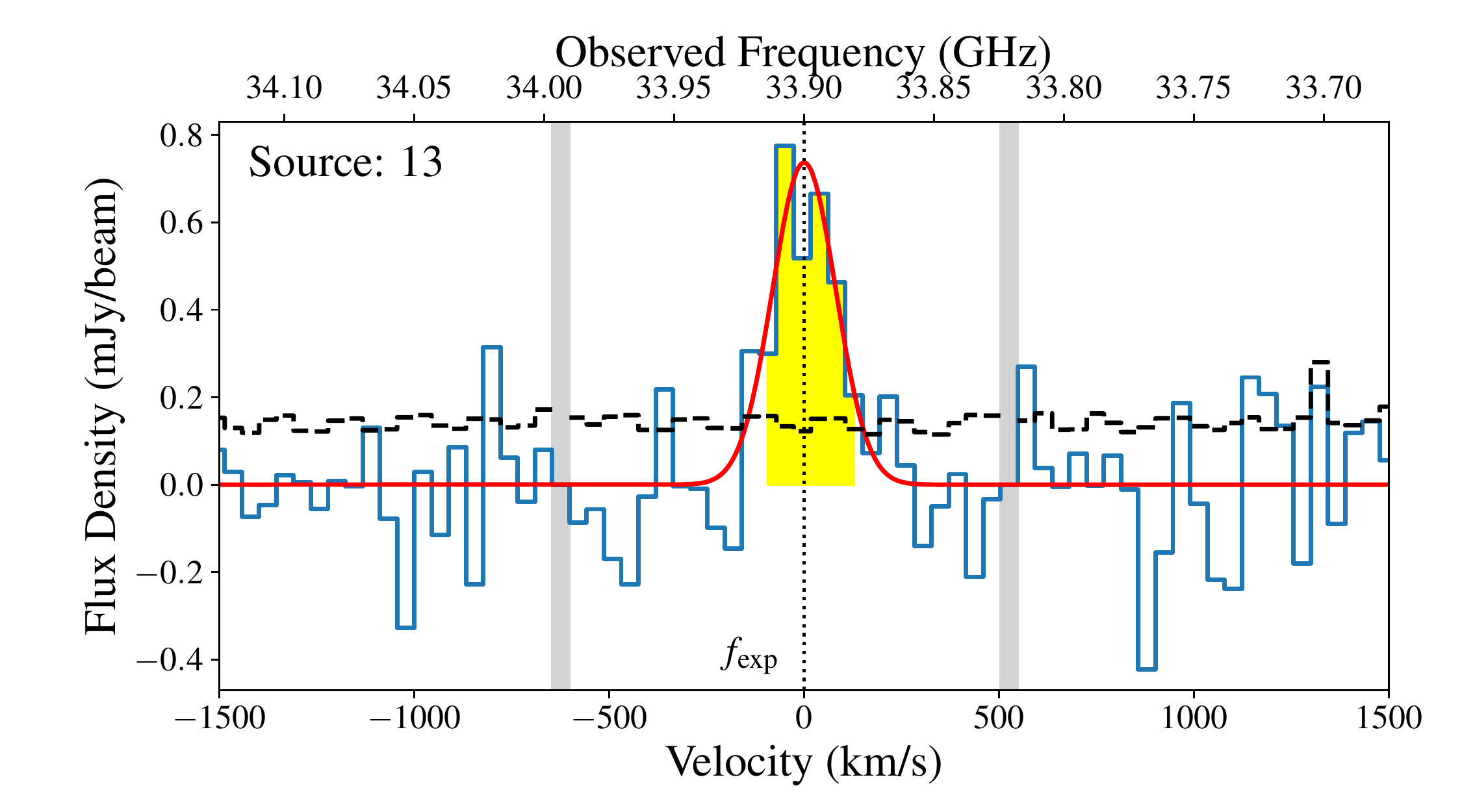}
		~
		\includegraphics[width=0.48\columnwidth, trim={0 0.4cm 0 0.4cm},clip]{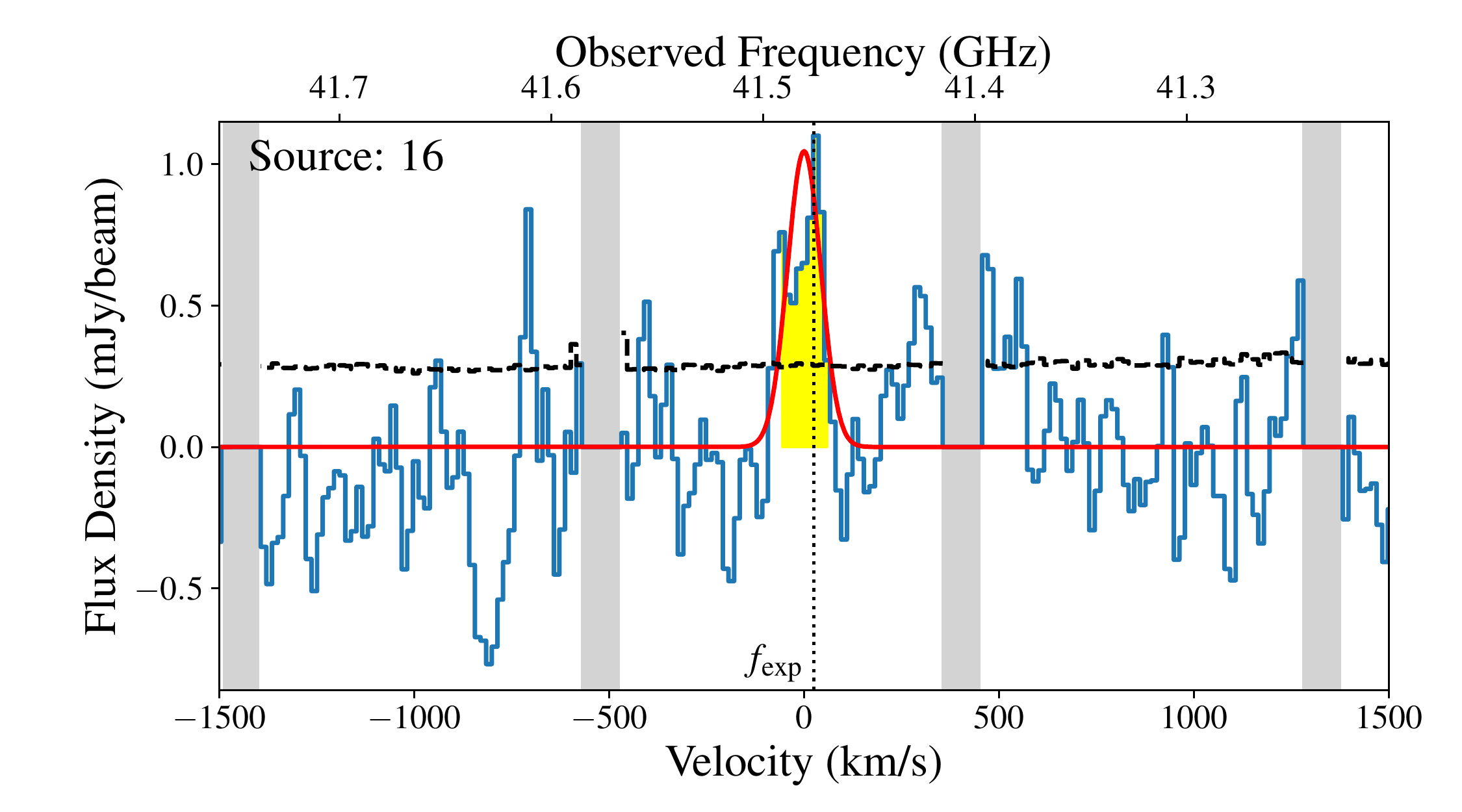}
		\caption{Spectra of the CO(1-0)-detected sources (blue). The Gaussian line fits are shown for comparison (red) with the spectral region used to create the integrated maps in Figure \ref{fig:chmap_comp} shaded in yellow ($1.2$ FWHM for all sources apart from 4 and 5). Flagged channels, not used for the line fits, are shaded in grey. The root-mean-square noise per channel is indicated by the black, dashed histogram. \label{fig:spectra}}
	\end{figure*}

We provide the data for our sources here. In Figure \ref{fig:non_detections}, we provide all 16 observed VLA spectra, extracted at the source position. The spectral line fits described in Section \ref{sub:co_dat} are shown in Figure \ref{fig:spectra} and the dust continuum and CO(1-0) emission maps of our final sample of CO-detected sources are shown in Figure \ref{fig:chmap_comp}. The SED fits, described in Sections \ref{sub:sed_based_method} and \ref{subsub:SED_fitting}, are shown along with the photometric measurements on which they are based in Figure \ref{fig:SEDs}.

	\begin{figure*}[h!]
		\centering
		\includegraphics[width=0.48\columnwidth, trim={0 0.4cm 0 0.4cm},clip]{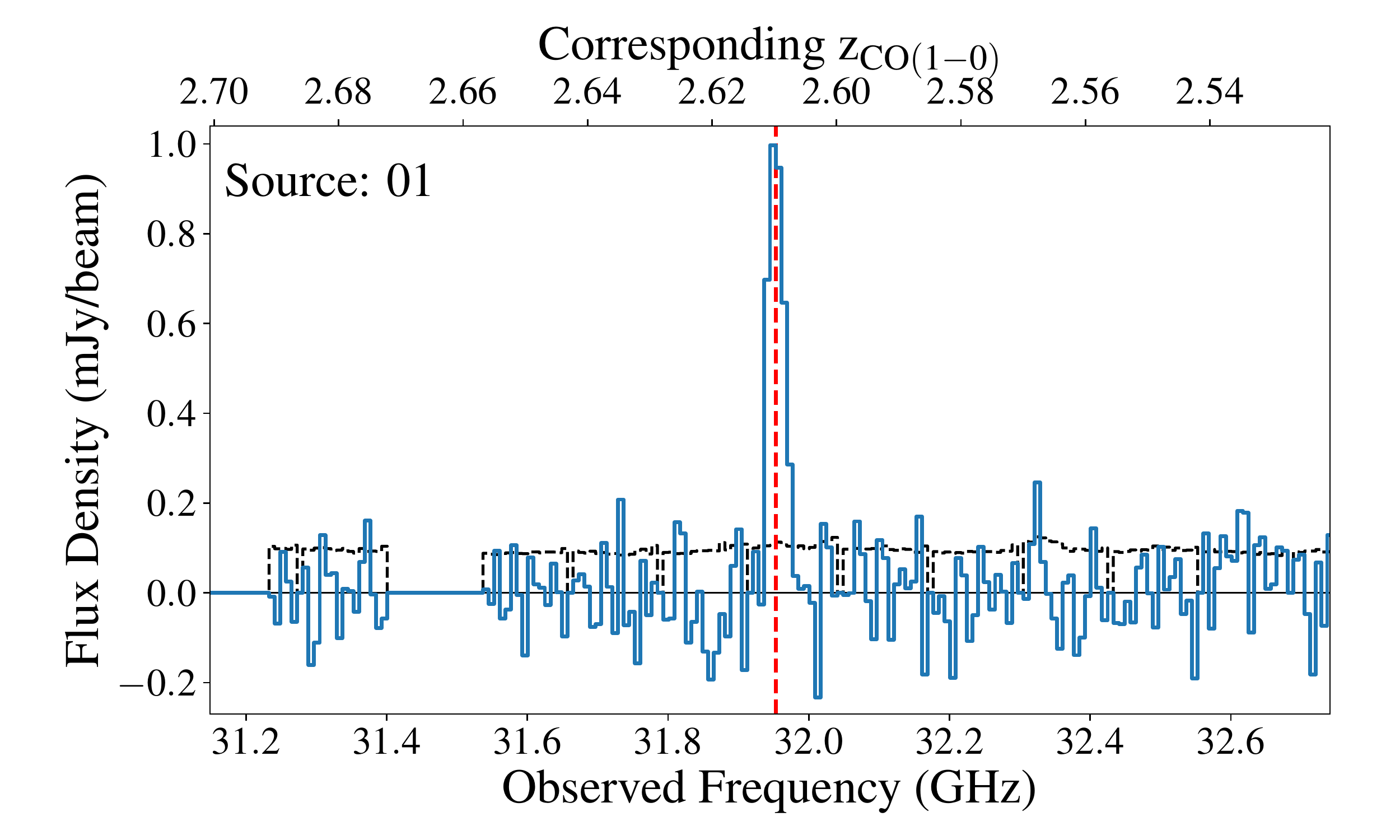}
		~
		\includegraphics[width=0.48\columnwidth, trim={0 0.4cm 0 0.4cm},clip]{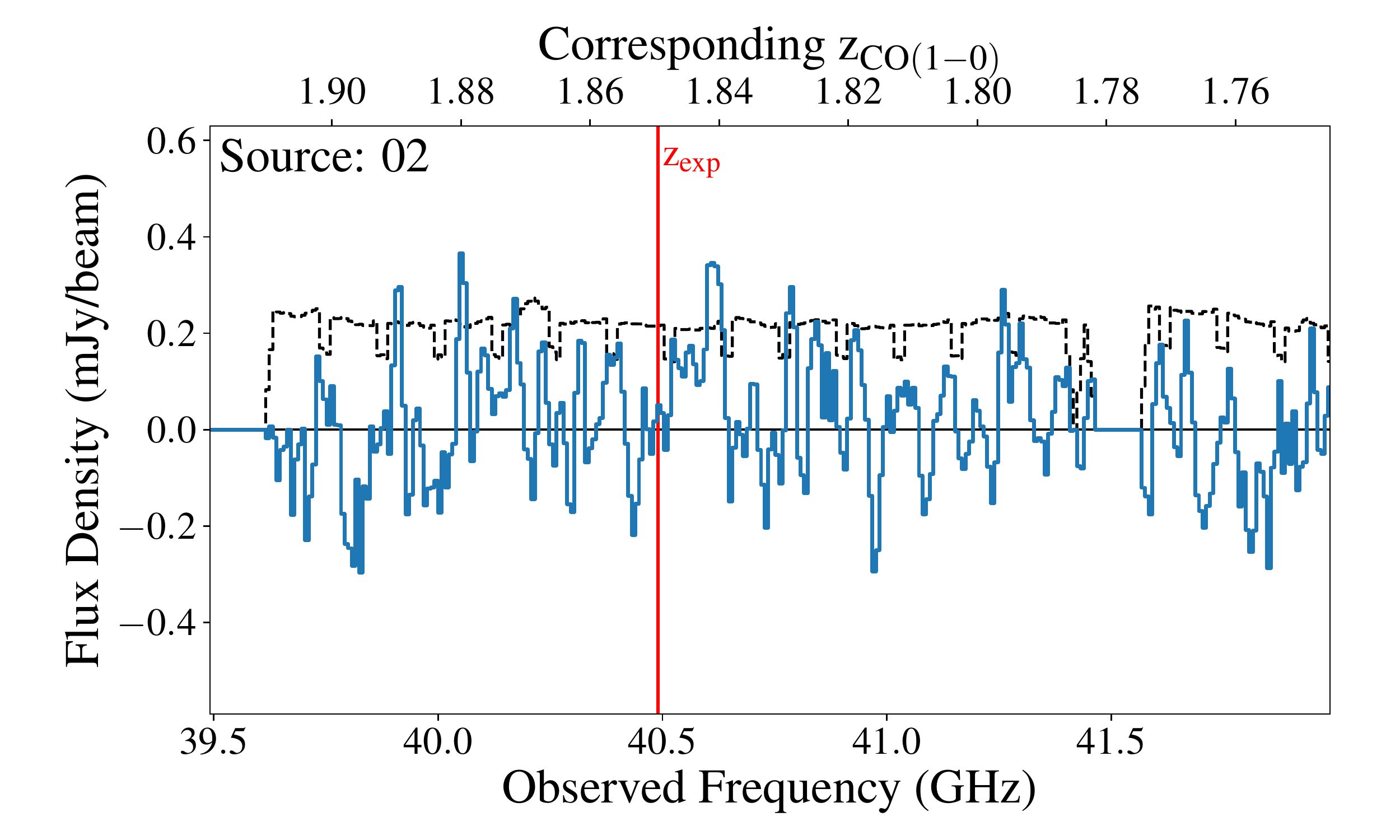}
		\\
		\includegraphics[width=0.48\columnwidth, trim={0 0.4cm 0 0.4cm},clip]{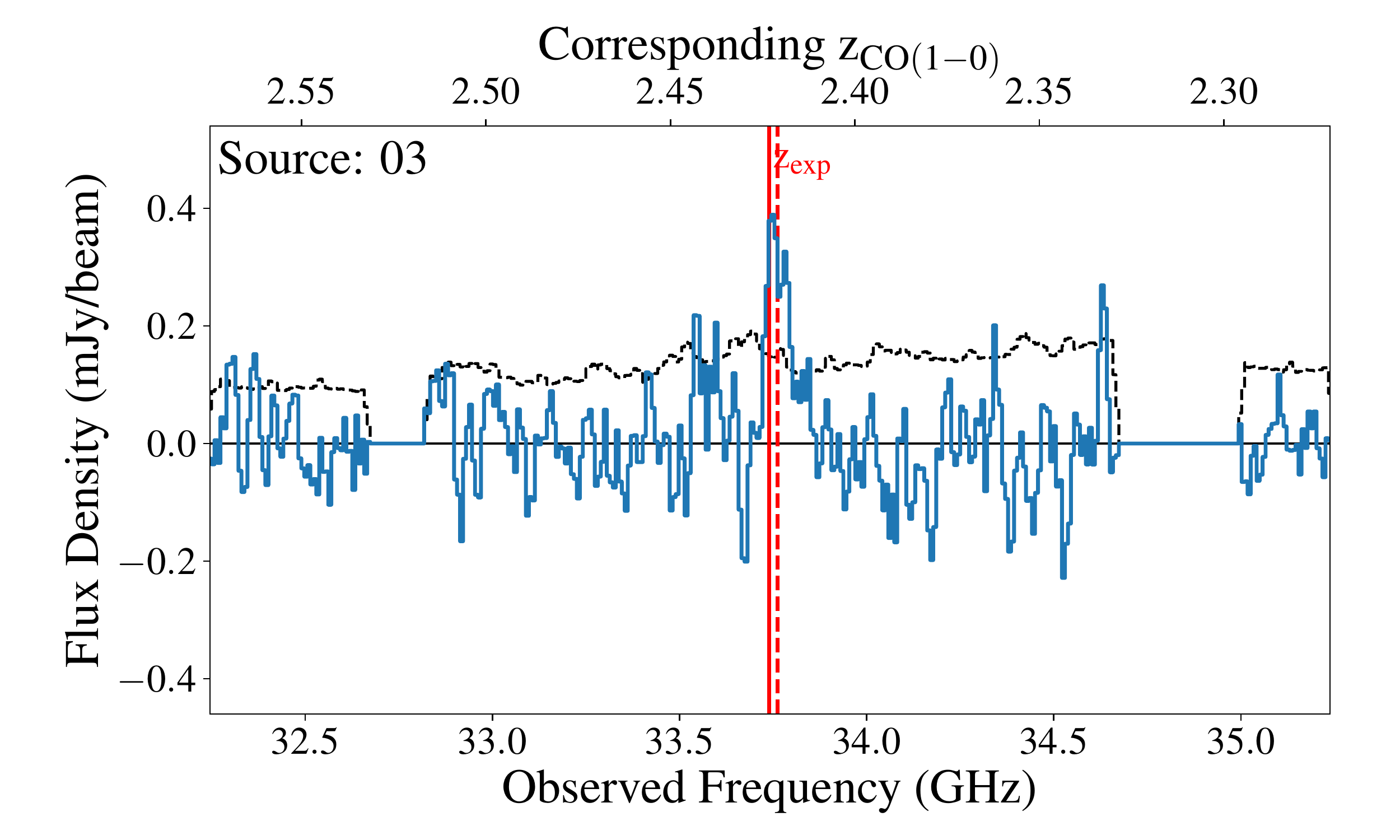}
		~
		\includegraphics[width=0.48\columnwidth, trim={0 0.4cm 0 0.4cm},clip]{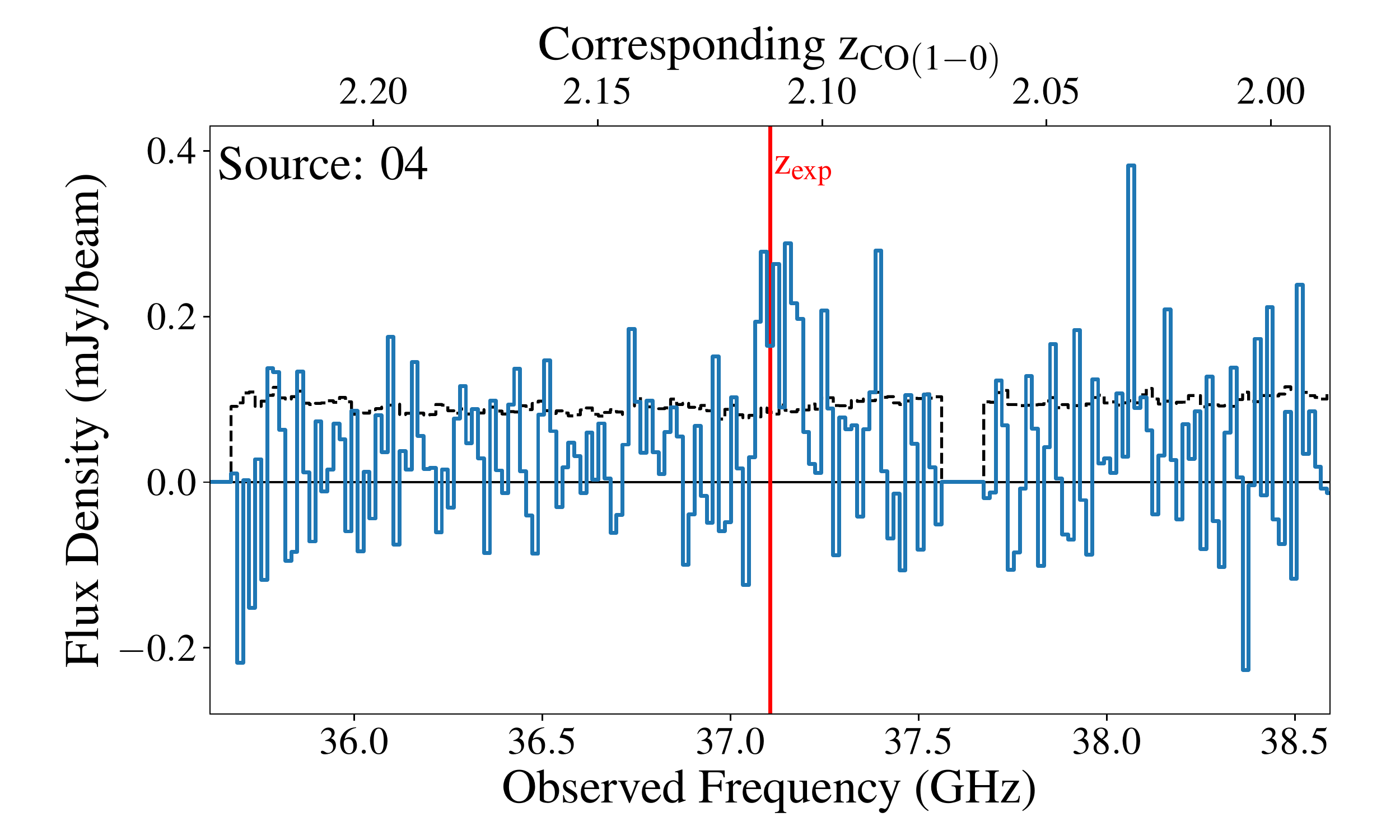}
		\\
		\includegraphics[width=0.48\columnwidth, trim={0 0.4cm 0 0.4cm},clip]{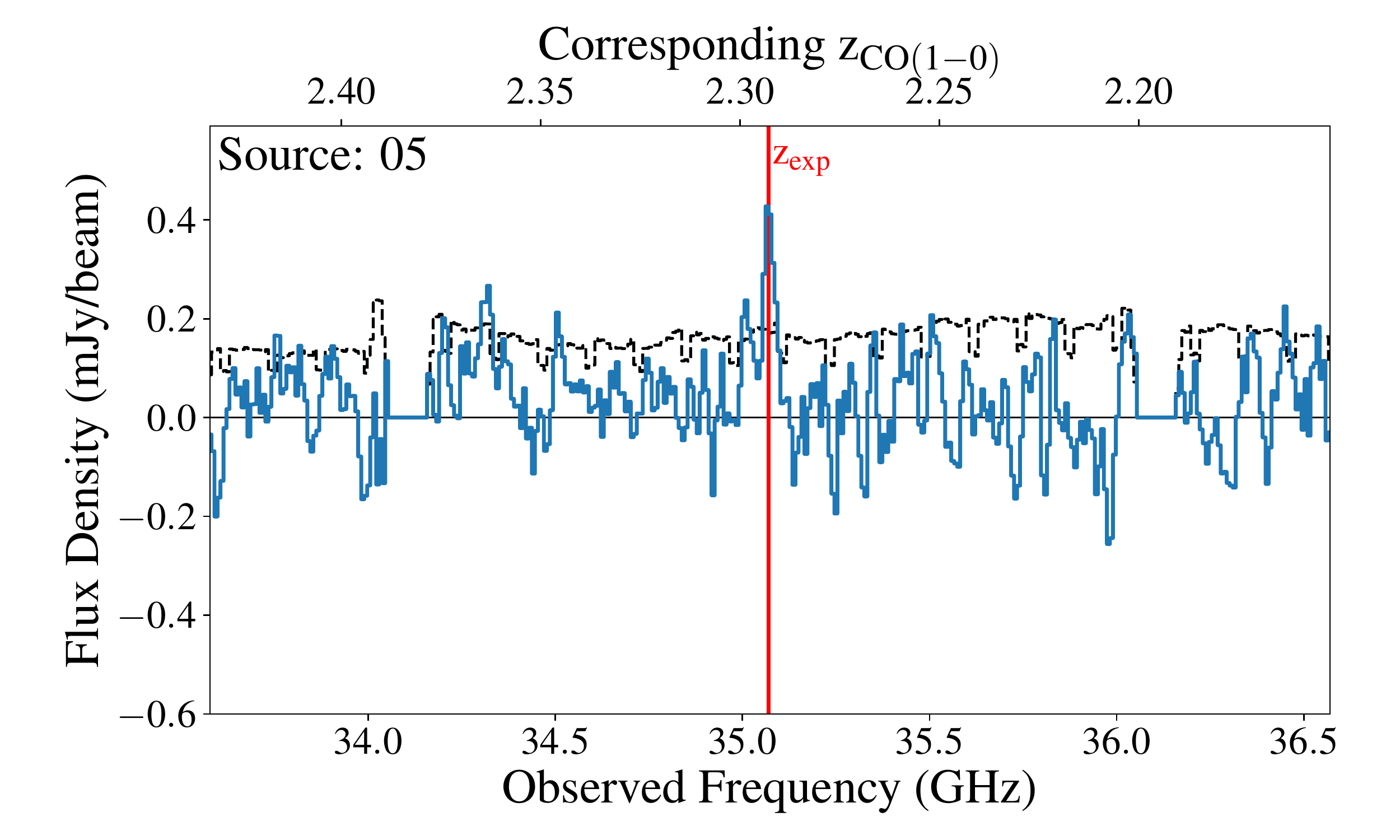}
		~
		\includegraphics[width=0.48\columnwidth, trim={0 0.4cm 0 0.4cm},clip]{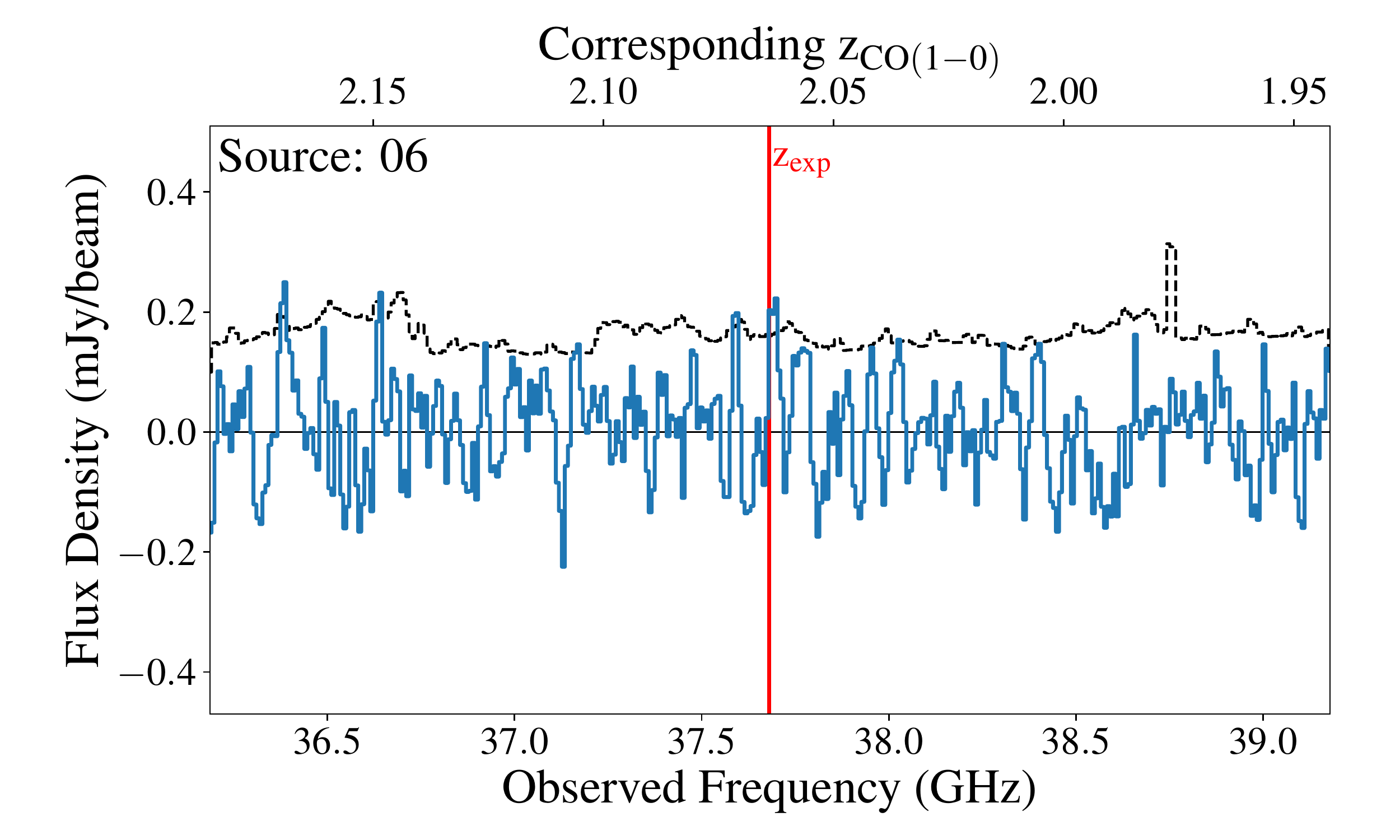}	
		\caption{VLA source spectra extracted at the expected source position. The catalogued redshift, used to design the observations, is indicated by the solid, red, vertical line and is labelled as $z_\mathrm{exp}$. For cases where the CO(1-0)-derived redshift differed from this expected redshift, the CO(1-0)-derived redshift is shown by a dashed, red, vertical line.  \label{fig:non_detections}}
	\end{figure*}

	\begin{figure*}[h!]
		\ContinuedFloat
		\includegraphics[width=0.48\columnwidth, trim={0 0.4cm 0 0.4cm},clip]{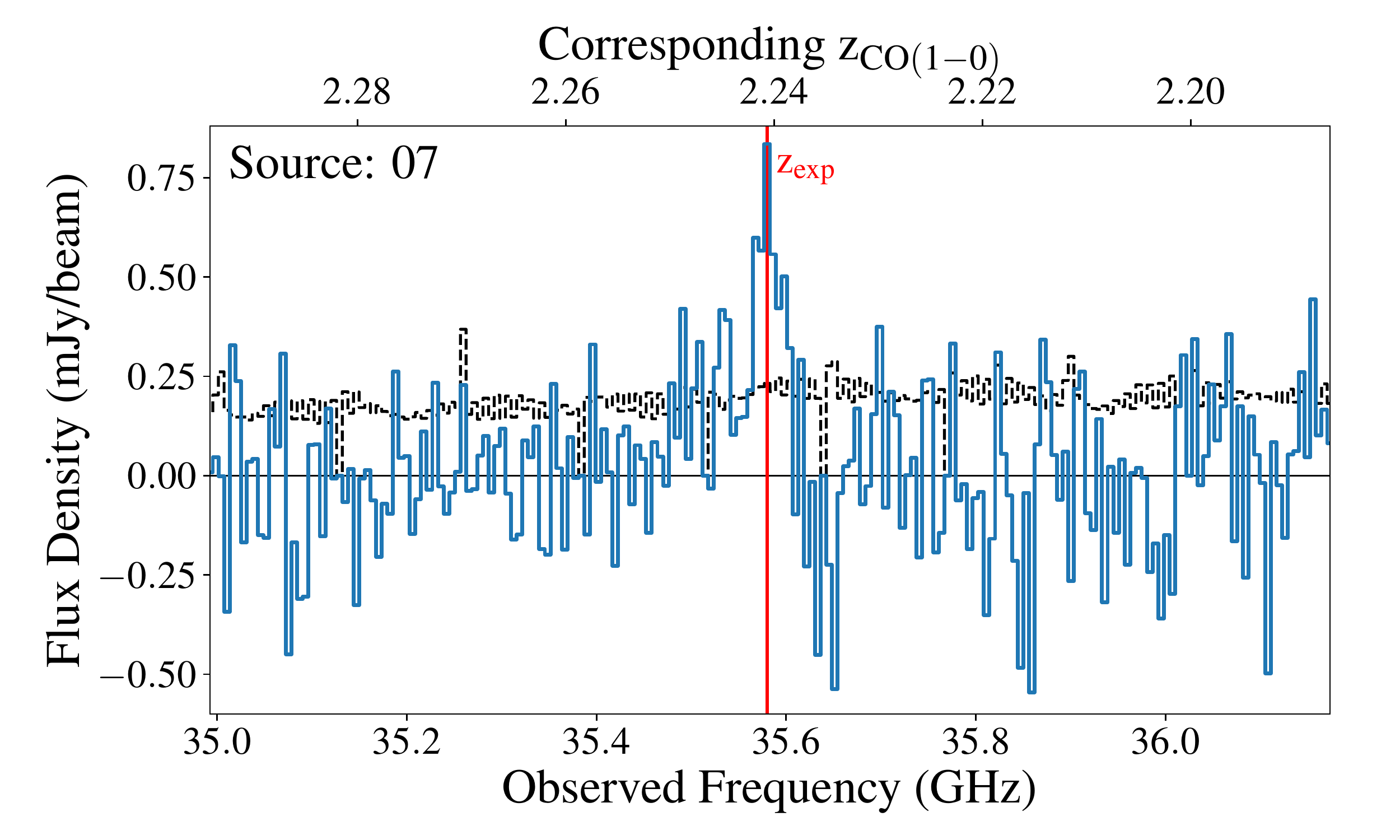}
		~
		\includegraphics[width=0.48\columnwidth, trim={0 0.4cm 0 0.4cm},clip]{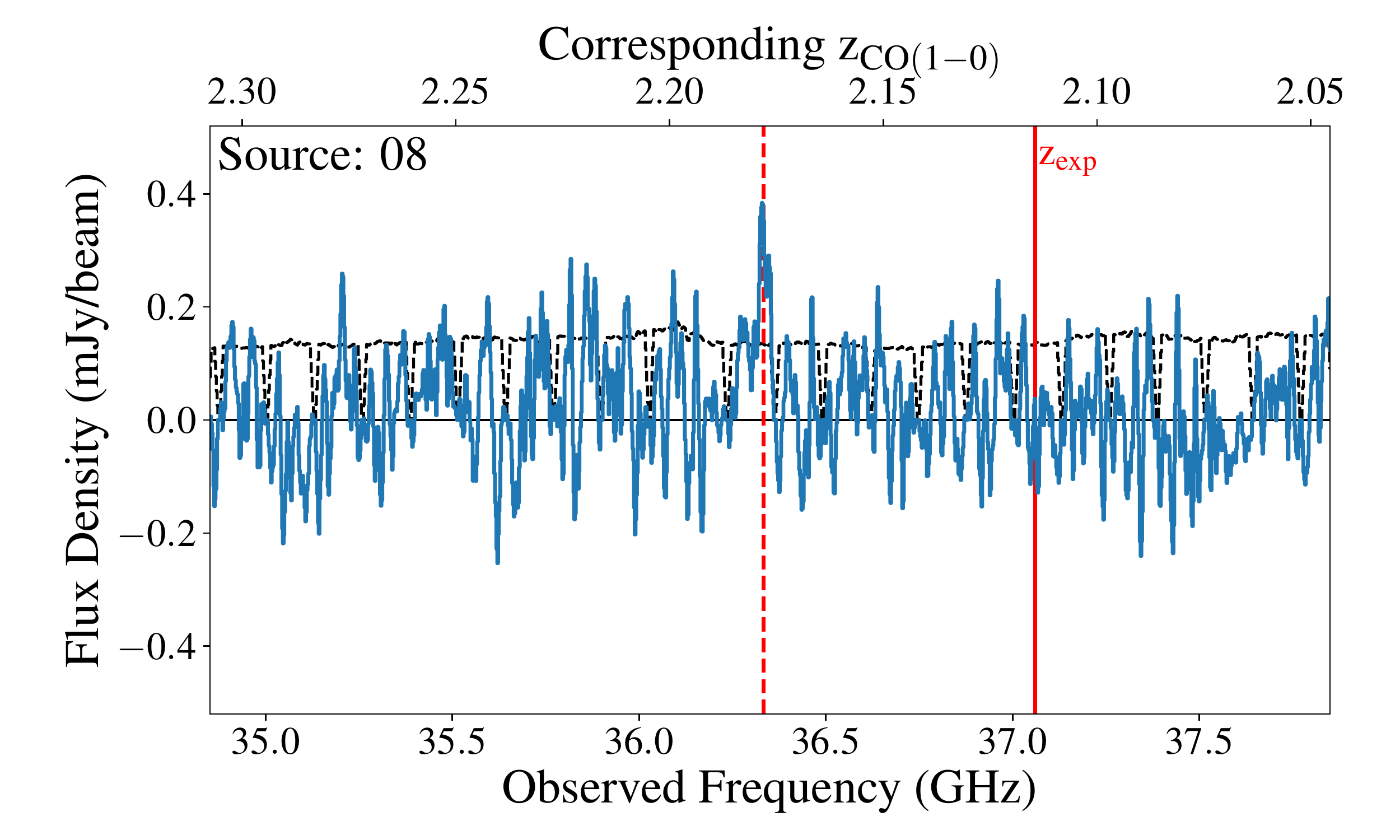}	
		\\
		\includegraphics[width=0.48\columnwidth, trim={0 0.4cm 0 0.4cm},clip]{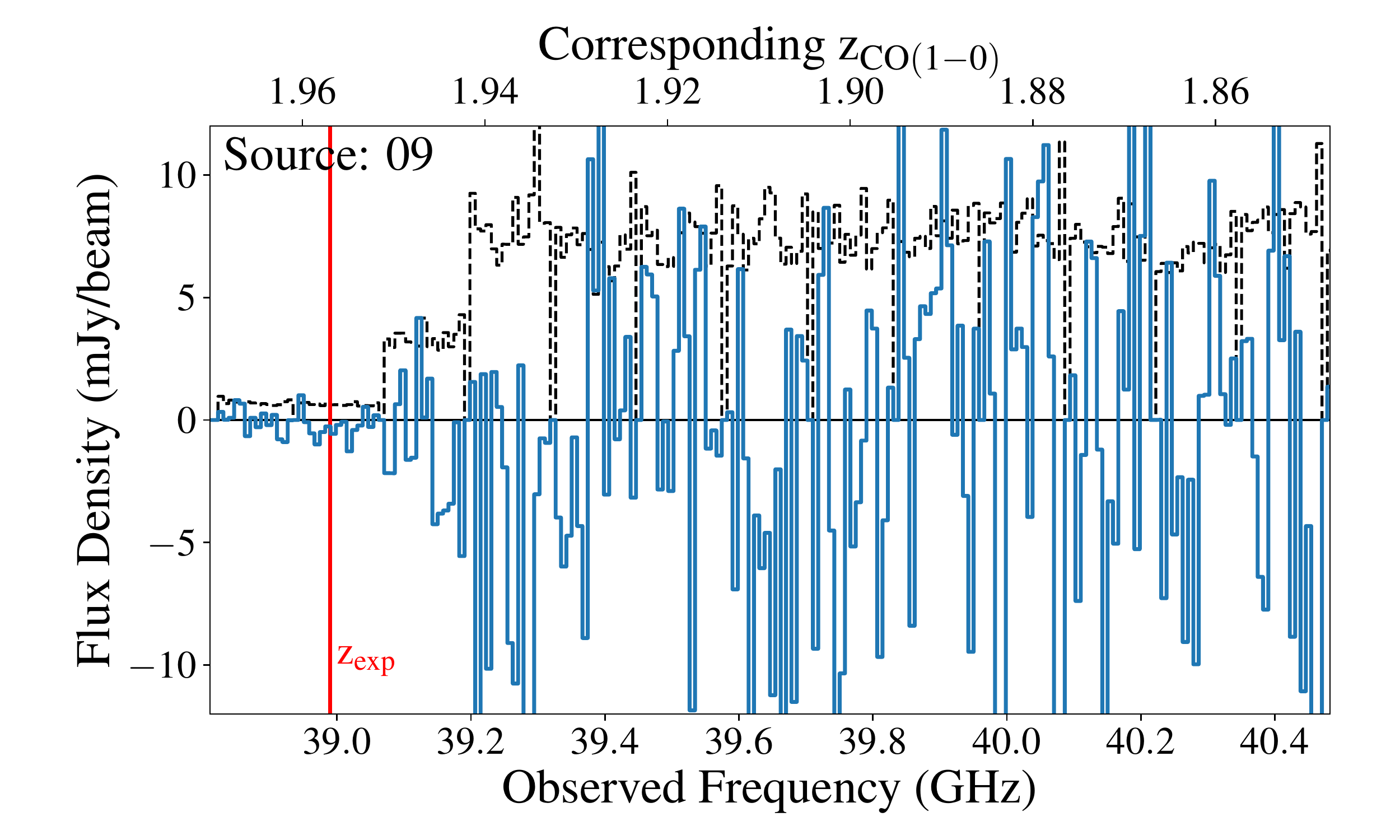}
		~
		\includegraphics[width=0.48\columnwidth, trim={0 0.4cm 0 0.4cm},clip]{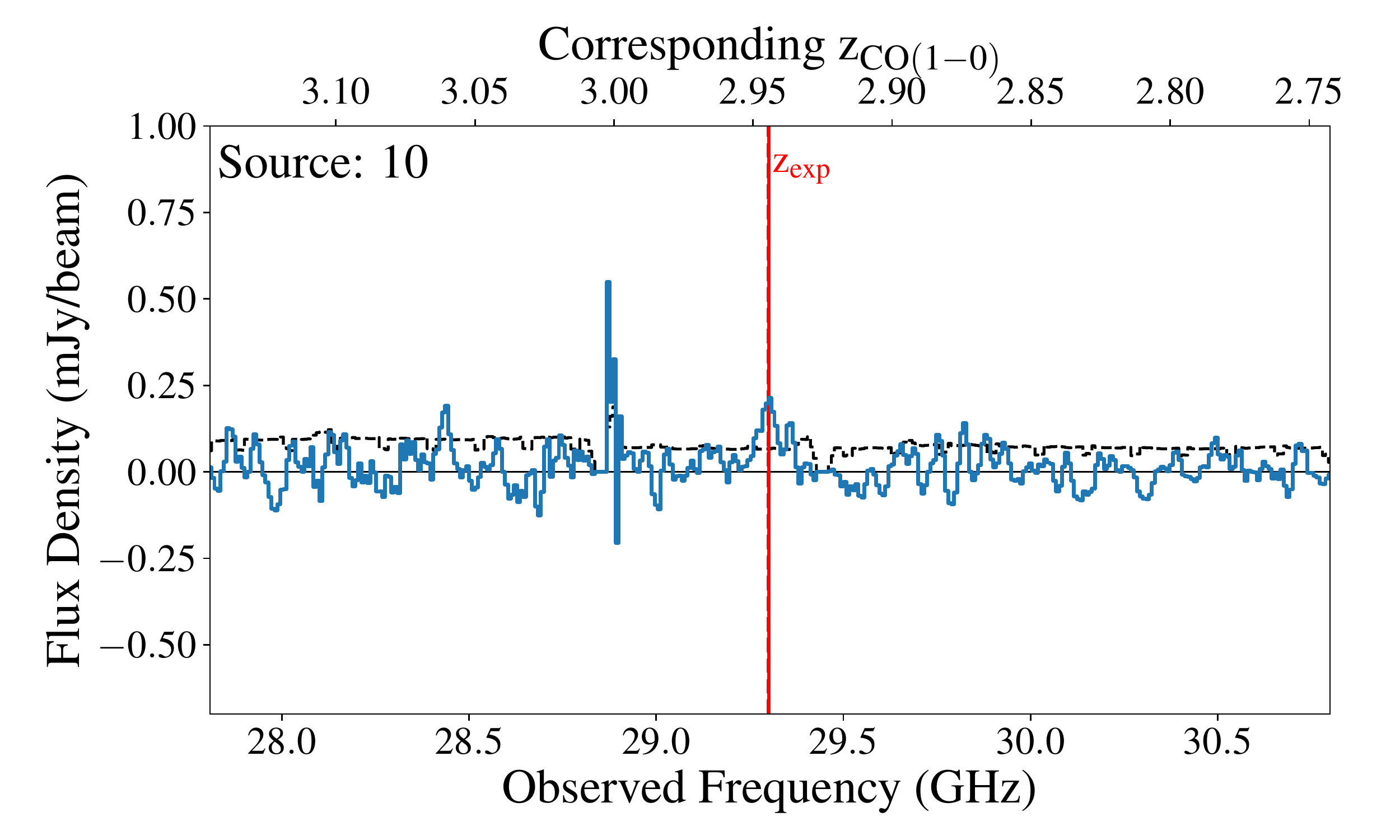}
		\\
		\includegraphics[width=0.48\columnwidth, trim={0 0.4cm 0 0.4cm},clip]{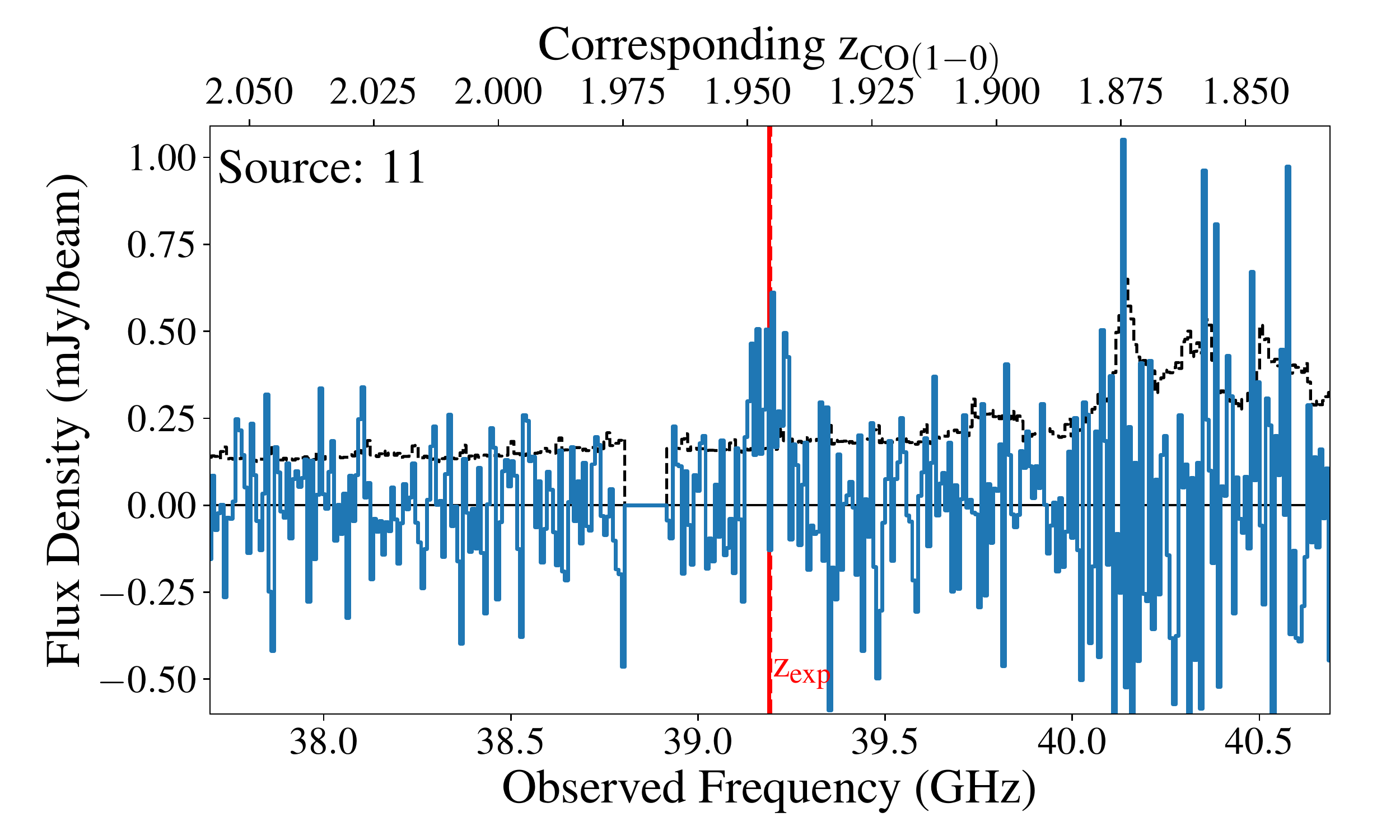}
		~
		\includegraphics[width=0.48\columnwidth, trim={0 0.4cm 0 0.4cm},clip]{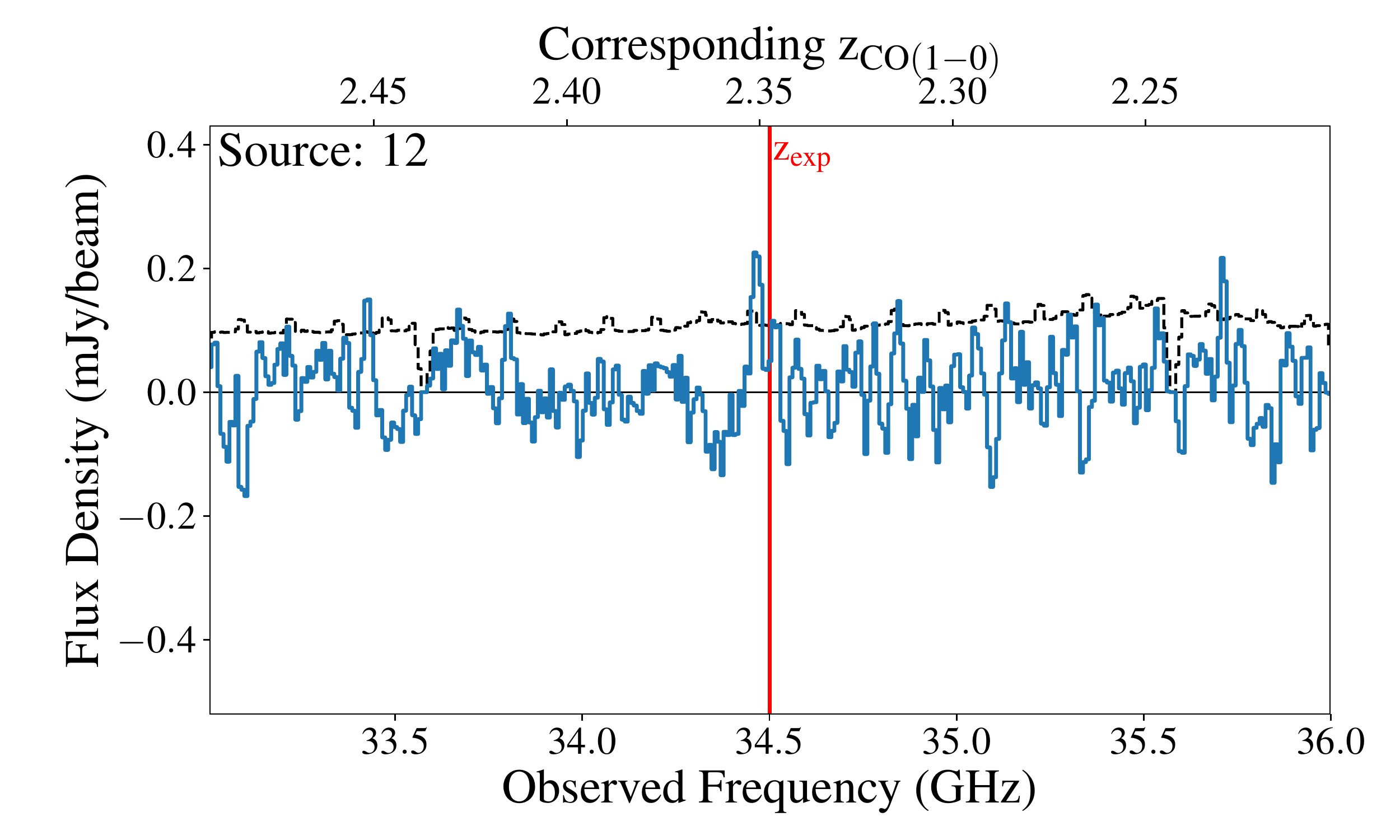}
		\\
		\includegraphics[width=0.48\columnwidth, trim={0 0.4cm 0 0.4cm},clip]{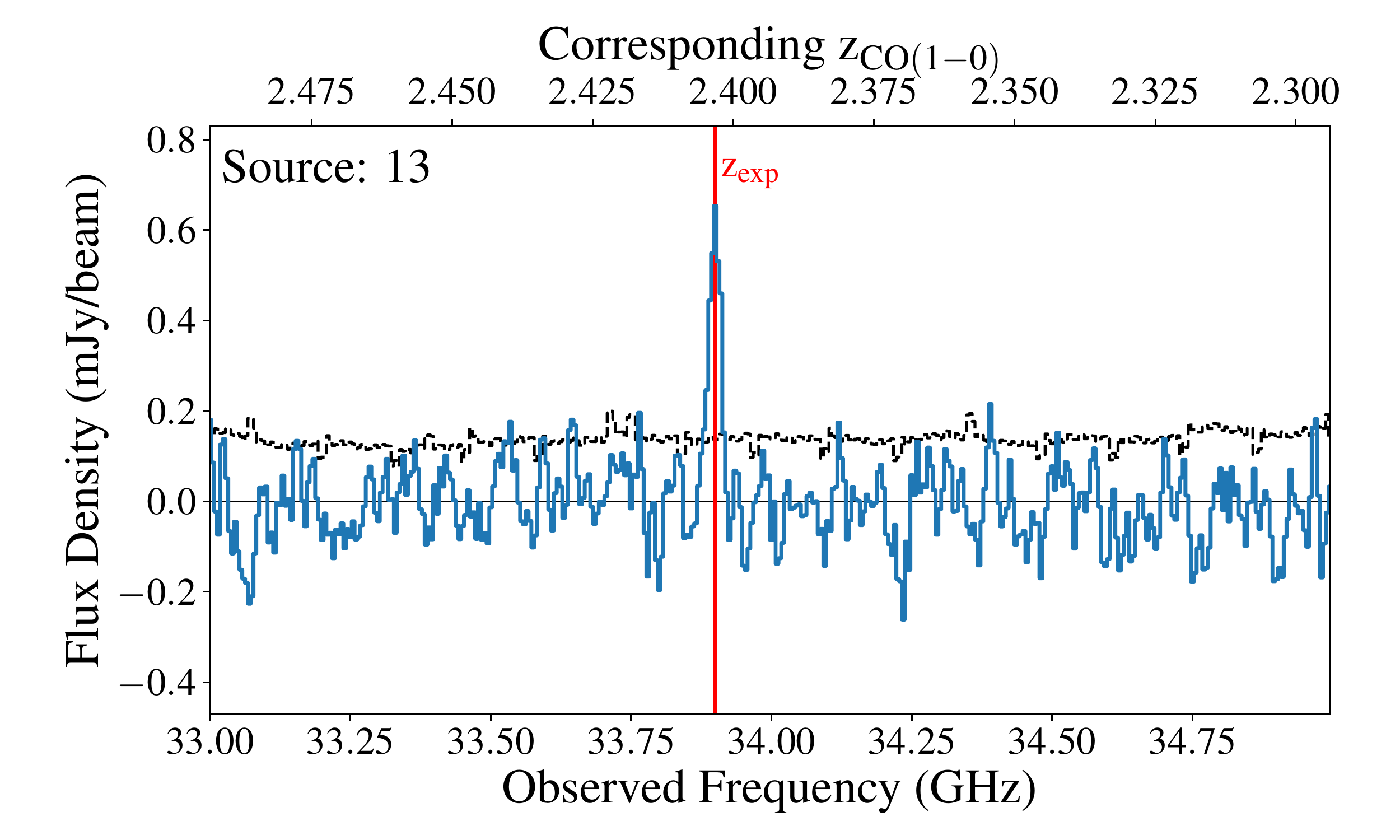}
		~
		\includegraphics[width=0.48\columnwidth, trim={0 0.4cm 0 0.4cm},clip]{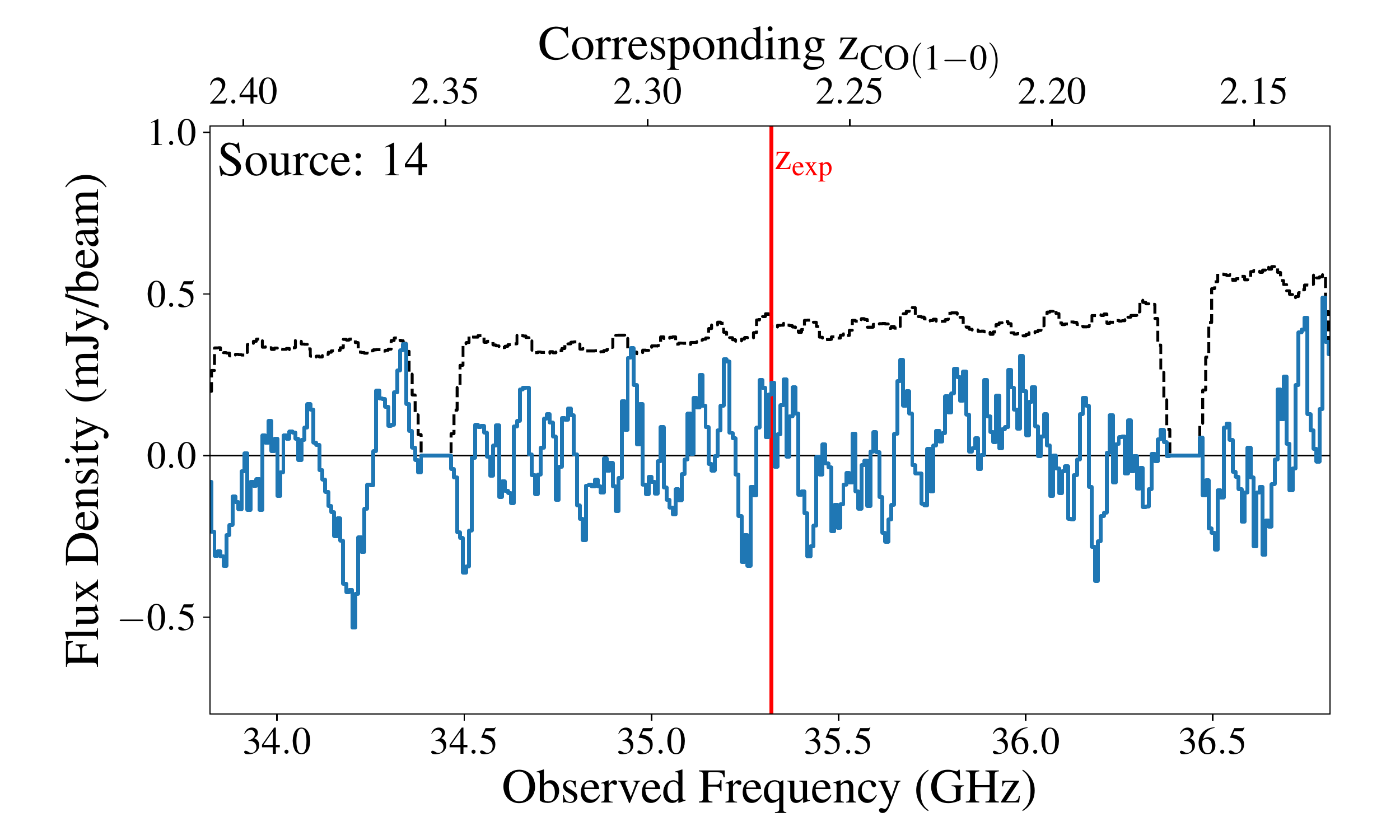}
		\\
		\includegraphics[width=0.48\columnwidth, trim={0 0.4cm 0 0.4cm},clip]{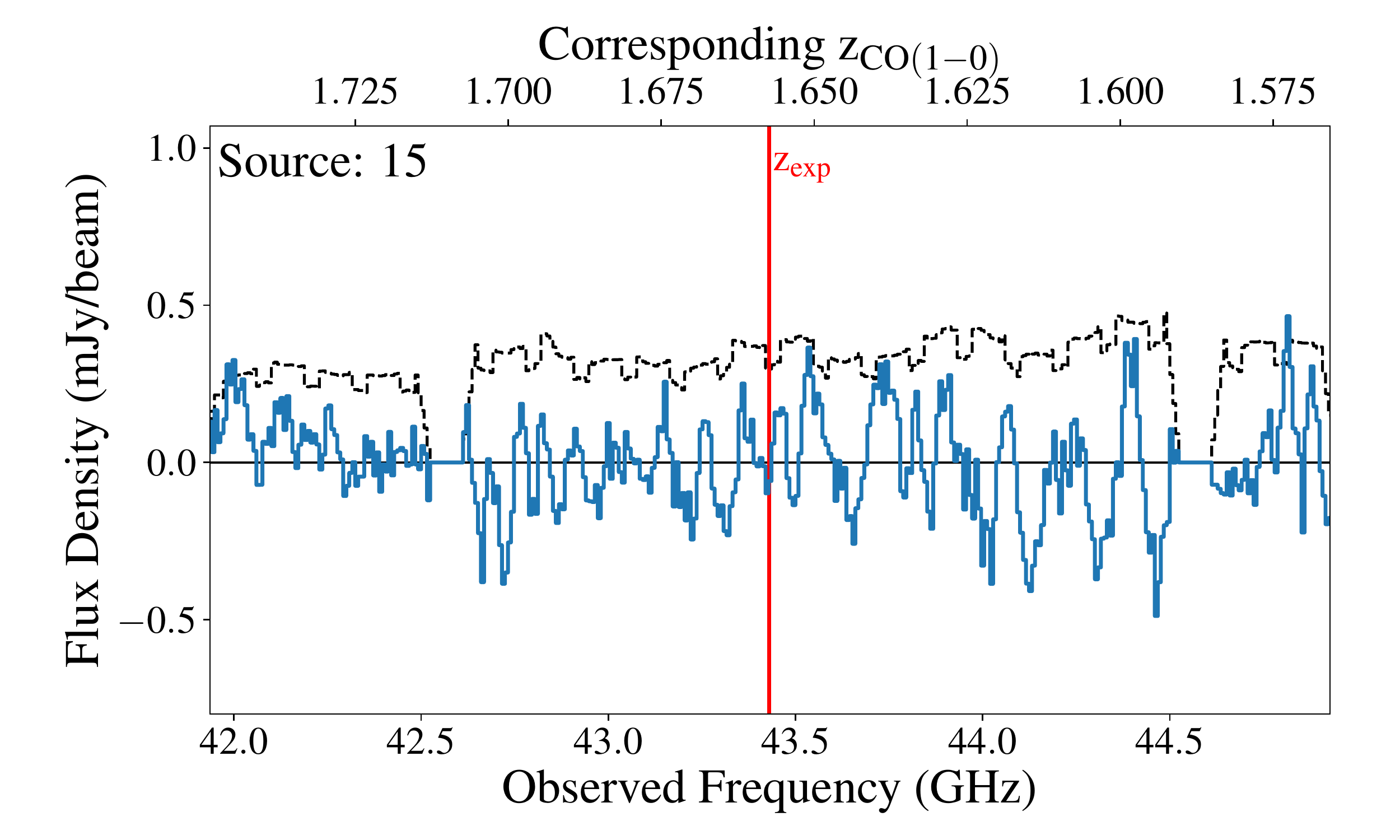}
		~
		\includegraphics[width=0.48\columnwidth, trim={0 0.4cm 0 0.4cm},clip]{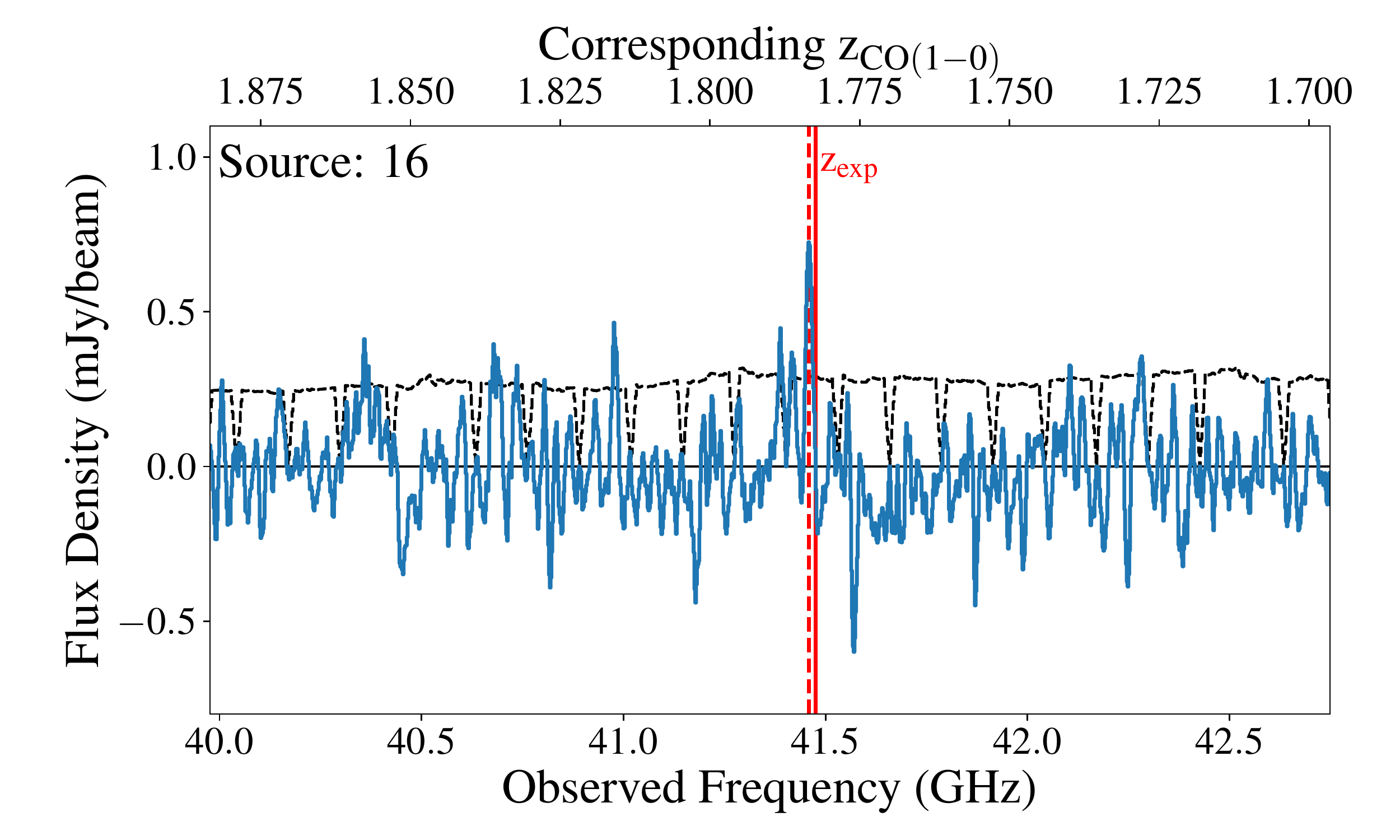}
		\caption{continued}
	\end{figure*}

	\begin{figure*}[h!]
		\centering
		\includegraphics[width=0.29\textwidth, trim={0 1.2cm 3.cm 0.4cm},clip]{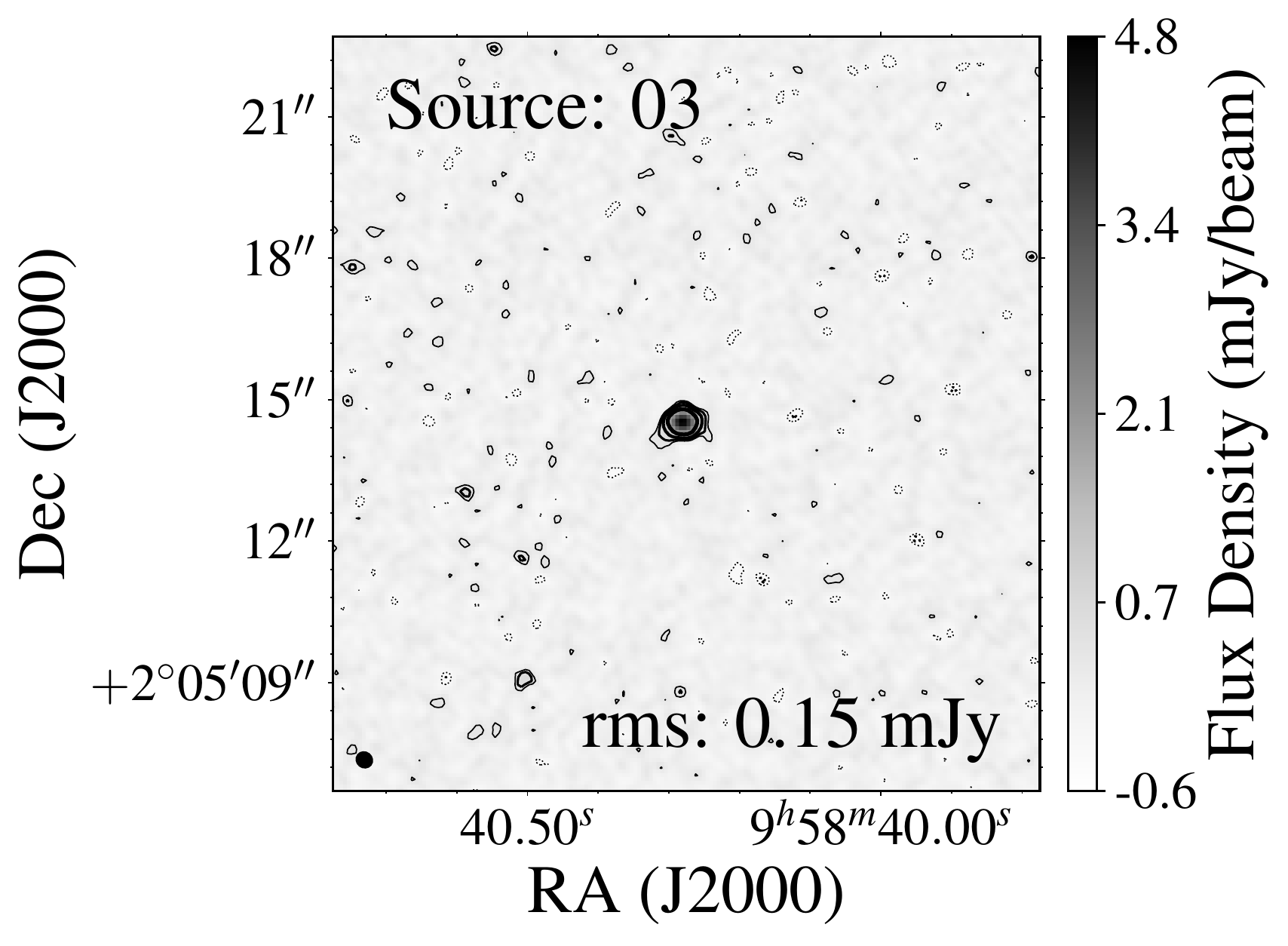}
		\includegraphics[width=0.63\textwidth, trim={2.7cm 0.95cm 2cm 0},clip]{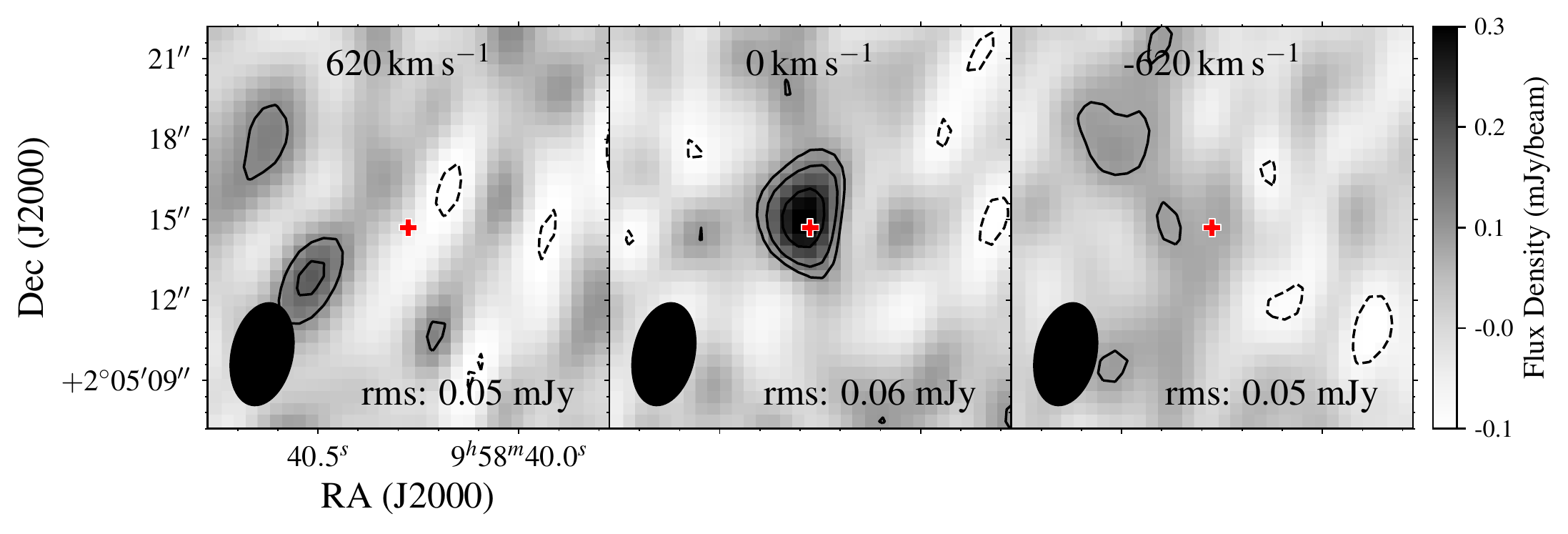}
		\\
		\includegraphics[width=0.29\textwidth, trim={0 1.2cm 3.cm 0.4cm},clip]{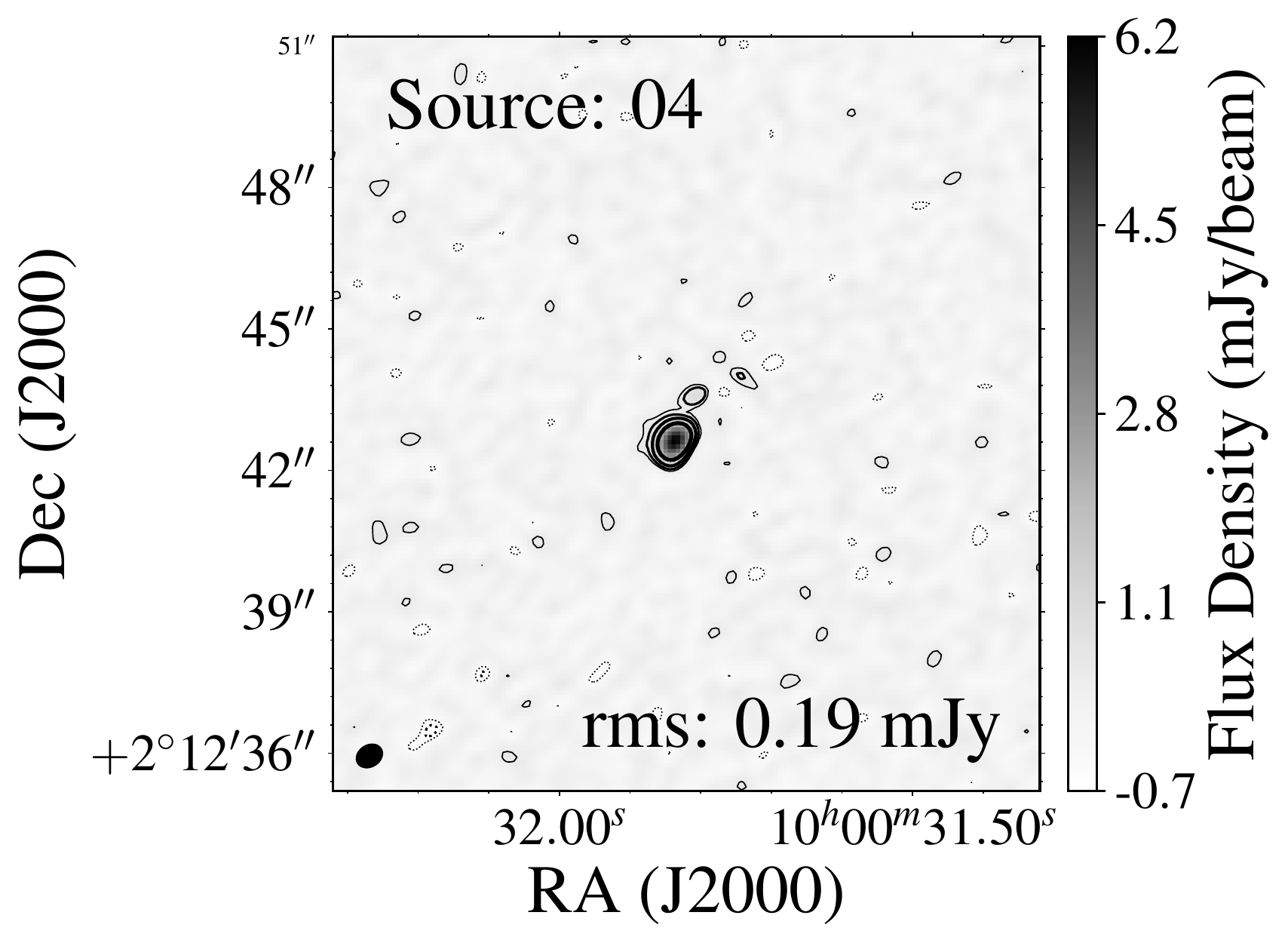}
		\includegraphics[width=0.63\textwidth, trim={2.7cm 0.95cm 2cm 0},clip]{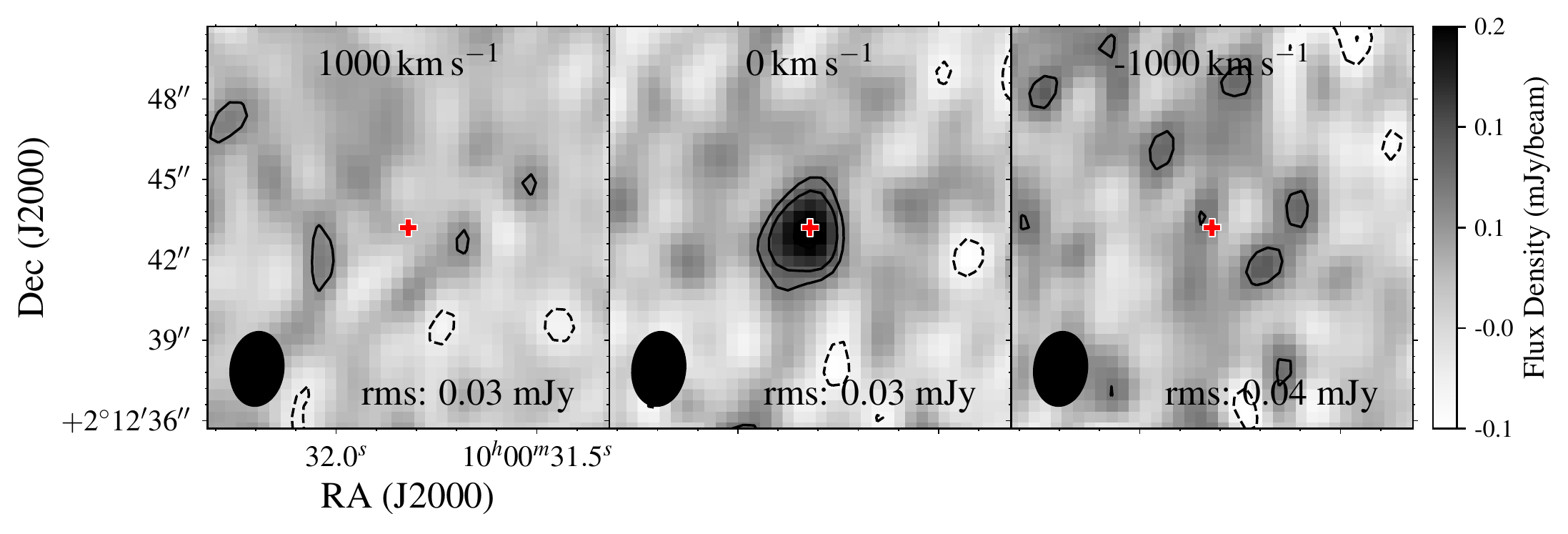}
		\\
		\includegraphics[width=0.29\textwidth, trim={0 1.2cm 3.cm 0.4cm},clip]{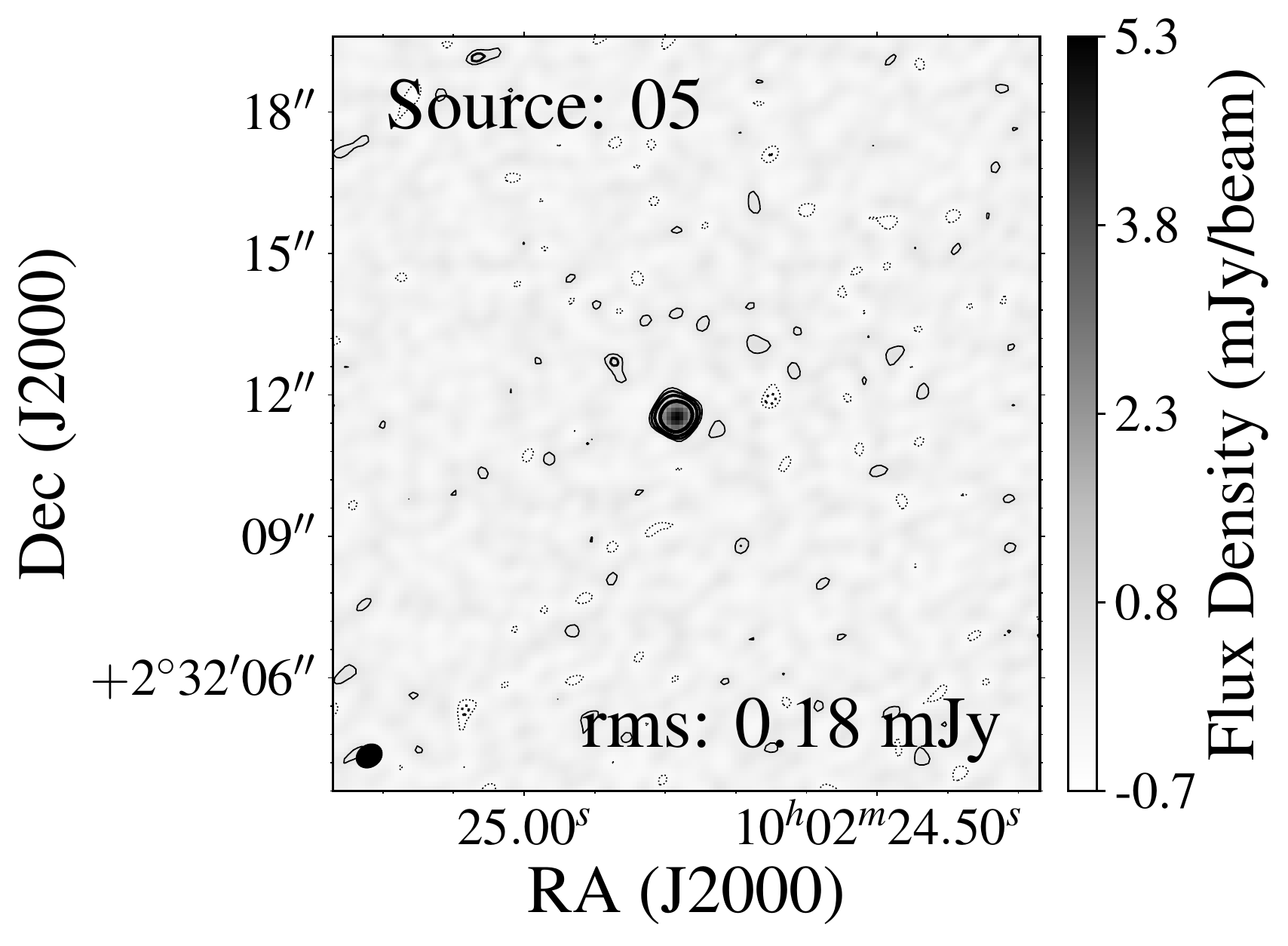}
		\includegraphics[width=0.63\textwidth, trim={2.7cm 0.95cm 2cm 0},clip]{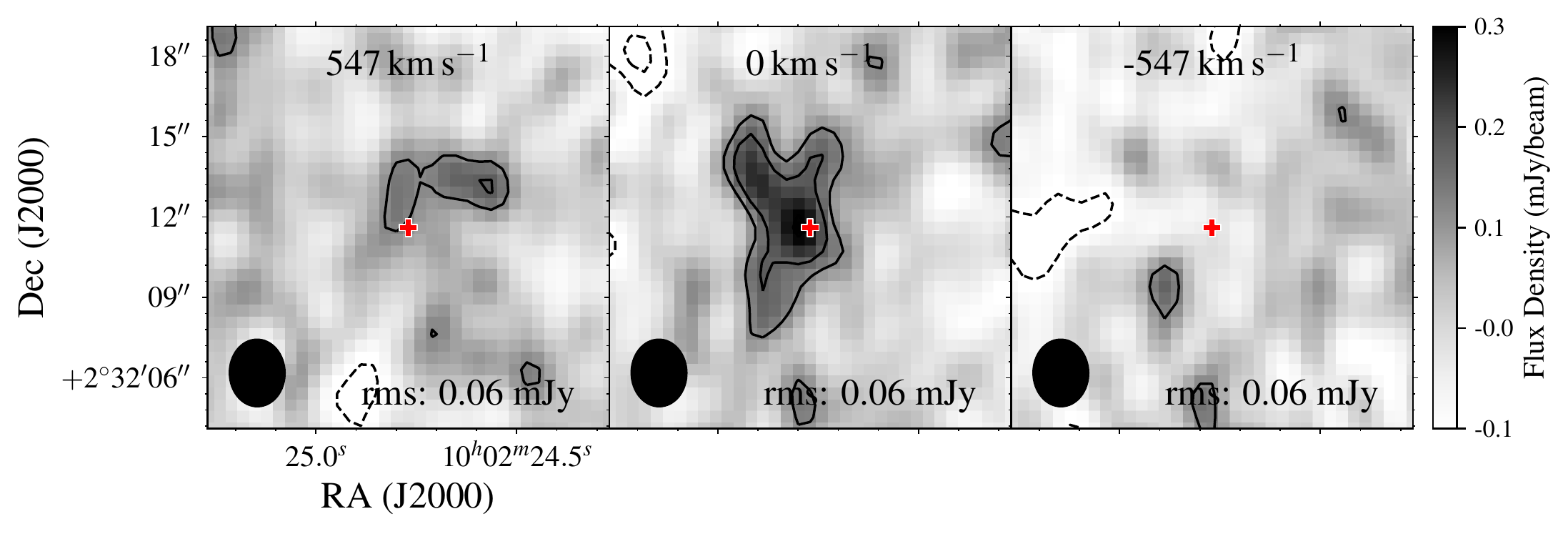}
		\\
		\includegraphics[width=0.29\textwidth, trim={0 -0.5cm 3.cm 0.cm},clip]{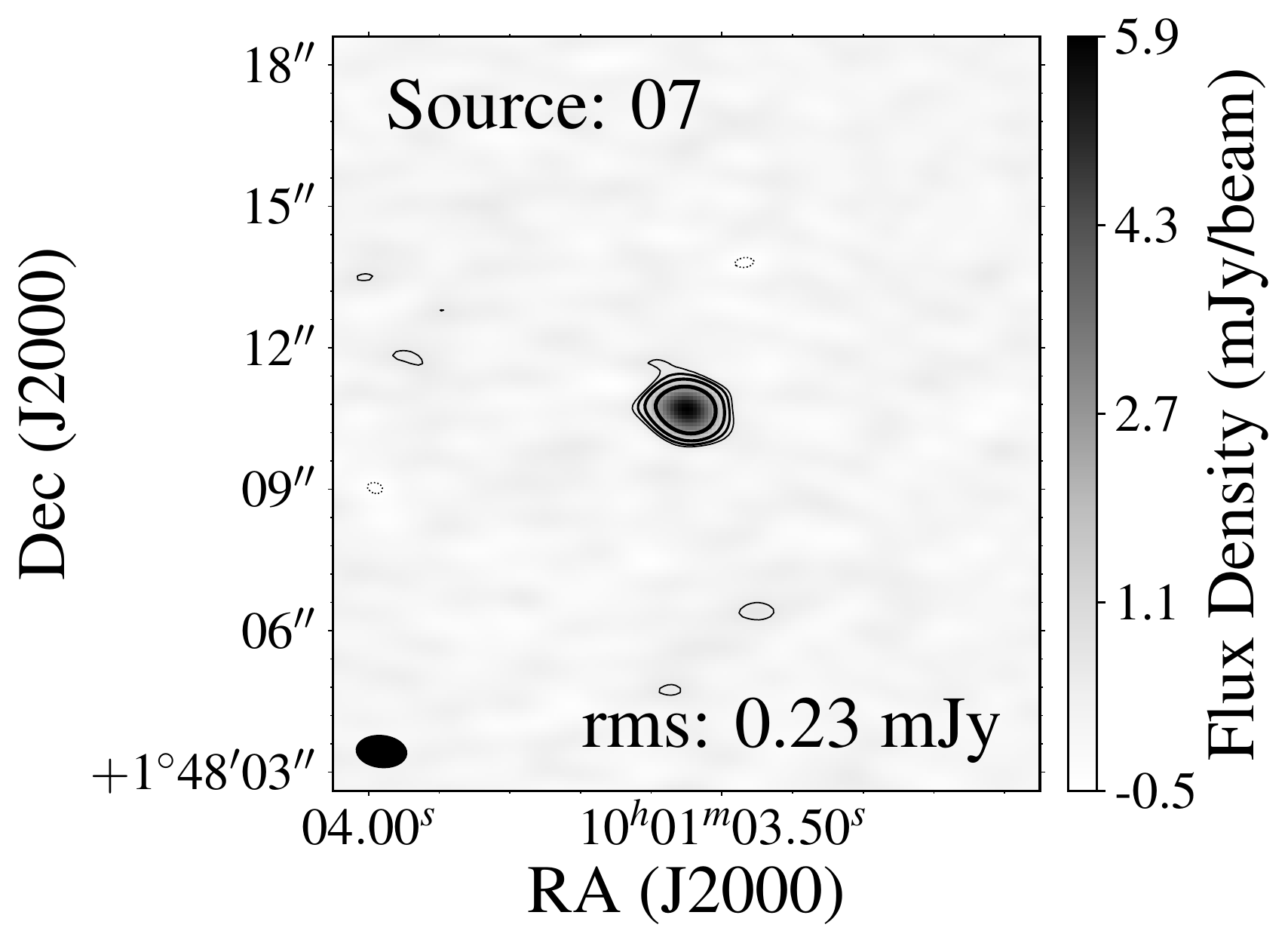}
		\includegraphics[width=0.63\textwidth, trim={2.8cm 0.cm 2cm 0},clip]{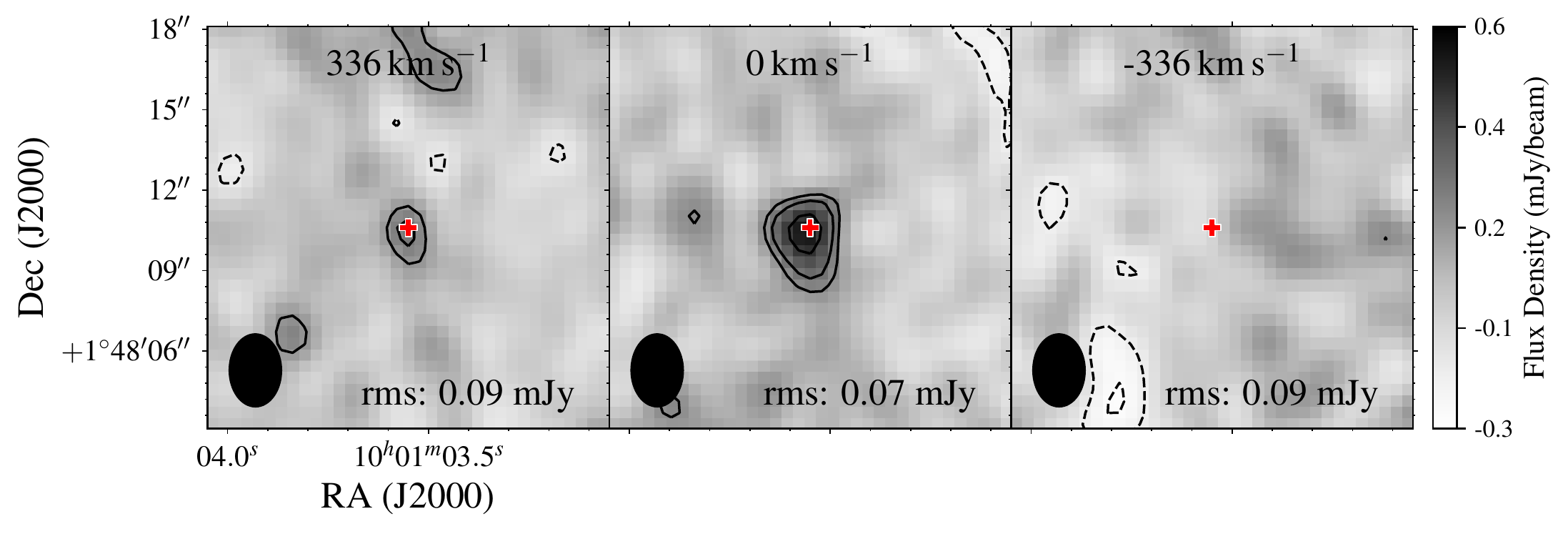}
		\caption{Comparison of the ALMA dust continuum (left map) and integrated VLA CO(1-0) maps for sources with CO(1-0) detections. The source number is labeled at the top left of the left hand panel in each row. The rms value is given in the bottom right corner of each map. Left column: ALMA observations at 343.5 GHz. Contours are shown for $\pm 2, 3, 5$ and $10\sigma$ (dashed contours for negative values). Right columns: Channel maps around the measured (expected) CO(1-0) line. For each source, the central panel of the VLA channel map represents the moment zero map and is centred at the central velocity of the CO spectrum. The velocity width of the integrated maps is chosen to encompass the full source emission (1.2$\times$FWHM of the CO(1-0) line for sources other than 4 and 5).  Contours are shown for $\pm 2, 3$ and $5\sigma$ (dashed contours for negative values). The red cross indicates the expected position of the source, at which the CO spectrum was extracted. The colour shading indicates the flux density in mJy/beam.\label{fig:chmap_comp}}
	\end{figure*}

	\begin{figure*}[h!]
		\ContinuedFloat
		\centering	
		\includegraphics[width=0.29\textwidth, trim={0 1.2cm 3.cm 0.4cm},clip]{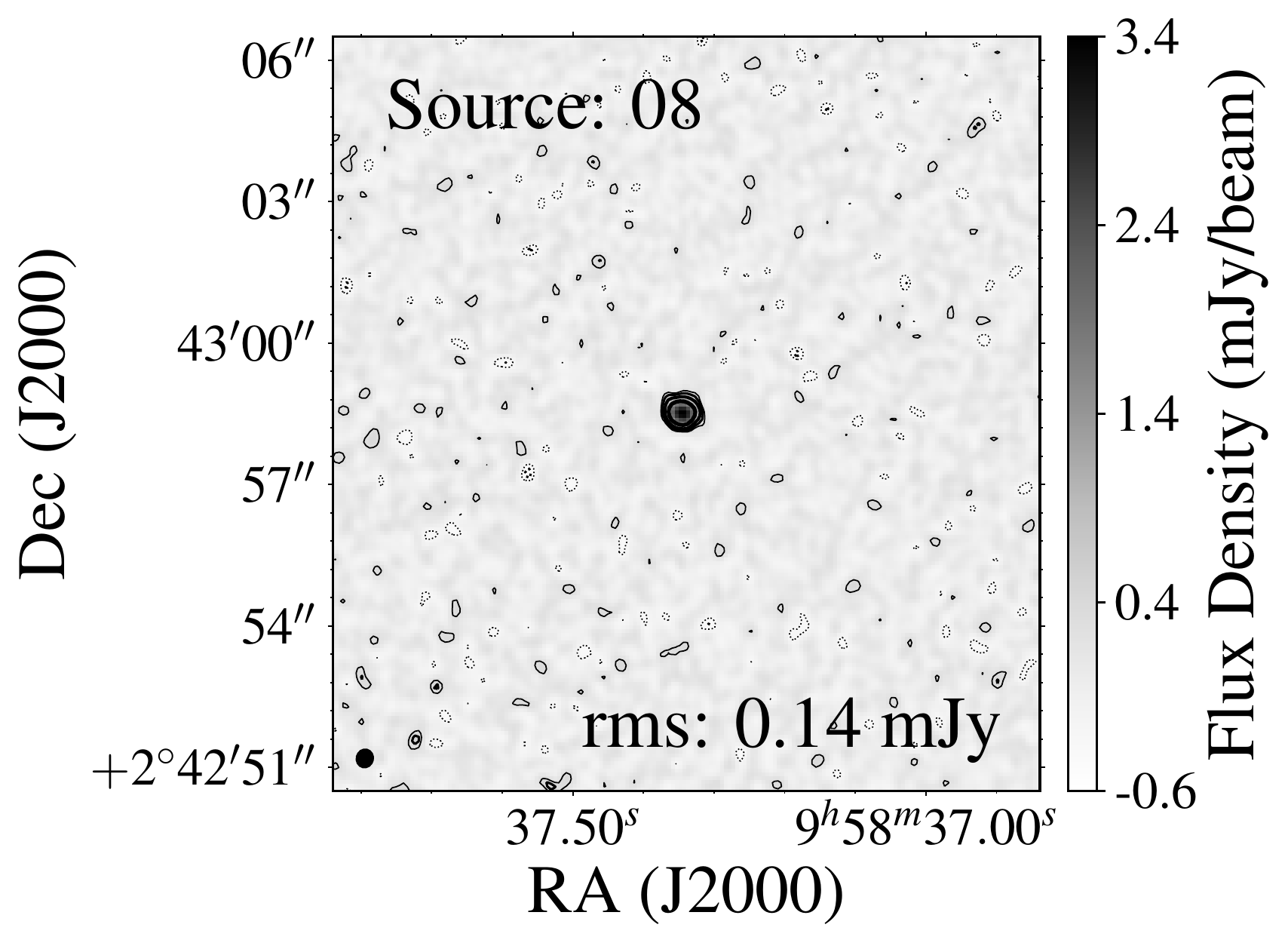}
		\includegraphics[width=0.63\textwidth, trim={2.7cm 0.95cm 2cm 0},clip]{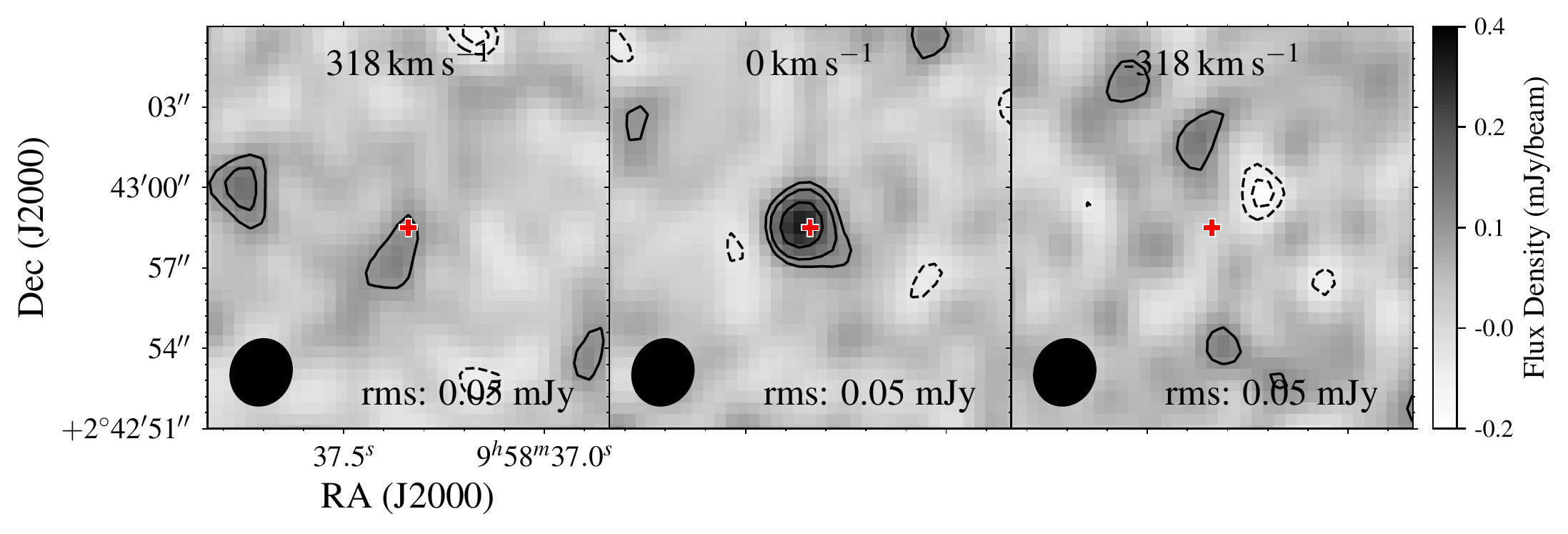}
		\\
		\includegraphics[width=0.29\textwidth, trim={0 1.2cm 3.cm 0.4cm},clip]{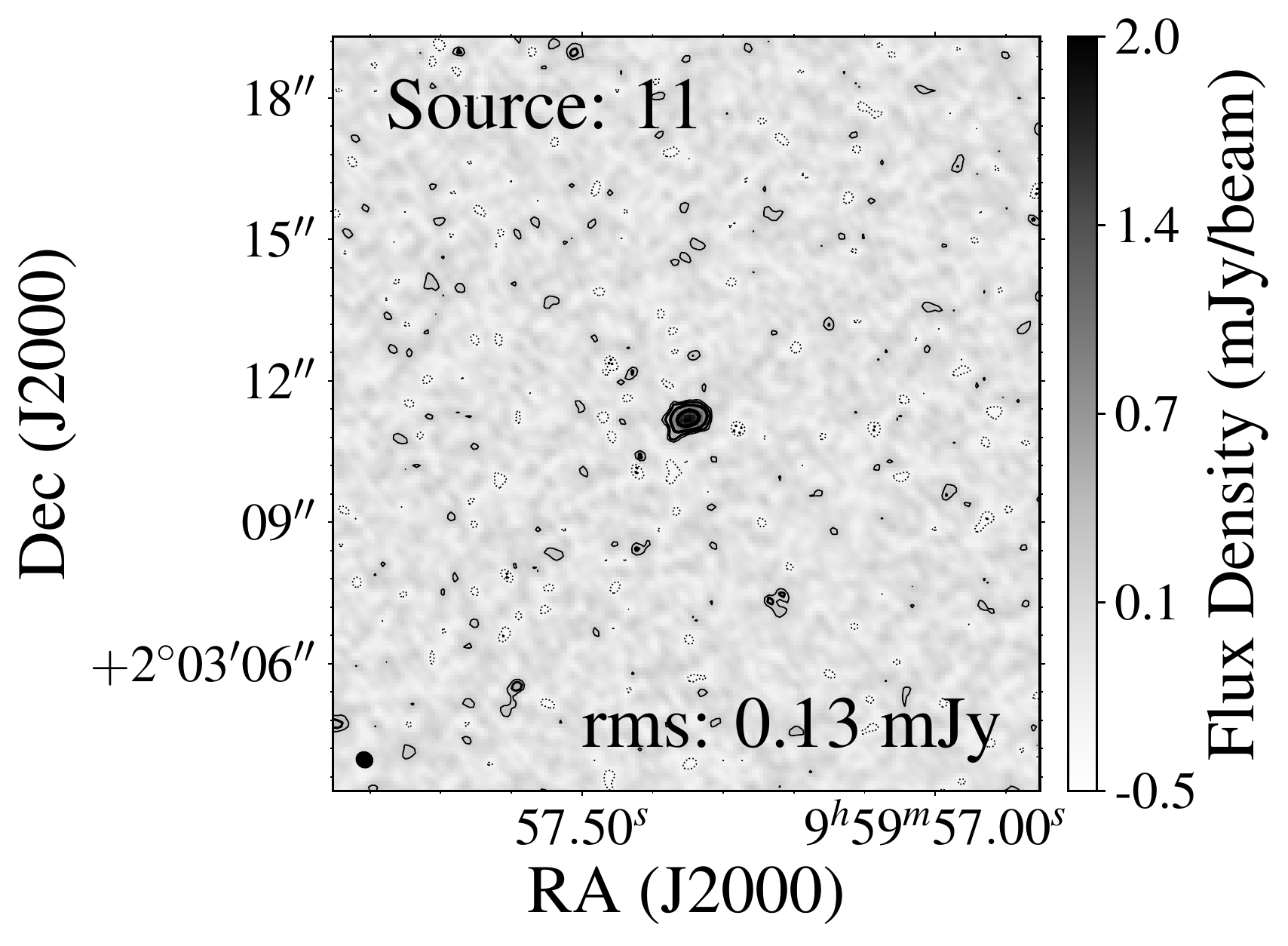}
		\includegraphics[width=0.63\textwidth, trim={2.7cm 0.95cm 2cm 0},clip]{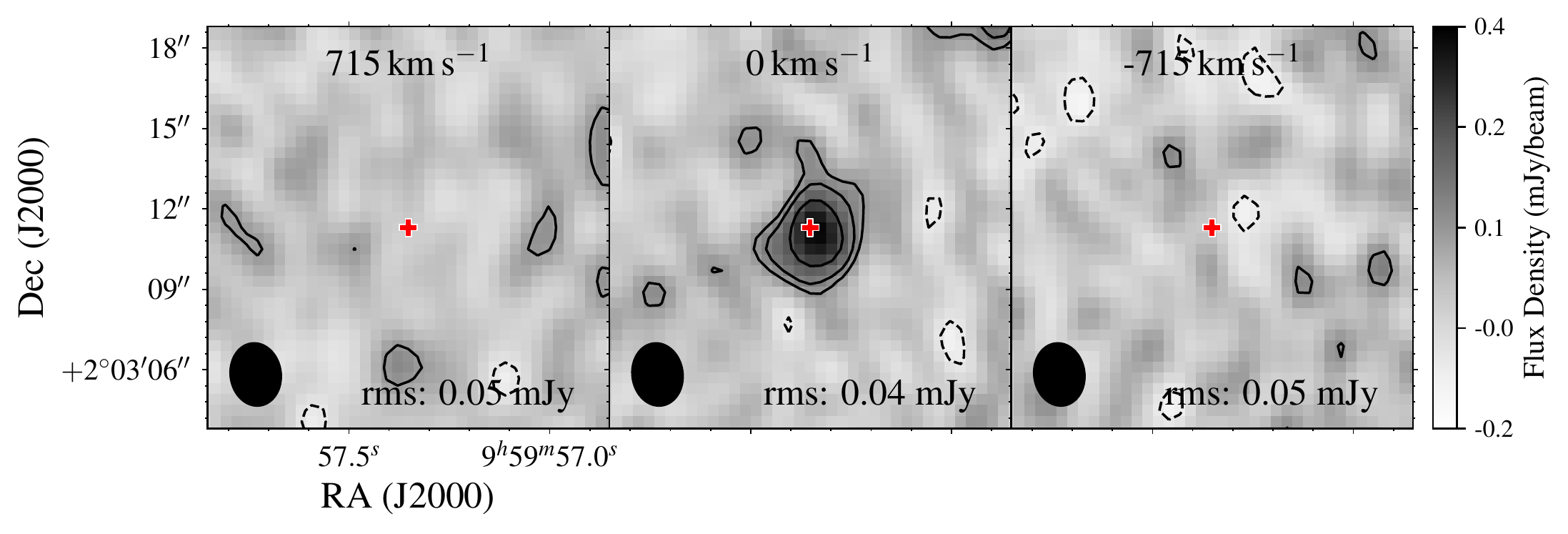}
		\\
		\includegraphics[width=0.29\textwidth, trim={0 1.2cm 3.cm 0.4cm},clip]{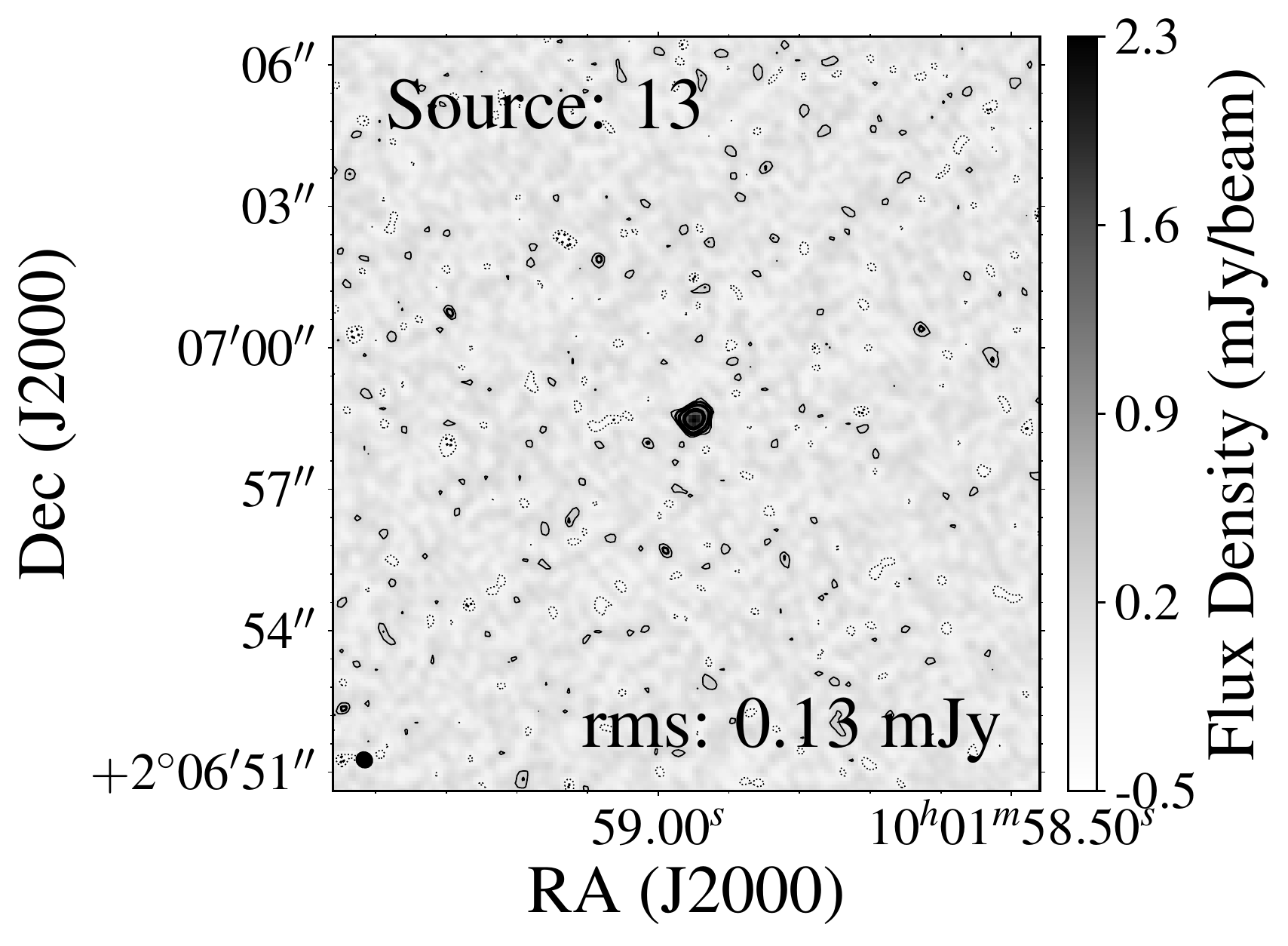}
		\includegraphics[width=0.63\textwidth, trim={2.7cm 0.95cm 2cm 0},clip]{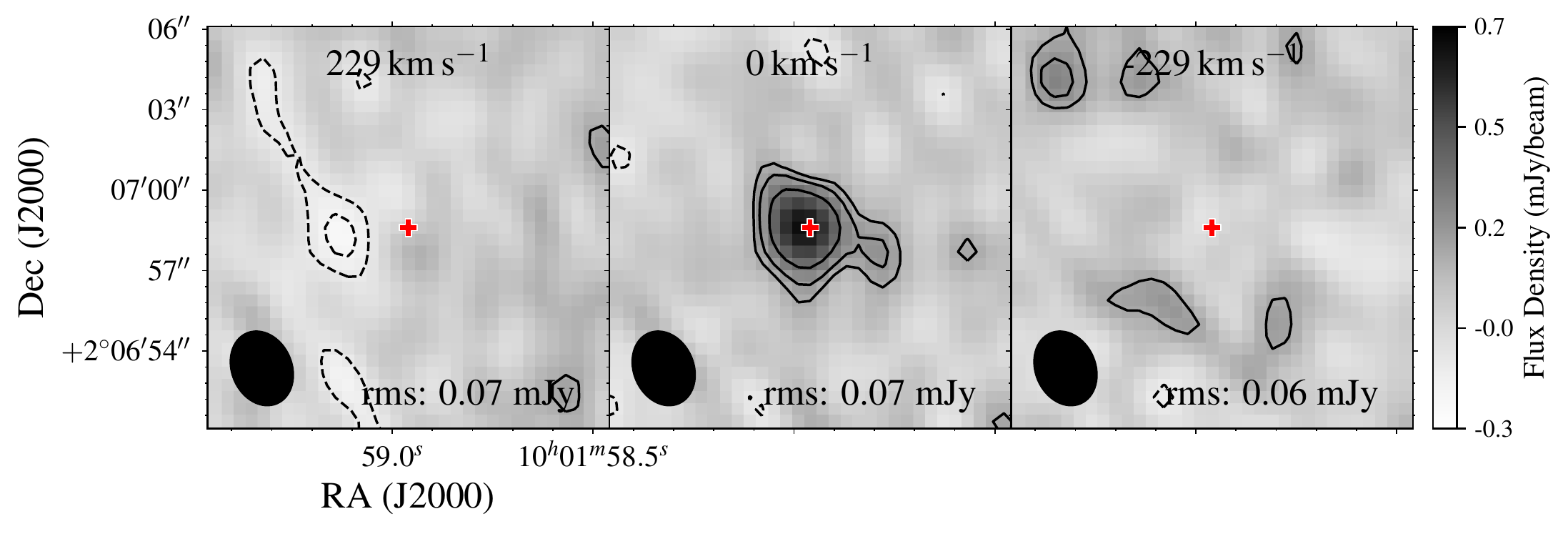}
		\\
		\includegraphics[width=0.29\textwidth, trim={0 -0.5cm 3.cm 0},clip]{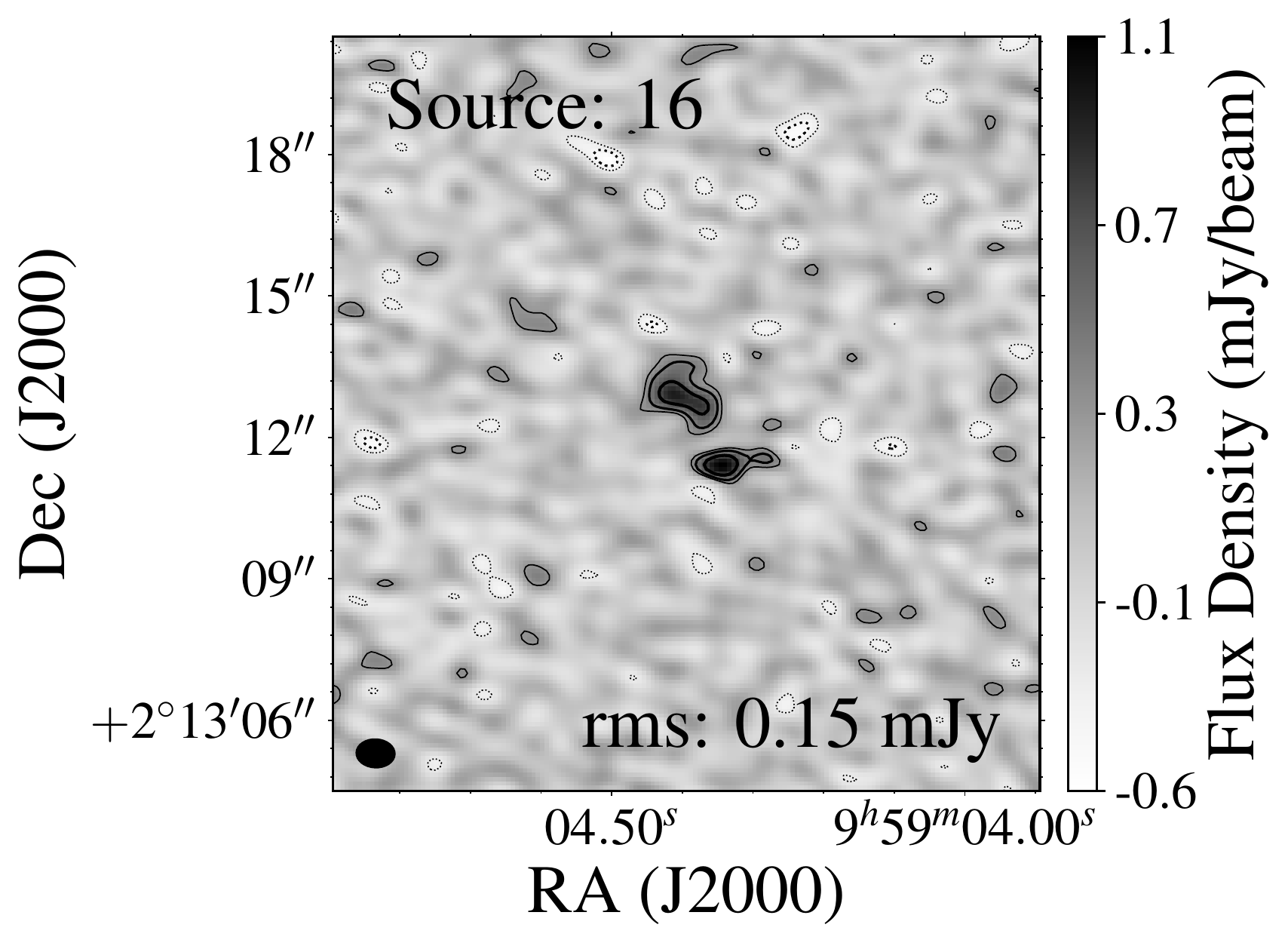}
		\includegraphics[width=0.63\textwidth, trim={2.8cm 0 2cm 0},clip]{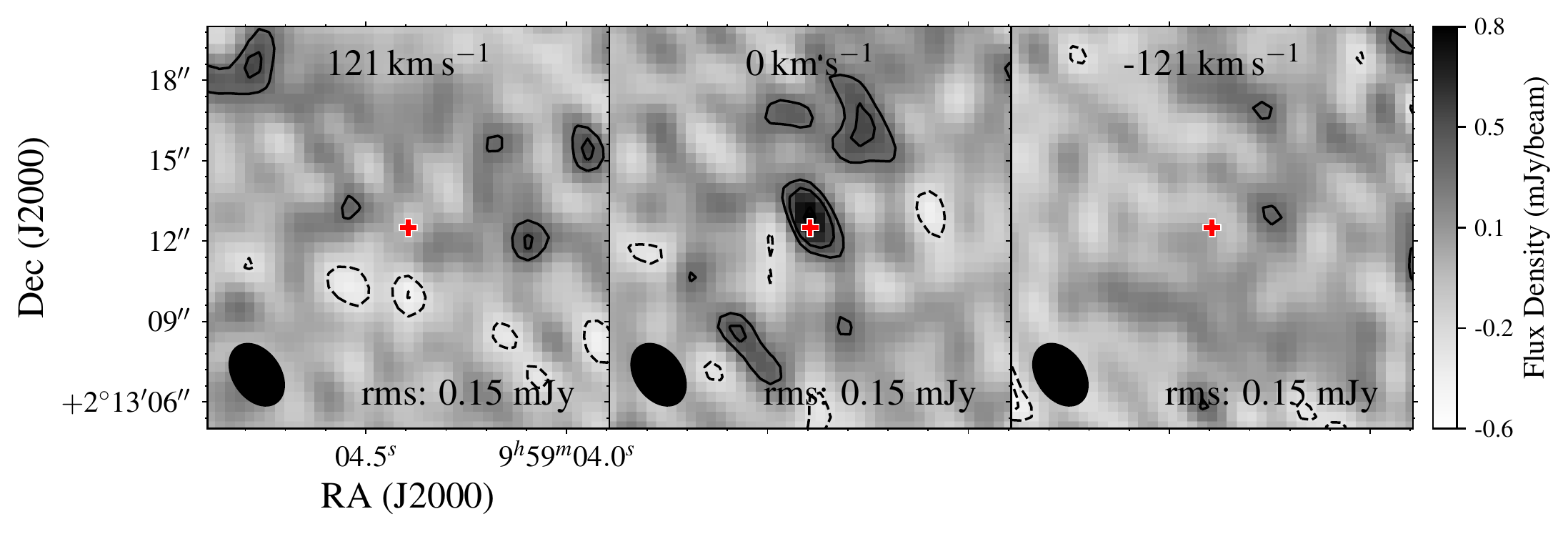}
		\caption{continued}
	\end{figure*}



	\begin{figure*}[h!]
		\centering
		\includegraphics[width=0.45\columnwidth]{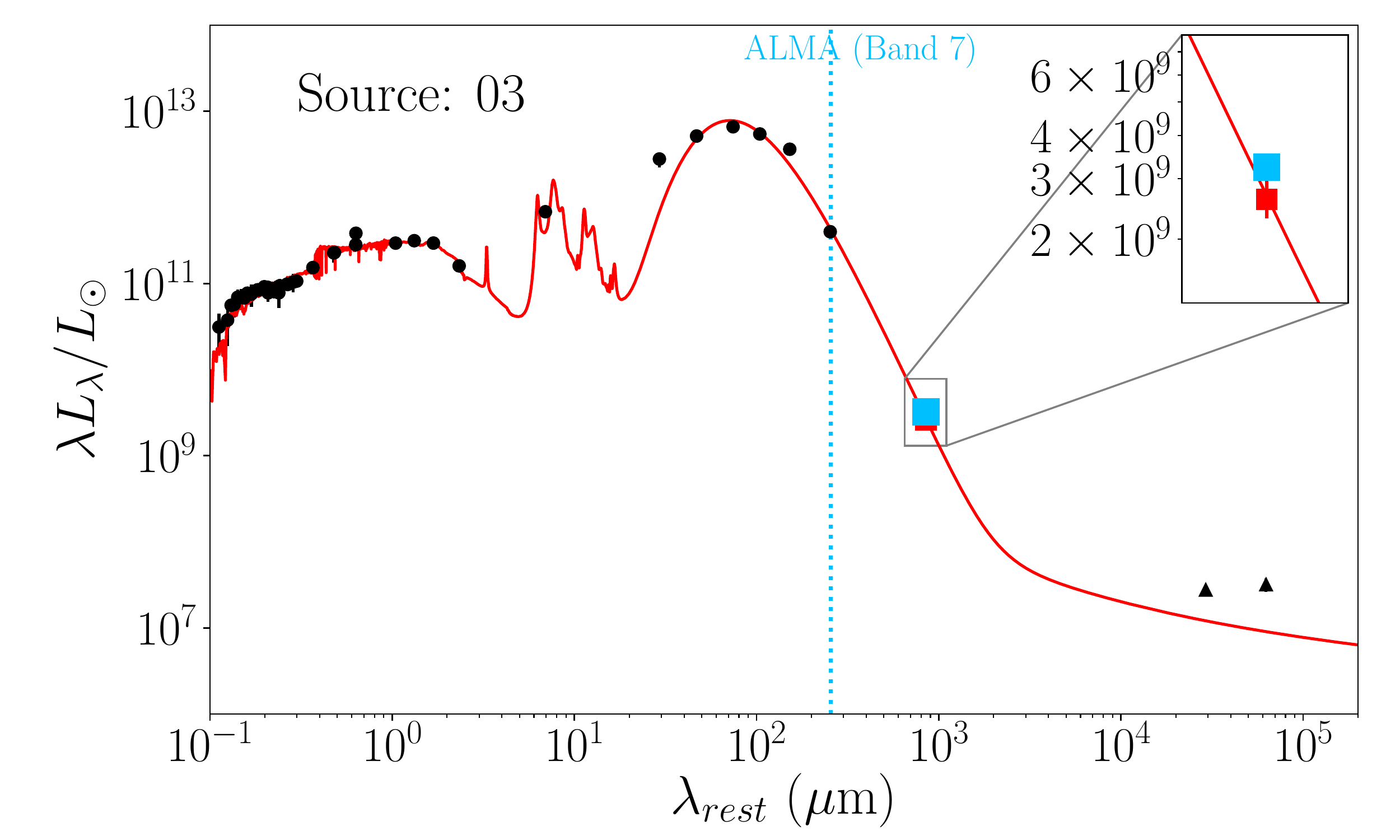}
		~
		\includegraphics[width=0.45\columnwidth]{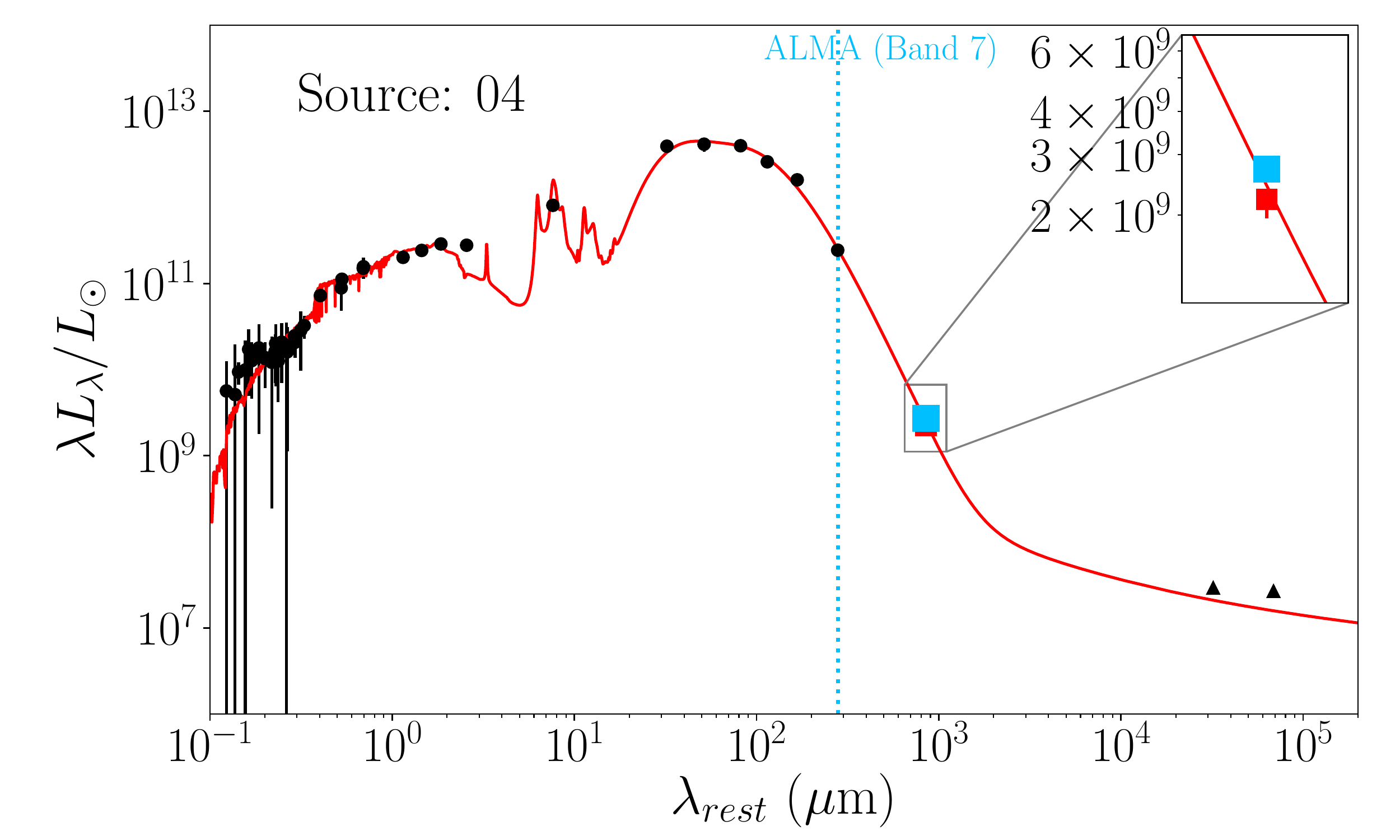}
		\\
		\includegraphics[width=0.45\columnwidth]{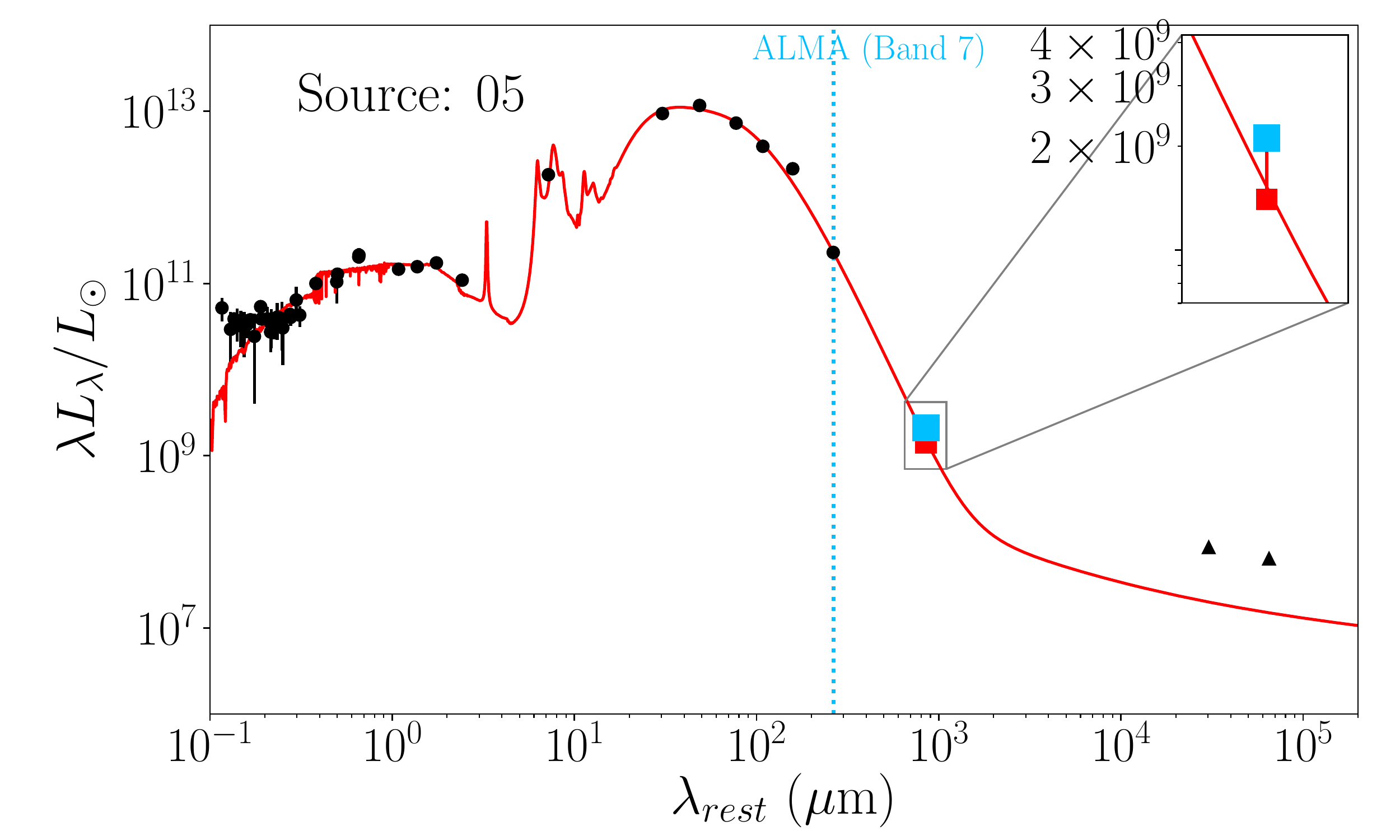}
		~
		\includegraphics[width=0.45\columnwidth]{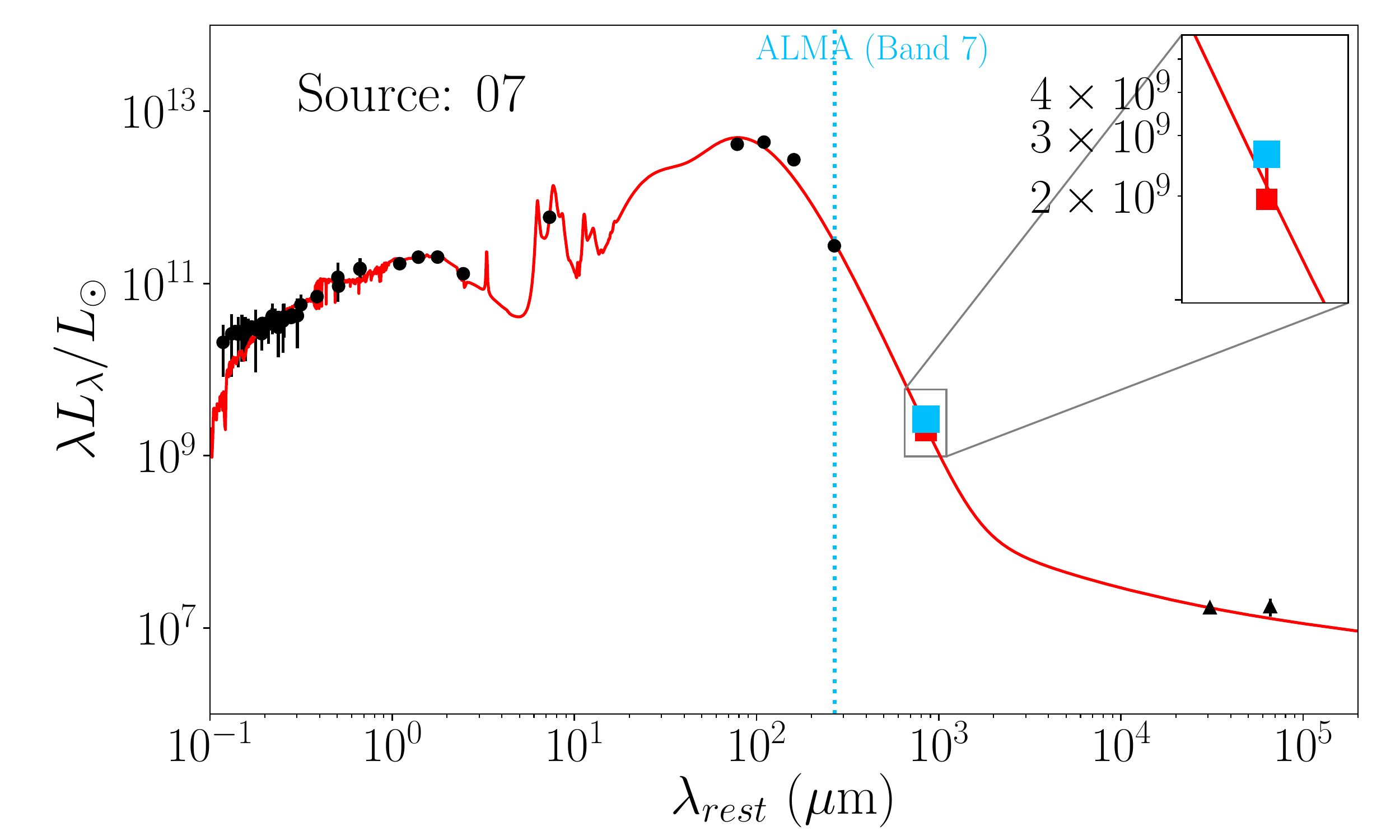}
		\\
		\includegraphics[width=0.45\columnwidth]{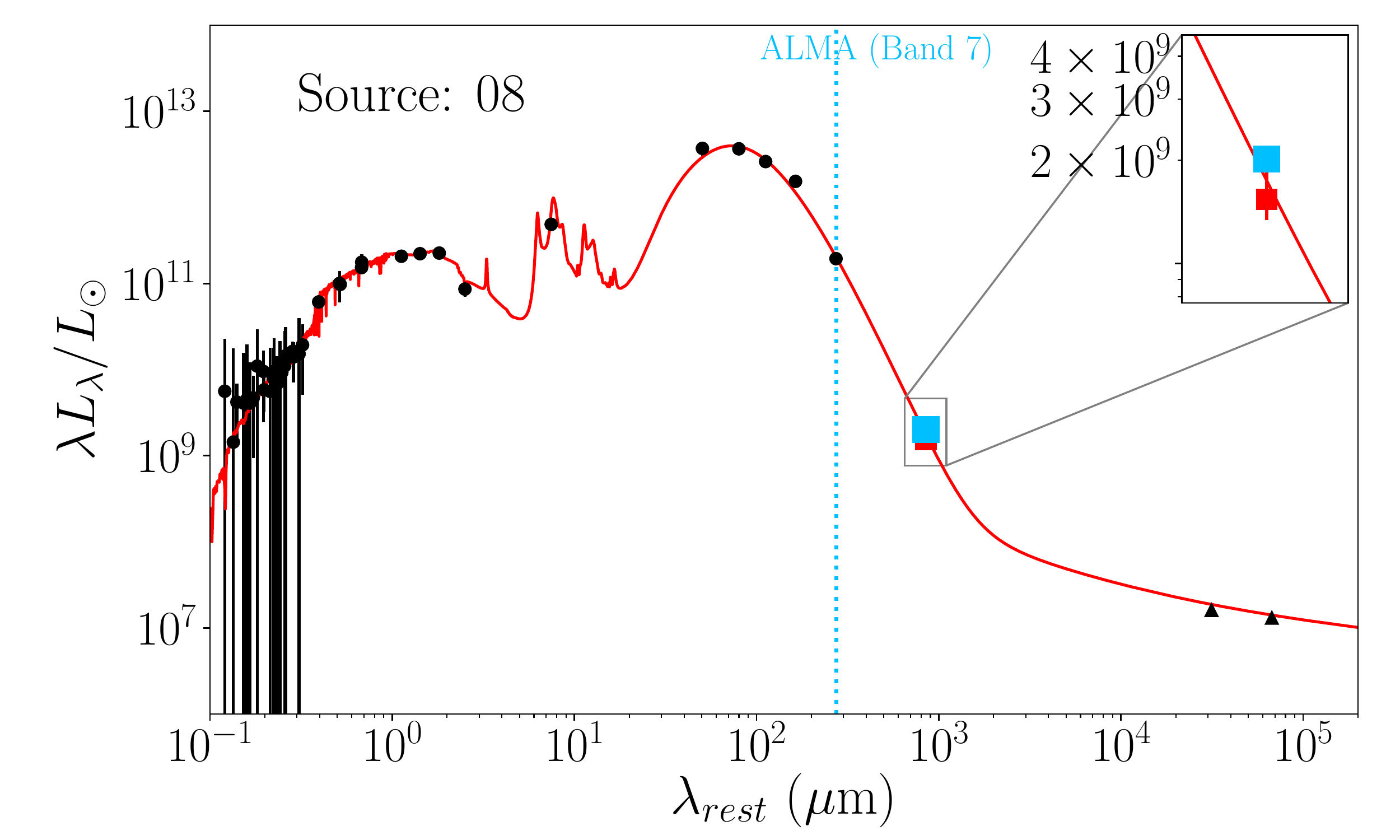}
		~
		\includegraphics[width=0.45\columnwidth]{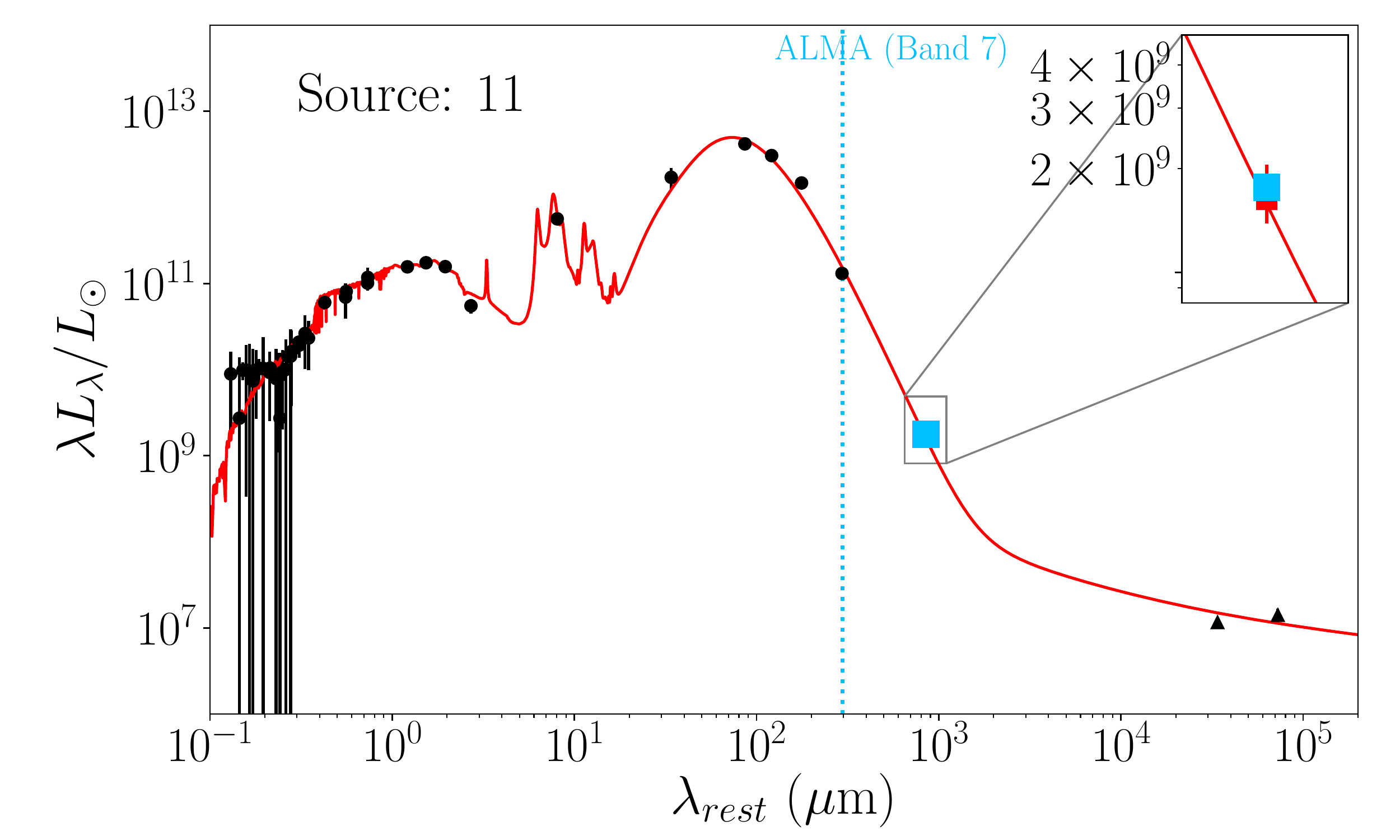}
		\\
		\includegraphics[width=0.45\columnwidth]{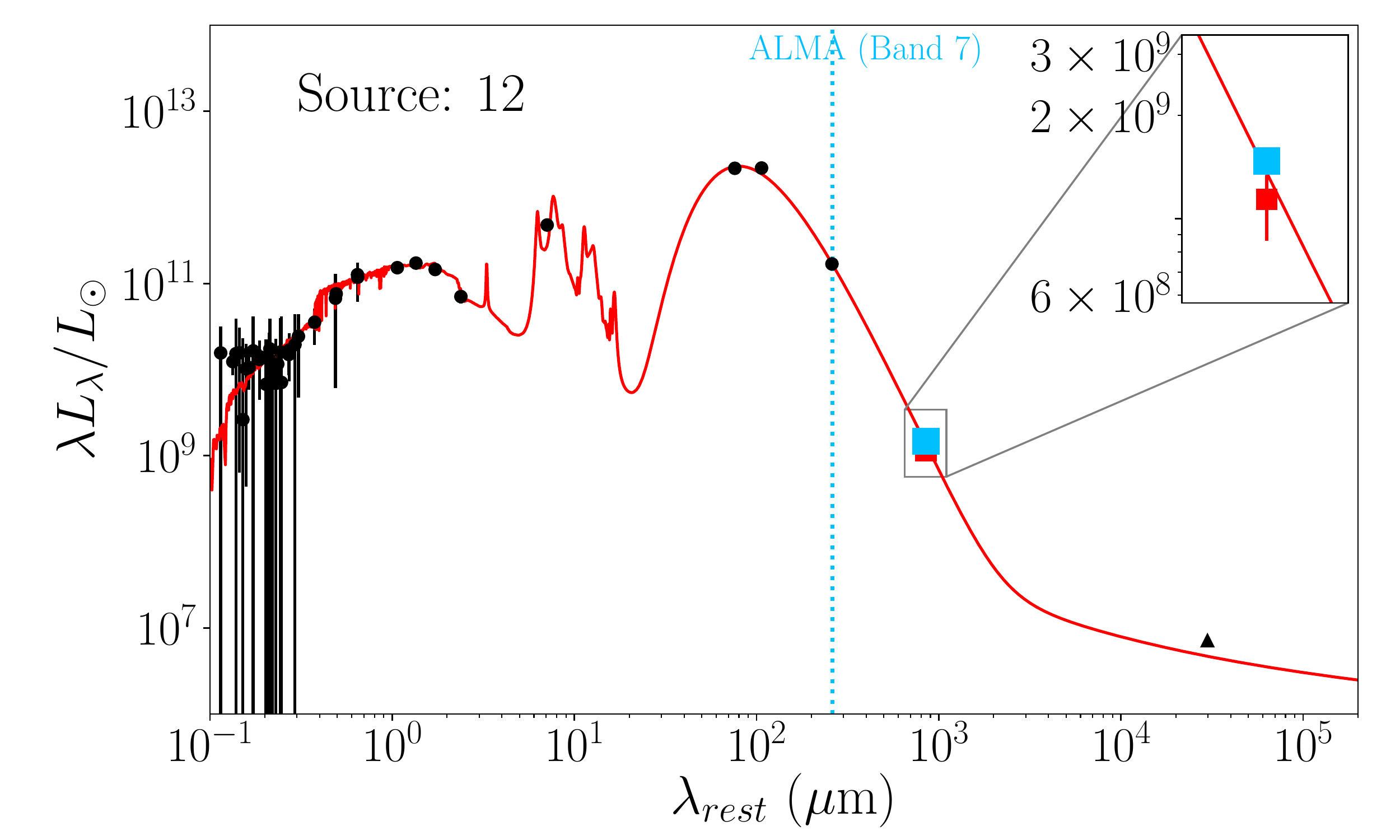}
		~
		\includegraphics[width=0.45\columnwidth]{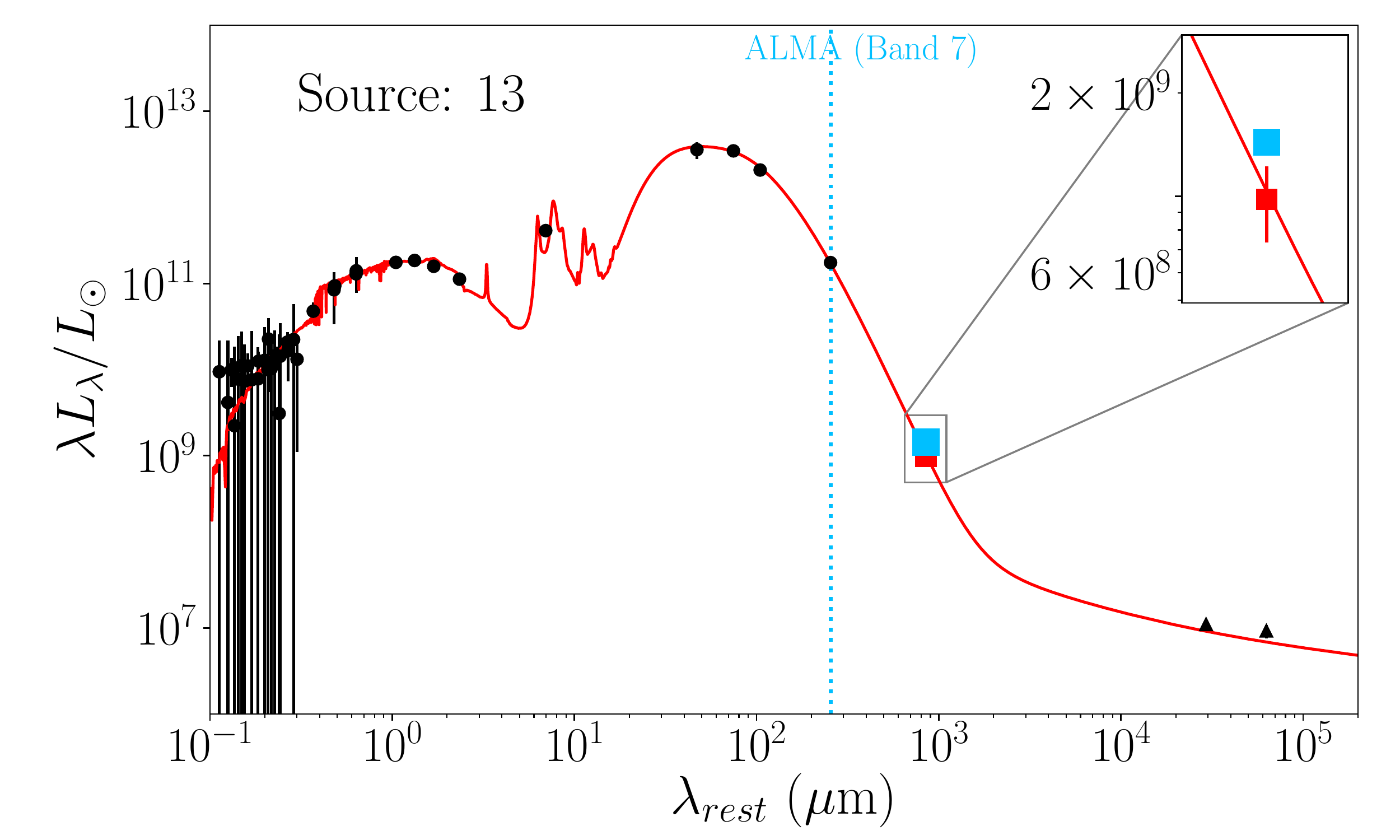}
		\caption{Rest-frame SEDs of the CO(1-0) detected galaxies. The inset focuses on the portion of the rest-frame spectrum around $850\mu m$, comparing the single-band (blue) and SED-derived (red) $\Lref$, where the single-band derived value is calculated from Equation \eqref{eq:L850}. Note that the best-fit SED and value of $\Lref$ shown are based on the standard \texttt{MAGPHYS} assumptions of $\beta=1.5$ and $2.0$ for the warm and cold components, respectively, with the temperature of the cold dust component as a free parameter. The black filled circles represent the photometry used to fit the SEDs whereas the black triangles show the radio fluxes, which were not used to fit the SEDs. \label{fig:SEDs}}		 
	\end{figure*}

	\begin{figure*}[h!]
		\ContinuedFloat
		\centering
		\includegraphics[width=0.45\columnwidth]{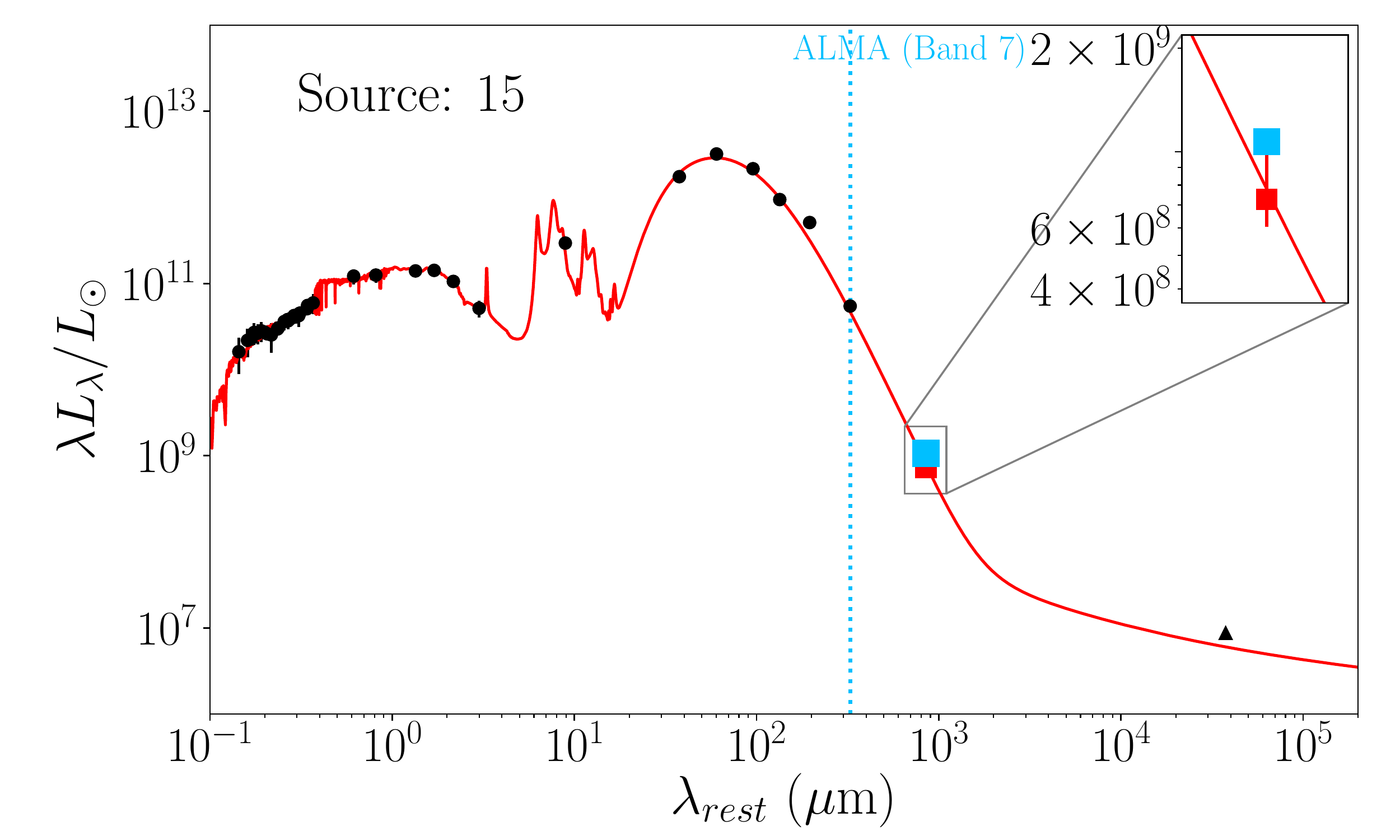}
		~
		\includegraphics[width=0.45\columnwidth]{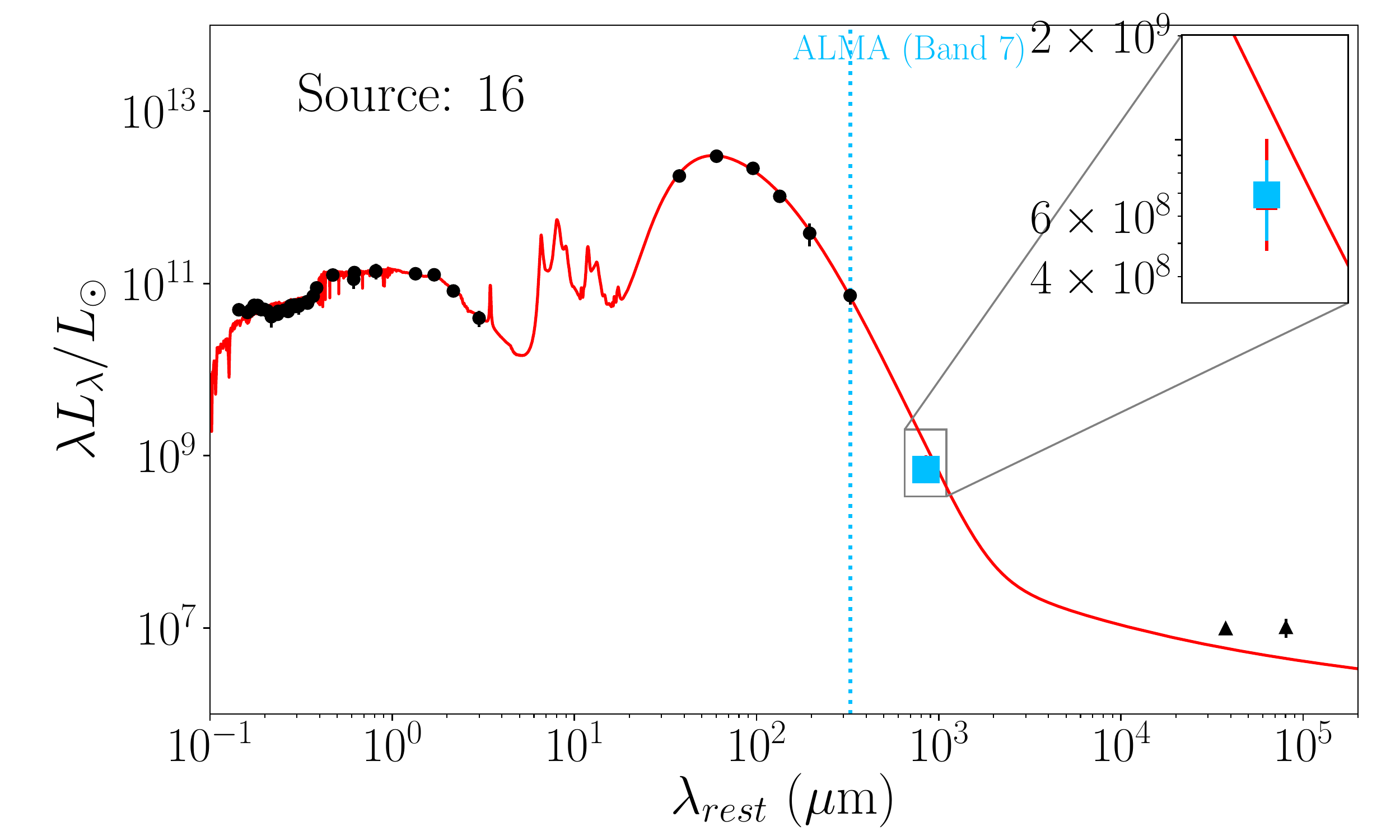}
		\caption{continued}
	\end{figure*}

\end{document}